\documentclass[twocolumn,tighten,times,floatfix,]{aastex631}
\usepackage{amsmath}
\usepackage{xspace,xcolor}
\usepackage{natbib}
\usepackage{outlines}
\usepackage{savesym}
\savesymbol{tablenum}
\usepackage{siunitx}
\restoresymbol{SIX}{tablenum}

\newcommand{\beq}{\begin{equation}}
\newcommand{\eeq}{\end{equation}}
\newcommand{\igm}{\texttt{IllinoisGRMHD}\xspace}
\newcommand{\etk}{\texttt{Einstein Toolkit}\xspace}
\newcommand{\carpet}{\texttt{Carpet}\xspace}
\newcommand{\newrad}{\texttt{Newrad}\xspace}
\newcommand{\baikal}{\texttt{Baikal}\xspace}

\newcommand{\lorene}{\texttt{LORENE}\xspace}
\newcommand{\Fdual}{{}^{*}\!F}

\newcommand{\mb}{\ensuremath{m_{\rm b}}\xspace}
\newcommand{\nb}{\ensuremath{n_{\rm b}}\xspace}
\renewcommand{\ne}{\ensuremath{n_{\rm e}}\xspace}

\newcommand{\ye}{\ensuremath{Y_{\rm e}}\xspace}
\newcommand{\Msun}{\ensuremath{M_{\odot}}\xspace}

\newcommand{\JHU}{Physics and Astronomy Department, Johns Hopkins University, Baltimore, MD 21218, USA}
\newcommand{\ISEF}{ISEF International Fellowship}
\newcommand{\STScI}{Space Telescope Science Institute, Baltimore, MD 21218, USA}
\newcommand{\Idaho}{Department of Physics, University of Idaho, Moscow, ID 83843, USA}

\newcommand{\Goddard}{Gravitational Astrophysics Lab, NASA Goddard Space Flight Center, Greenbelt, MD 20771, USA}
\newcommand{\NASAFS}{NASA Einstein fellow}
\def\eqref#1{Eq.\;(\ref{#1})\xspace}
\newcommand{\sfho}{\texttt{SFHo}\xspace}

\begin{document}
\author[0000-0002-0632-8897]{Yossef Zenati}
\altaffiliation{\ISEF}
\affiliation{\JHU}
\affiliation{\STScI}

\author[0000-0002-2995-7717]{Julian~H.~Krolik}
\affiliation{\JHU}

\author[0000-0002-4541-8553]{Leonardo~R.~Werneck}
\affiliation{\Idaho}

\author[0000-0002-6838-9185]{Zachariah~B.~Etienne}
\affiliation{\Idaho}
\affil{Department of Physics and Astronomy, West Virginia University, Morgantown, WV 26506}
\affil{Center for Gravitational Waves and Cosmology, West Virginia University, Chestnut Ridge Research Building, Morgantown, WV 26505}

\author[0000-0003-3547-8306]{Scott~C.~Noble}
\affiliation{\Goddard}

\author[0000-0003-2333-6116]{Ariadna~Murguia-Berthier}
\altaffiliation{\NASAFS}
\affil{Center for Interdisciplinary Exploration and Research in Astrophysics (CIERA), 1800 Sherman Ave., Evanston, IL 60201, USA}

\author[0000-0002-2942-8399]{Jeremy~D.~Schnittman}
\affiliation{\Goddard}

\title{The Dynamics of Debris Disk Creation in Neutron Star Mergers}

\correspondingauthor{Yossef Zenati}
\email{yzenati1@jhu.edu}

\date{March 2024}

\begin{abstract}
The detection of GW170817/AT2017gfo inaugurated an era of multimessenger astrophysics, in which gravitational wave and multiwavelength photon observations complement one another to provide unique insight on astrophysical systems.
A broad theoretical consensus exists in which the photon phenomenology of neutron star mergers largely rests upon the evolution of the small amount of matter left on bound orbits around the black hole or massive neutron star remaining after the merger. Because this accretion disk is far from inflow equilibrium, its subsequent evolution depends very strongly on its initial state, yet very little is known about how this state is determined. Using both snapshot and tracer particle data from a numerical relativity/MHD simulation of an equal-mass neutron star merger that collapses to a black hole, we show how gravitational forces arising in a non-axisymmetric, dynamical spacetime supplement hydrodynamical effects in shaping the initial structure of the bound debris disk. The work done by hydrodynamical forces is ${\sim}10$ times greater than that due to time-dependent gravity. Although gravitational torques prior to remnant relaxation are an order of magnitude larger than hydrodynamical torques, their intrinsic sign symmetry leads to strong cancellation; as a result, hydrodynamical and gravitational torques have comparable effect. We also show that the debris disk's initial specific angular momentum distribution is sharply peaked at roughly the specific angular momentum of the merged neutron star's outer layers, a few $r_g c$, and identify the regulating mechanism.

\end{abstract}


\section{Introduction}

Neutron star mergers have long attracted significant interest for a variety of reasons. For several decades, they have been considered the most plausible sources of both short gamma-ray bursts and r-process nucleosynthesis \citep{Lattimer_Schramm74, Eichler+89, Rosswog_Ramirez-Ruiz02, Nakar07, Metzger+10_KN, Baiotti&Rezzolla17, Cowan2021}, and the discovery of GW170817 \citep{Abbott17:gw, Abbott17a, Abbott17b, Villar17} has dramatically increased interest in these aspects of them \citep{Goldstein17, Baym+18, Alford_Han_Schwenzer19, Baiotti2019, Beniamini+22}.

The prospect of kilonovae associated with neutron star mergers \citep[e.g,][]{Li98, Kulkarni05, Metzger+10_KN, Roberts+11_KN, Foucart+16, Hotokezaka_Sari_Piran17, Korobkin+20}---and the actual observation of one associated with GW170817---prompted further work 
\citep {Kulkarni05, Kilpatrick17:gw, Troja+17Nat, Coulter17, Guidorzi17, Margutti+17, Breschi+21, Pian2023}.
In addition, the gravitational wave signal from such mergers has provided strong constraints on the properties of nuclear matter \citep{Douchin&Haensel01, Balberg+99, Akmal+98, Haensel2003, Oechslin_Janka06, OechslinJanka07PRL, LattimerPrakash07_EoS, Hotokezaka+11PRD, Damour_Nagar_Viilain+12, Bauswein+12_EoS, Bauswein+13, Hotokezaka+13a, Radice+18_NSEoS, Kiuchi+2019, Dietrich+21, Breschi+21, Werneck+23, Rosswog&Korobkin24}.
Beyond all of these reasons to study them closely, strong radio and X-ray emission has been observed for a number of years following GW170817 \citep{Troja+17Nat, Hallinan17, Alexander2017, Alexander2018, Mooley+18a_Nat, Mooley2018b, Margutti18, Margutti_Chornock21, Balasubramanian+21_radio}.

Although calculations of neutron star merger dynamics all agree that the great majority of the neutron stars' mass is retained within the black hole or long-lived neutron star remnant, and a small fraction of the mass outside the remnant is propelled outward at mildly relativistic speeds \citep{Davies+94, Rosswog2002, Rosswog&Davies02, Hotokezaka+13a, Bauswein+13, Kiuchi+14, Fernandez+15, Sekiguchi+16, Foucart+16, Lehner+16, Radice+16, Ruiz+16, Dietrich+17, Fujibayashi+18, Radice_Bernuzzi+18, Ruiz+21, Vincent+20, Ciolfi&Kalinani20, Sarin_Lasky21, AriM+21, Nedora+21, Most-Quataert2023, zenati+23BNS, Combi_Siegel23, Camilletti+2024}, a large fraction of all the observable properties---the $\gamma$-ray burst, the nucleosynthesis, and at least part of the kilonova, as well as possibly the long-term radio and X-ray emissions---have their origins in the mass left in bound orbits close to the remnant. Because there is no particular reason to assume that this material is distributed with a radial profile corresponding to a state of inflow equilibrium, its subsequent evolution depends strongly on this initial state. Nonetheless, despite the importance of this bound debris to so many aspects of these events, little is known about the dynamics that transport it from the neutron stars, establishing its initial state. It is the goal of this paper to elucidate this process.

In \cite{zenati+23BNS} we showed how the material escaping from neutron star merger remnants is selected, demonstrating, among other things, that it originates from a wide range of radii within the original neutron stars.
This effort was significantly aided by the use of simulation data describing both the motion of tracer particles and snapshots of fluid properties. In the present work, we will extend this technique.

In particular, we will follow the history of the two quantities characterizing orbits in stationary axisymmetric spacetimes (the conserved specific energy $u_t$ and the conserved angular momentum $u_\phi$) in order to identify the mechanisms resulting in their values at the time the debris disk is formed.%
\footnote{The thermal state of the matter determines the disk's vertical thickness, but we ignore it here because thermal equilibrium is established much faster than inflow equilibrium, so the effects of the initial mass profile are much longer-lasting than the effects of the initial thermal profile.}
When the merger remnant is first formed, the surrounding spacetime is neither time-steady nor axisymmetric; during that period, gravity can exert torques and change the energy of orbits. Additionally, hydrodynamic forces, i.e., pressure gradients, can also influence the matter's motion. Much of this paper will be about the competition between these two mechanisms.  A few milliseconds after the black hole formed, the magnetic field grew rapidly in strength (as seen, e.g., in \citet{Kiuchi+18}), reaching a pressure ${\sim}0.1$ times the gas pressure in many places; we leave analysis of its role at this and later times for further work.

\section{Calculation}
\label{sec:model}

\subsection{Overview}
\subsection{Physics treated and equations solved}

We select a merger scenario in which both neutron stars have a baryonic (gravitational) mass of $1.550 M_\odot$ ($1.348 M_\odot$) and follow a circular orbit with an initial separation of \qty{45}{\km} (for further details, see Table~\ref{table_mhd_torus}). To trace their evolution, we employ \igm to solve the equations of MHD,
\begin{align}
  \nabla_{\mu}\left(\nb u^{\mu}\right) &= 0\;, \label{eq:baryon_cons}\\
  \nabla_{\mu}\left(\ne u^{\mu}\right) &= {\cal R}\;, \label{eq:lepton_cons} \\
  \nabla_{\mu}T^{\mu\nu} &= \mathcal{Q} u^{\nu}\;, \label{eq:enmom_cons}\\
  \nabla_{\mu}\Fdual^{\mu\nu} &= 0\;, \label{eq:maxwell}
\end{align}
corresponding to the conservation of the baryon number, conservation of Lepton number, conservation of energy-momentum, and the two homogeneous Maxwell's equations, respectively.
In coordination with the solution of these equations, the BSSN evolution thorn \texttt{Baikal} solves the Einstein Field Equations.

In these equations, $\nb$ ($\ne$) represents the baryon (lepton) number density and $u^{\mu}$ the fluid four-velocity. The net rate of change in the lepton number, ${\cal R}$, is the difference between the creation rates of electron anti-neutrinos and neutrinos. The cooling rate, ${\cal Q}$, quantifies the rate at which neutrinos (and anti-neutrinos) carry energy away from the gas. It is determined using a local emission model combined with a leakage rate formalism
(see \cite{AriM+21} and \cite{Werneck+23} for details about the calculation of ${\cal R}$ and ${\cal Q}$).

The energy-momentum tensor is assumed to be that of a perfect fluid plus an EM contribution,
\beq
  T^{\mu\nu} = \left(\rho h + b^{2}\right)u^{\mu}u^{\nu} + \left(P + \frac{b^{2}}{2}\right)g^{\mu\nu} - b^{\mu}b^{\nu},
\eeq
where \mbox{$\rho = \mb\nb$} is the baryon density, $\mb$ is the mean baryon mass, \mbox{$\Fdual^{\mu\nu}=(1/2)\tilde{\epsilon}^{\mu\nu\rho\sigma}F_{\rho\sigma}$} is the dual of the Faraday tensor $F^{\mu\nu}$, and $\tilde{\epsilon}^{\mu\nu\rho\sigma}$ is the Levi-Civita tensor. The parameter
\mbox{$h = 1 + \epsilon + P/\rho$} is the specific enthalpy, $\epsilon$ is the specific internal energy, and $P$ is the fluid pressure, which is given through a tabulated form of the \sfho EOS for hot degenerate matter~\citep{OConnor:2009iuz}.%
\footnote{The \sfho EOS table was downloaded from \url{http://stellarcollapse.org}.}
The magnetic field is given by $b^{\mu}=(4\pi)^{-1/2}B^{\mu}_{(u)}$, the rescaled magnetic 4-vector in the fluid frame, where
\begin{align}
  B^{0}_{(u)} &= u_{i}B^{i}/\alpha\;,\\
  B^{i}_{(u)} &= \bigl(B^{i}/\alpha + 
               B^{0}_{(u)}u^{i}\bigr)/u^{0}\;.
\end{align}
Here $B^{i}$ is the magnetic field in the frame normal to the hypersurface, $b^2 \equiv b^\mu b_\mu$ is the magnetic energy density, $g_{\mu \nu}$ is the spacetime metric, and $\alpha$ is the lapse function. The lapse, the shift vector, and the metric are all defined on spatial hypersurfaces of constant coordinate time $t$. Ideal MHD (\mbox{$u_{\mu}F^{\mu}=0$}) is assumed throughout.

The update algorithm for \igm is intrinsically conservative. In its formulation, the conserved energy density is
\begin{equation}
\tau = \alpha \sqrt{\gamma} \left(\alpha T^{00} - \rho u^0\right),
\end{equation}
where $\gamma$ is the determinant of the ADM 3-metric. Similarly, the conserved spatial momentum density is
\begin{equation}
S_i = \alpha\sqrt{\gamma} \left[ \left(\rho h + b^2\right)u^0 u_i - b^0b_i \right].
\end{equation}
The corresponding source terms are

\begin{equation}\label{eq:tausource}
\begin{split}
\alpha\sqrt{\gamma}\Bigl[\bigl(T^{00} \beta^i \beta^j &+ 2 T^{0i} \beta^j + T^{ij}\bigr) K_{ij} \\
&-\bigl(T^{00} \beta^i + T^{0i}\bigr) \partial_i \alpha \Bigr]\;.
\end{split}
\end{equation}
for the conserved energy and
\begin{equation}
(1/2) \alpha \sqrt{\gamma} T^{\mu\nu} \partial_i g_{\mu\nu}
\end{equation}
for the conserved 3-momentum. Here $\beta^i$ is the ADM shift vector, and $K_{ij}$ is the extrinsic curvature, $-{\cal L}_n \gamma_{ij}/2$, where ${\cal L}_n$ is the Lie derivative along the time-like normal to the spatial sub-space and $\gamma_{ij}$ is the spatial metric.

In addition to solving these equations, we also follow the locations and velocities of a large number of tracer particles. The initial positions of the tracer particles are selected by randomly choosing points \mbox{$p=(x,y,z)$} within a radius \mbox{$R_{\rm seed}=\qty{75}{\km}$} from the system center of mass and accepting the particle if
\begin{equation}
  \frac{\rho(x,y,z)}{\max\rho} > \zeta\;,
\end{equation}
where $\zeta\in[0,1]$ is a random number. The process is repeated iteratively until all $N_{\rm particles}=50,000$ have been seeded.

The fact that the neutron stars are extremely compact and have central densities many orders of magnitude larger than that of the surrounding gas ensures that the vast majority of the tracer particles are seeded within each neutron star.

Once the positions \mbox{$p_{n} = (x_{n}, y_{n}, z_{n})$} of the particles are known, where \mbox{$n=0,1,\ldots,N_{\rm particles}-1$}, we update them in time using
\begin{equation}
    \frac{dp^{i}_{n}}{dt} = v^{i}(p_{n})\;,
\end{equation}
where $v^{i}(p_{n})$ is the fluid three-velocity $v^{i} = u^{i}/u^{0}$ interpolated to the particle's position $p_{n}$. The tracer particle positions are updated every 16 local time steps, and the three velocities are obtained using fourth-order Lagrange interpolation.
We will use relativistic units throughout this paper. Length is then measured in terms of the gravitational radius $r_g \equiv \frac{GM}{c^2}$, the units of specific angular momentum $u_{\phi}$ are $r_g\ c$, and the units of specific energy $u_t$ are $c^2$.

\subsection{Numerical setup}

We performed our simulation using \igm, which is part of the \etk~\citep{Loffler:2011ay,roland_haas_2022_7245853},%
\footnote{See \url{https://github.com/zachetienne/nrpytutorial}.}
applying it to a Cartesian grid with adaptive mesh refinement (AMR) provided by \carpet~\citep{Schnetter:2003rb}. Initially, the grid consists of eight refinement levels, differing by factors of two, such that the resolution at the finest refinement level is ${\approx}185$\,m. Once the minimum value of the lapse function drops below 0.1, indicating that black hole formation is imminent, two additional refinement levels are added, bringing the resolution of the finest refinement level to ${\approx}\qty{46}{\m}$.

At the outer boundary, located at ${\approx}\qty{5670}{\km}$, we apply radiation boundary conditions to the metric quantities using \newrad~\citep{Alcubierre:2002kk, Loffler:2011ay} and a simple copy boundary condition for the MHD fields.

The spacetime is evolved using \baikal~\citep{roland_haas_2022_7245853}, while the MHD fields are evolved using a newly developed version of \igm~\citep{Werneck+23} that supports finite-temperature, microphysical EOSs, and neutrino physics via a leakage scheme. In addition to the MHD fields, we evolve the entropy by assuming that it is conserved~\citep[see e.g.,][for a similar strategy]{Noble:2008tm}, an approximation that, while poor at shocks, allows us to use the entropy as a backup variable during primitive recovery~\citep[see][for more details]{Werneck+23}.

Initial spacetime data for our equal-mass binary are constructed using \lorene~\citep{Gourgoulhon+01, Feo+16,lorene_website}.  The stars' internal structure is, as usual, computed from a solution of the TOV equation assuming our equation of state (SFHo).
Each neutron star is also seeded with a strong poloidal magnetic field~\citep[see e.g., Appendix A of][]{Etienne+15} such that $\max\bigl(P_{\rm mag}/P\bigr) = \max\bigl(b^{2}/2P\bigr) = 10^{-4}$, corresponding to $\max\sqrt{b^2}=\qty{5.05e15}{G}$.

\begin{table}[!htb]
\begin{center}
\begin{tabular}{c|c} 
 \hline\hline
 \textbf{Parameter} & \textbf{Value} \\
 \hline
 \multicolumn{2}{c}{\textbf{Initial data}} \\
 \hline
 EoS & \sfho \\
 NS baryonic mass & $1.550\Msun$ \\
 NS gravitational mass & $1.348\Msun$ \\
 NS radius $R_*$ & \qty{9.3}{\km} \\
 Initial separation & \qty{45}{\km} \\
 $\max\sqrt{b^{\mu}b_{\mu}}$ & $\qty{5.05e15}{G}$ \\
 Number of tracer particles & $50000$ \\
 \hline
 \multicolumn{2}{c}{\textbf{Post-merger}} \\
 \hline
 $\chi = a/M$ & $0.795$ \\
 BH irreducible mass, $M_{\rm irr}$ & $2.444\Msun$ \\
 BH mass, $M_{\rm BH}$  & $2.726\Msun$ \\
 $\langle\ye\rangle$ & $0.122$ \\
 [0.5ex] 
 \hline\hline
 \end{tabular}
 \caption{Simulation parameters for an equal-mass, magnetized binary neutron star merger performed with \igm using a microphysical, finite-temperature EOS and a neutrino leakage scheme. Here, $\chi$ is the BH dimensionless spin parameter and $\langle\ye\rangle$ is the mass-weighted mean \ye at $r=\qty{60}{\km}$ and $t\approx\qty{17}{ms}$.}
 
  \label{table_mhd_torus}
  \end{center}
\end{table}
Where polar coordinates are more easily interpretable, we define such a system through the simple coordinate transformation
$r^2= (x^2 + y^2 + z^2)$, $\cos\theta = z/r$, $\tan\phi = y/x$, and the $xy$-plane is identical to the initial binary orbital plane. In these coordinates, the polar component of angular momentum is $u_\phi \equiv xu_y - yu_x$, which becomes a conserved quantity when the spacetime is axisymmetric.

\subsection{Diagnostics of tracer particle dynamics }

\igm records the spatial positions and the contravariant and covariant 4-velocities for all the tracer particles every \qty{1.23e-3}{\ms}, writing them to a file every \qty{9.856e-3}{\ms}. Because the probability density for their initial conditions is proportional to density, they can be thought of as, on average, representing equal amounts of mass.

Of the 50,000 total tracer particles, only 2416 can be found outside the black hole immediately after its horizon forms, and only 287 survive all the way to $t=30$~ms, $\approx 13$~ms after creation of the black hole.  We call these 287 "survivor particles".

When we plot dynamical properties of the tracer particles, we do so at one of the particle output times. The coordinates used by \igm are 3+1 ADM Cartesian; they are translated into spherical coordinates by the coordinate transformation $r = (x^2 + y^2 + z^2)^{1/2}$, $\cos\theta = z/r$, $\tan\phi = y/x$ with appropriate quadrant distinctions.

When we consider the forces acting on these tracer particles at a certain time, we need to situate the tracer particles relative to the ``fluid" data produced by \igm. This data comprises maps of its several grid functions: the metric elements, the rest-mass density, the pressure, the fluid coordinate velocities, etc. However, interpolation of the 3D AMR grid used by \igm into a uniform grid is available only on three planes: $x$-$y$, $x$-$z$, and $y$-$z$. 
We are fortunate, given this constraint, that the great majority of the particles ($\gtrsim 80\%$) have displacements from the plane less than the merged neutron star's mass scale height. We therefore project their positions into the $x$-$y$ plane and use the interpolated fluid data in that plane. This procedure incurs errors at the tens of percent level, so results resting on it have an uncertainty $\sim 10\%$. Because the distinctions we make are based on much larger contrasts, this is an acceptable error level.

In particular, we define the gravitational torque density by
\begin{equation}
{\partial_t j}_{\rm grav} \rho = \alpha \sqrt{\gamma} T^{\mu\nu} \left(x \partial_y - y \partial_x\right)  g_{\mu\nu},
\end{equation}
where $\alpha$ is the lapse function and $\gamma$ is the determinant of the spatial 3-metric. We also make the approximation that the dominant term in $T^{\mu\nu} \partial_i g_{\mu\nu}$ is the one corresponding to $\mu=\nu=0$ because $u^t \gtrsim 1$ is generally several times larger than $u^i$. The gravitational torque per unit rest mass is then
\begin{equation}
{\partial_t j}_{\rm grav} =  \left(x \partial_y - y \partial_x\right)  g_{tt},
\end{equation}
while the torque per unit rest mass
associated with pressure forces is
\begin{equation}
{\partial_t j}_{\rm hydro} = \left(y \partial_x P - x \partial_y P\right)/\rho.
\end{equation}
In both cases, we approximate the relevant inertia by $\rho$ on the grounds that the quantities multiplying it in the stress-energy tensor, $h$ and $u^0$, are different from unity by at most ${\sim}\qty{10}{\percent}$.

The source term for the energy-like primitive variable in \igm (see eqn.~\ref{eq:tausource}) is rather complex. In order to represent its principal features, we study a proxy, $\partial g_{tt}/\partial t$, for the rate of change of energy per unit rest-mass. This is effectively present within the source term through the relation $g_{tt} = -\alpha^2 + \gamma_{ij}\beta^i \beta^j$.
The rate of energy change, per unit mass, due to hydrodynamic forces is $-(u^x\partial_x P + u^y\partial_y P)/\rho$. Just as for the torque, we evaluate these quantities at the position of each tracer particle projected into the binary orbital plane. 

\subsection{Physical interpretation of coordinate-dependent quantities}

The origin of the coordinate system created by \igm is set to the center-of-mass of the binary at $t=0$, but the slicing of the spacetime changes throughout the calculation due to the gauge evolution. Formally, the relationship between these coordinates and, for example, the Cartesian analog of a Boyer-Lindquist system corresponding to the black hole created, is therefore difficult to define. However, in practice this is less of a problem than it might seem. We have measured the relative velocity between the origin and the black hole, and it indicates a steady coordinate drift in which the black hole has a velocity relative to the origin $\simeq - 7 \times 10^{-4}c \hat y$. Departures from axisymmetry do exist (see Fig.~\ref{fig:dj_after}), but their illustration in this figure also demonstrates that these departures have a fractional magnitude less than $\sim 1\%$ everywhere outside $r \sim 5r_g$.

The most pertinent question regarding gauge choices is whether they obscure the interpretation of the quantities $u_t$ and $u_\phi$. Time-steadiness of the spacetime guarantees that $u_t$ is conserved; axisymmetry of the spacetime guarantees that $u_\phi$ is conserved. These quantities can be interpreted as the values of energy and angular momentum as $r\to\infty$ provided the spacetime approaches Minkowski at large $r$. To verify that this condition is satisfied, we have measured the value of the conformal factor necessary to translate a Schwarzschild spacetime to the gauge of our spacetime at large distance from the black hole: at such distances it agrees with the conformal factor of the analytic ``trumpet"
solution~\citep{BaumgarteNaculich2007} to within a fraction of a percent.

\section{Results} \label{sec:results}

\subsection{Overview}

\begin{figure*}
\includegraphics[width=0.99\linewidth]{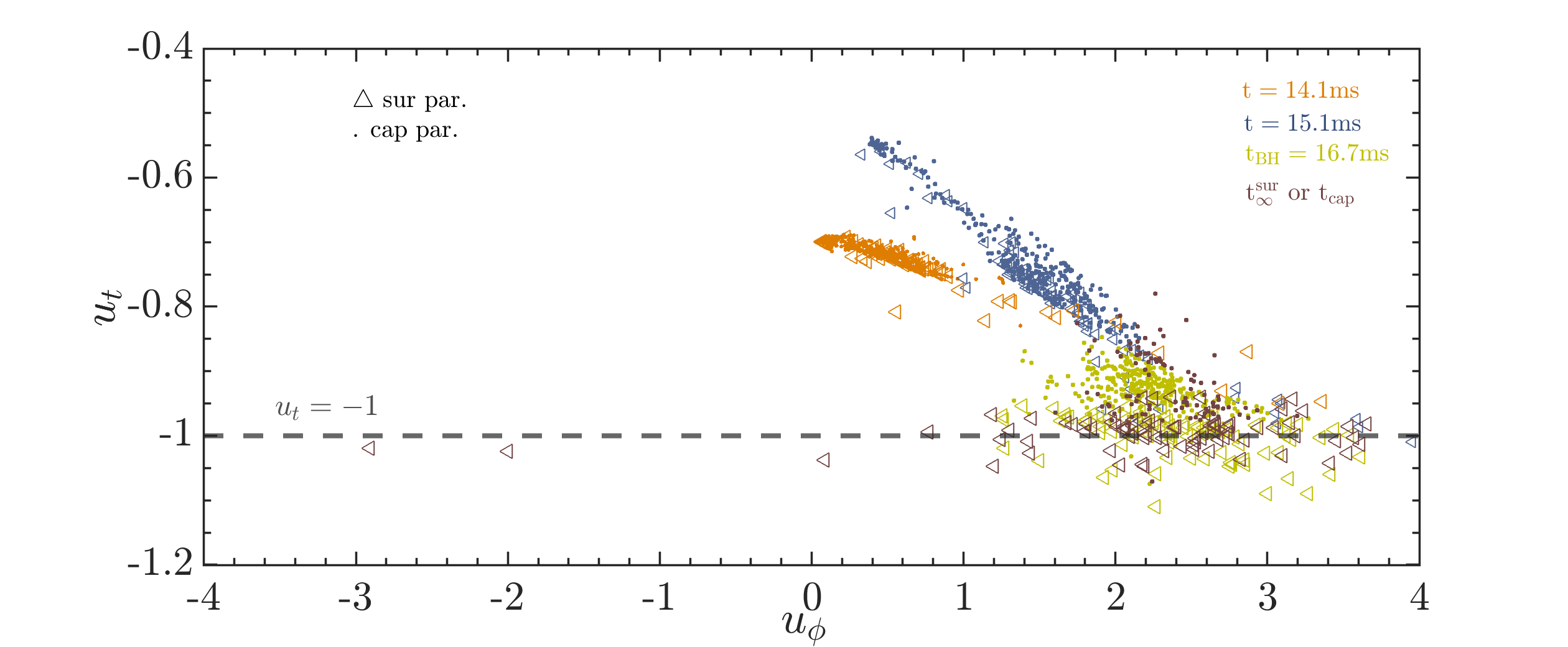}
\caption{The tracer particles' phase space distribution at \mbox{$t=\qty{14.1}{\ms}$} (orange; shortly after the neutron stars first touch), \qty{15.1}{\ms} (blue; while the merged neutron star is collapsing), \qty{16.7}{\ms} (olive; when the event horizon forms with total survivor particles 2416), and at late time (purple). For particles that remain outside the black hole (triangles), ``late time'' refers to the end of the simulation at $t=\qty{30}{\ms}$
for particles that are captured (dots), it is the moment at which they pass through the event horizon. Note that with our metric signature, $u_t = -1$ (dash dark gray line) indicates total energy exactly equal to the rest mass, and $u_t < -1$ indicates the particle is gravitationally unbound. For the sake of image clarity, we plot a randomly selected subsample comprising approximately \qty{28}{\percent} of the tracer particles.}

\label{fig:ut_uphi_compare_distribution}
\end{figure*}

As has been demonstrated by many previous simulations of neutron star mergers \citep{Balasubramanian+21_radio, Hotokezaka18, Hajela+22}, even before the two stars touch, tidal gravity accelerates small amounts of mass to speeds ${\gtrsim}0.4 c$; this gas is generally referred to as the ``fast ejecta''. Once the stars are in contact, pressure forces along the contact surface squeeze matter outward, roughly parallel to the orbital axis. However, only a small fraction of the matter expelled from the merged star's interior escapes the remnant, whether it results in a long-lived neutron star or a black hole \citep{Most+21_FastEjecta, Hajela+22, zenati+23BNS, Combi_Siegel23}. In fact, in the simulation we analyze here, most of the matter that evades the quick collapse to a black hole (only ${\sim}\qty{4}{\ms}$ after the neutron stars touch, at simulation time \qty{16.7}{\ms}) is nonetheless captured within a few more milliseconds; only ${\approx}0.016 M_\odot$ remains in orbit for a longer duration. It is this matter that is the focus of our study, as it forms the debris disk responsible for all the observable phenomena other than gravitational waves: $\gamma$-ray bursts, the optical/IR kilonova, persistent X-ray and radio emissions, and heavy-element nucleosynthesis.

\begin{figure*}[t!]
\centering
\includegraphics[width=0.99\linewidth]{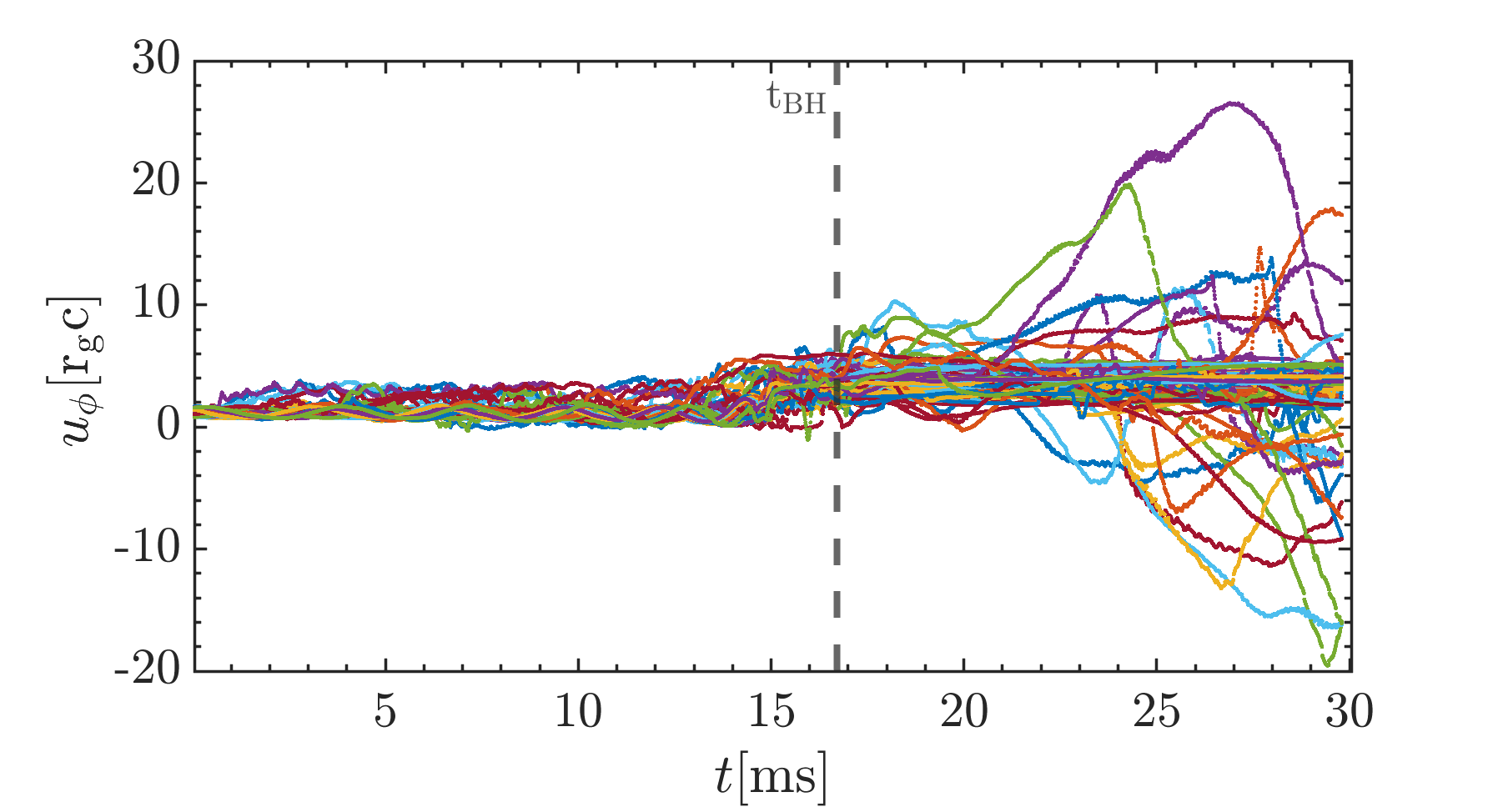}
\includegraphics[width=0.99\linewidth]{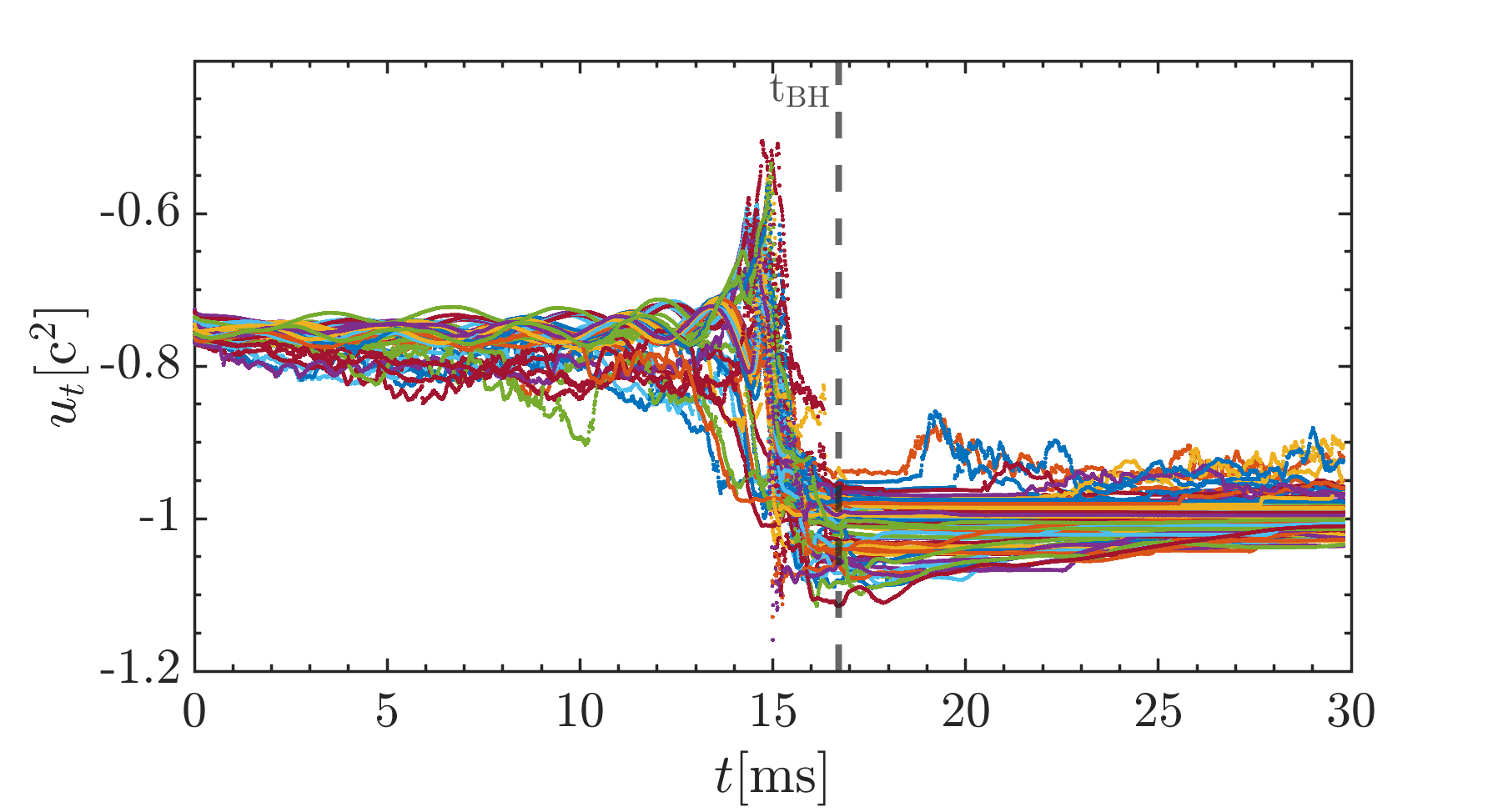}
\caption{Evolution of $u_{\phi}$ (A: upper panel) and $u_t$ (B: lower panel) for the 287 survivor particles, including both bound and unbound particles. The vertical dashed line marks the moment the black hole forms.}
\label{fig:uphi_ut_all}
\end{figure*}

The dynamical properties of the debris are encapsulated in two quantities that both become conserved for test-particle motion when the remnant's final state is reached (in the simulation we report, a Kerr spacetime containing a black hole with spin parameter $a/M=0.795$). These are the orbital energy in rest-mass units $u_t$ and the angular momentum component parallel to the spin axis $u_\phi$. With our metric signature, the gravitational binding energy of a particle is $1 + u_t$. As we will see, $u_t$ is the critical determinant of whether a particle remains outside the black hole and enters the debris disk, while $u_\phi$ (for surviving particles)
determines the scale of the particle's orbit.

Figure~\ref{fig:ut_uphi_compare_distribution} shows four snapshots of the distribution of our tracer particles in $u_\phi \times u_t$ phase space. In very coarse terms, the entire population of particles follows a similar path. In the first 1 -- 2 milliseconds after the two stars touch ($\approx 13.3 \rm ms$ \citet{zenati+23BNS}) the particles having low angular momentum ($u_\phi \lesssim 1.5$) become more deeply bound, particularly for those with $|u_\phi| \ll 1$. The sense of evolution reverses sharply ${\approx}\qty{1.5}{\ms}$ before the event horizon forms; from then until the creation of the black hole, most of the particles increase in angular momentum and decrease in binding energy. After the black hole is formed, the energy of the remaining particles changes relatively little, but their distribution in angular momentum is considerably broadened. 

\begin{figure*}

\includegraphics[width=0.99\linewidth]{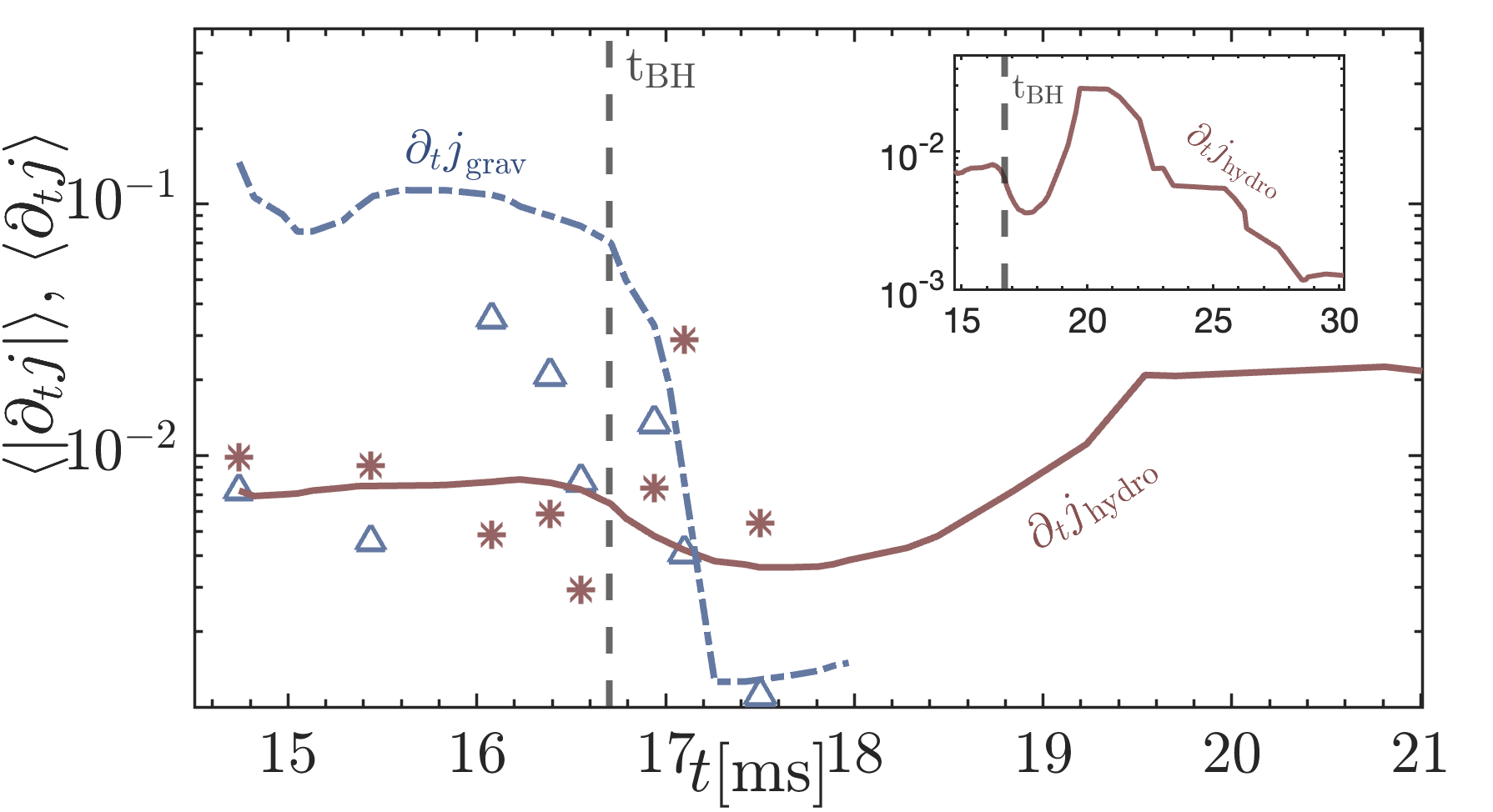}
\caption{Total gravitational and hydrodynamic torques per rest-mass (in units of $c^2$) as a function of time from shortly after the neutron stars touch to ${\approx}\qty{4}{\ms}$ after the black hole is formed. Curves show the absolute value of the gravitational torque (blue dashed line) and the hydrodynamic torque (solid brown line) integrated over the entire volume. Sums of the {\it signed} gravitational and hydrodynamic torques acting on the survivor tracer particles are shown by blue triangles and brown stars, respectively. All of these signed values are positive. The inset shows the longer-term behavior of the integrated magnitude of ${\partial_t j}_{\rm hydro}$. The vertical dashed line marks the moment the black hole forms.}.
\label{fig:J_evolution}
\end{figure*}

\begin{figure*}
\includegraphics[width=0.99\linewidth]{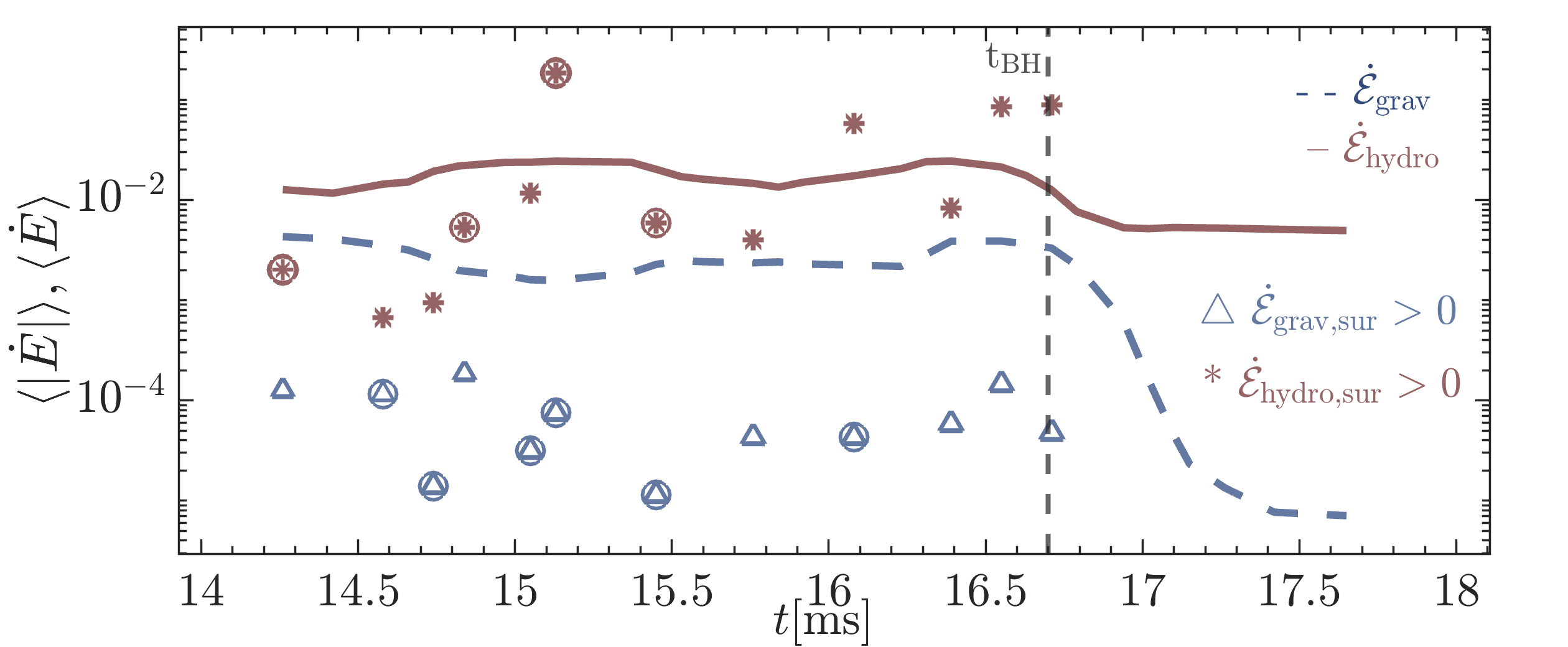}

\caption{Total gravitational and hydrodynamic rates of energy change per rest-mass (in units of $c^3/r_g$) from shortly after the neutron stars touch to ${\approx}\qty{1}{\ms}$ after the black hole is formed. As in Fig.\ref{fig:J_evolution}, curves are integrated absolute values, triangles and stars are the summed signed values. 
Positive signs are shown with simple points; circles around points denote negative sign. Although not shown in this plot, the absolute magnitude of $\dot E_{\rm hydro}$ at later times evolves similarly to $\partial_t j_{\rm hydro}$. The vertical dashed line marks the moment the black hole forms.} 
\label{fig:Edot_evolution}
\end{figure*}

Despite the general trends affecting all the neutron stars' mass, there are clear distinctions between the debris particles and those captured by the black hole from the start. At the beginning of the merger, the binding energies of the entire population of particles range from ${\approx}0.5$ to ${\approx}0.1$, and their angular momenta range from ${\approx}0$ to ${\approx}3$, but with a strong linear correlation between the two, indicating that higher angular momentum is associated with less binding energy. At this stage, although there is overlap between the phase space locations of particles that join the debris disk and those quickly captured into the black hole, there is a strong statistical separation: a large majority of the debris particles have higher angular momentum and are more weakly bound than nearly all the particles destined to be quickly captured by the black hole. By the time the black hole forms, the distinction between escaping particles and captured particles is primarily one of binding energy: all escaping particles have a binding energy ${\lesssim}0.05$, while nearly all captured particles have a binding energy ${\gtrsim}0.05$.

The significant role in binary neutron star phenomenology played by the small fraction of the mass in the debris disk makes examining the evolution of $u_\phi$ and $u_t$ for the small subset of particles (287 out of the initial 50,000) that remain outside the black hole for more than a few milliseconds after collapse particularly worthwhile. The evolution of this subset is portrayed in Figure~\ref{fig:uphi_ut_all}. Consistent with what is shown in Figure~\ref{fig:ut_uphi_compare_distribution}, on average, these particles gain ${\approx}2$--$3 r_g c$ in $u_\phi$ when the neutron stars have merged but not yet collapsed. During this same period, most, but not all, first sharply increase in binding energy and then even more sharply decrease (a minority decrease in binding energy without first increasing). After the black hole forms, the mean values of both the angular momentum and the binding energy remain nearly constant, but the angular momentum distribution widens steadily, including the emergence of a noticeable minority of particles on \textit{retrograde} orbits.

In the remainder of this paper, we will focus on identifying the forces driving the evolution of $u_t$ and $u_\phi$ for the surviving mass, the mass destined to join the debris disk. Departures from axisymmetry in the spacetime can alter $u_\phi$. Time-dependence in the spacetime (which may or may not relate to its degree of axisymmetry) can lead to changes in $u_t$. Beyond these gravitational mechanisms, ordinary hydrodynamic forces, i.e., pressure gradients, can affect both $u_\phi$ and $u_t$. We will distinguish which parts of the phase space evolution are accomplished by each of these mechanisms.

To begin, we illustrate, at a very coarse-grained level, their relative influence on $u_\phi$ in Figure~\ref{fig:J_evolution}. In this figure, the absolute value of the gravitational torque integrated over the orbital plane is compared to the similarly integrated absolute value of hydrodynamic torque as a function of time. 

Until approximately $\qty{0.3}{\ms}$ after the black hole horizon forms, both gravitational and hydrodynamic forces participate, but the integrated absolute magnitude of the gravitational torque is roughly an order of magnitude greater than that of the torque produced by hydrodynamic forces. This sort of comparison underlies several remarks in the literature suggesting that gravitational forces dominate the dynamics of mass ejected from the merging neutron stars \citep{Lovelace+08, Bauswein+12_EoS, Radice+20_AnnRev, Shibata_Kenta19, Shibata+23}. As gravitational wave emission during black hole ringdown rapidly symmetrizes the black hole spacetime, the gravitational torque swiftly declines over the ${\sim}\qty{0.3}{\ms}$ following the black hole's formation. At later times, gravitational forces are negligible, and hydrodynamics governs the evolution of the gas's angular momentum. It is important to note, however, that these conclusions about integrated forces do not necessarily apply locally, and integrating the magnitude of the force hides the impact of sign changes; as we will demonstrate in the next subsection, this last caveat is significant.

On the basis of this sharp change in the character of the dominant forces very shortly after black hole formation, we divide our account of debris dynamics into two epochs, pre- and post-collapse.

\subsection{Pre-collapse}

\begin{figure*}
  \begin{tabular}{ccc}
    \includegraphics[width=0.44\textwidth]{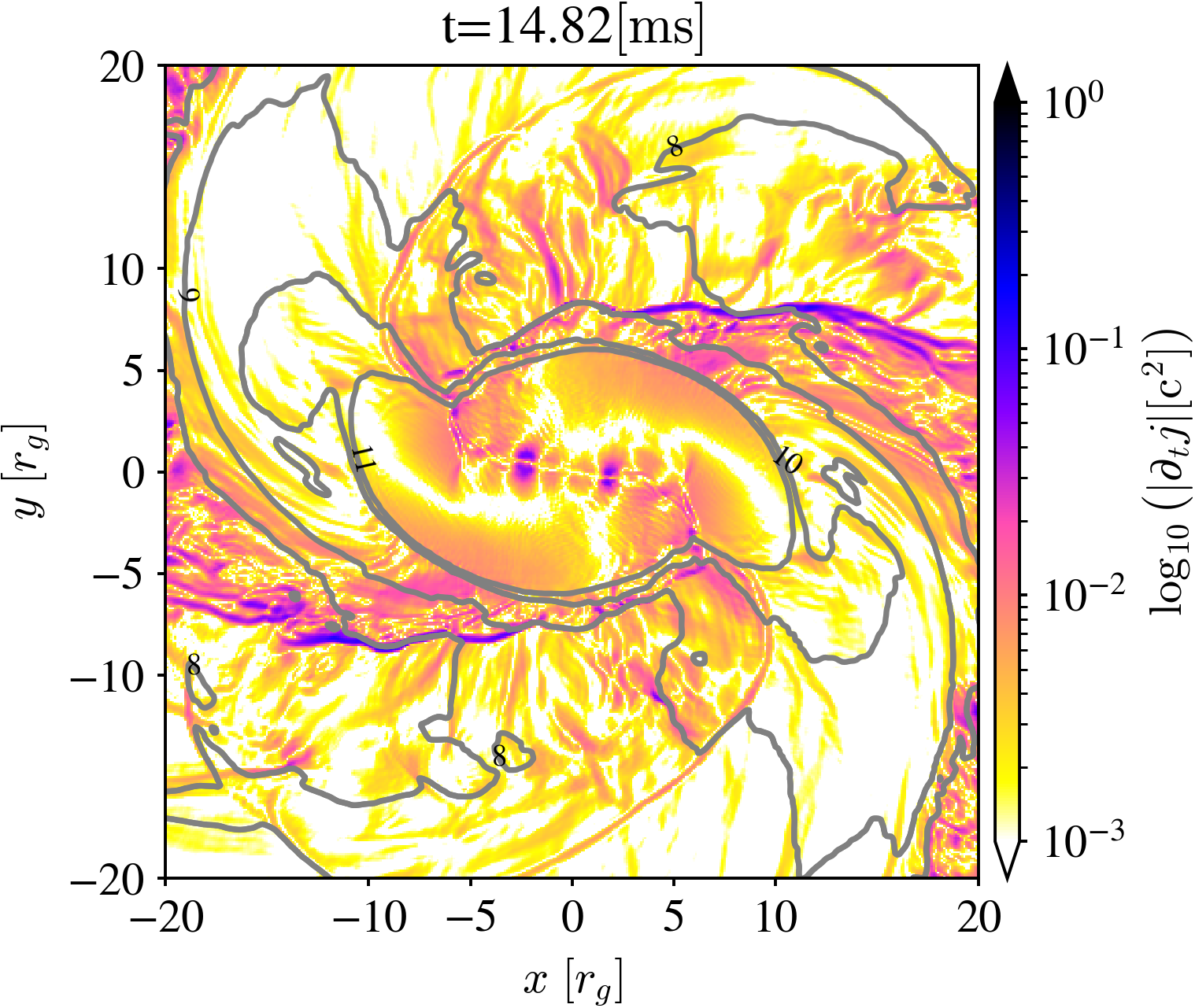} &%
    \includegraphics[width=0.44\textwidth]{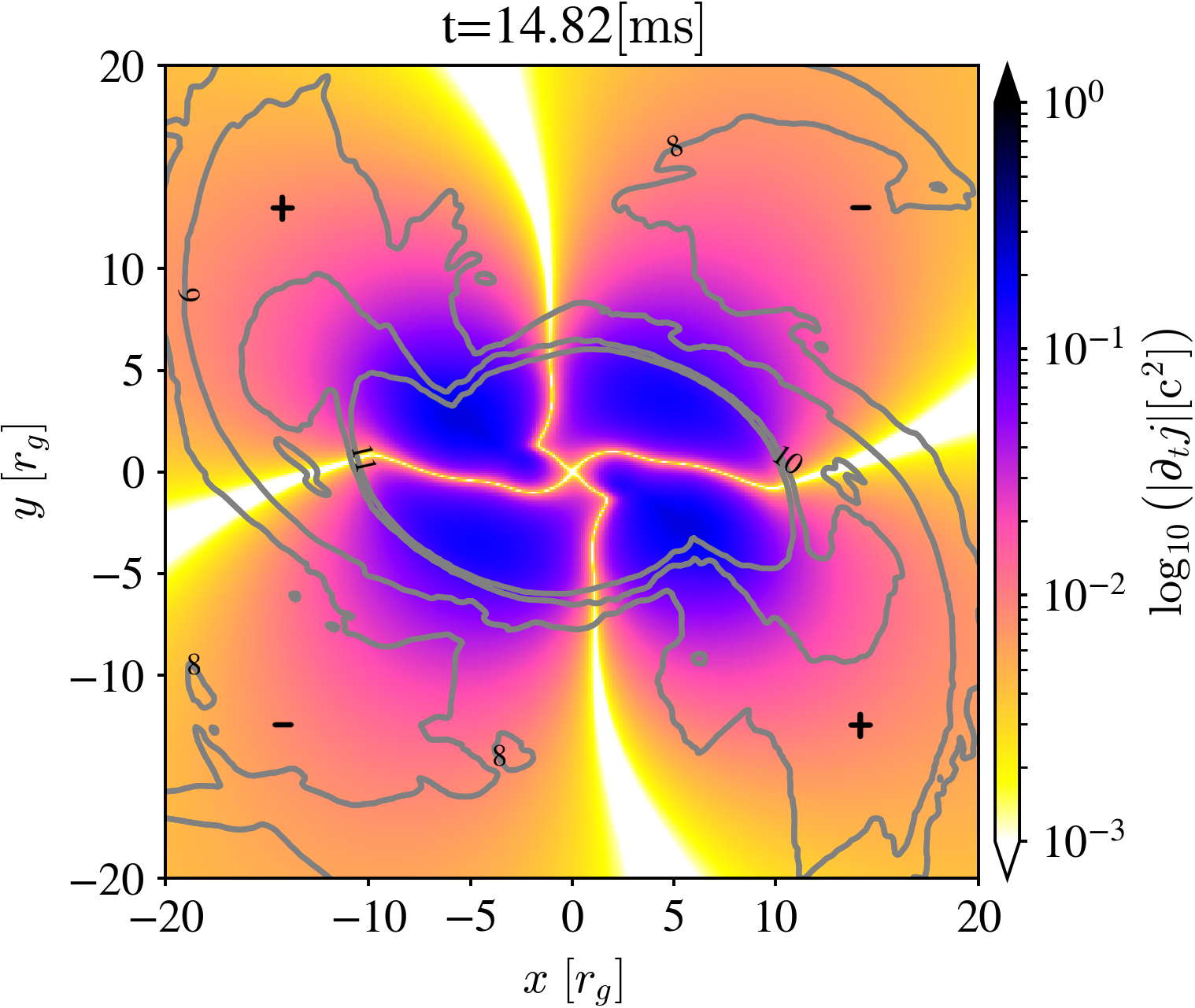} \\ 
    
   \includegraphics[width=0.44\textwidth]{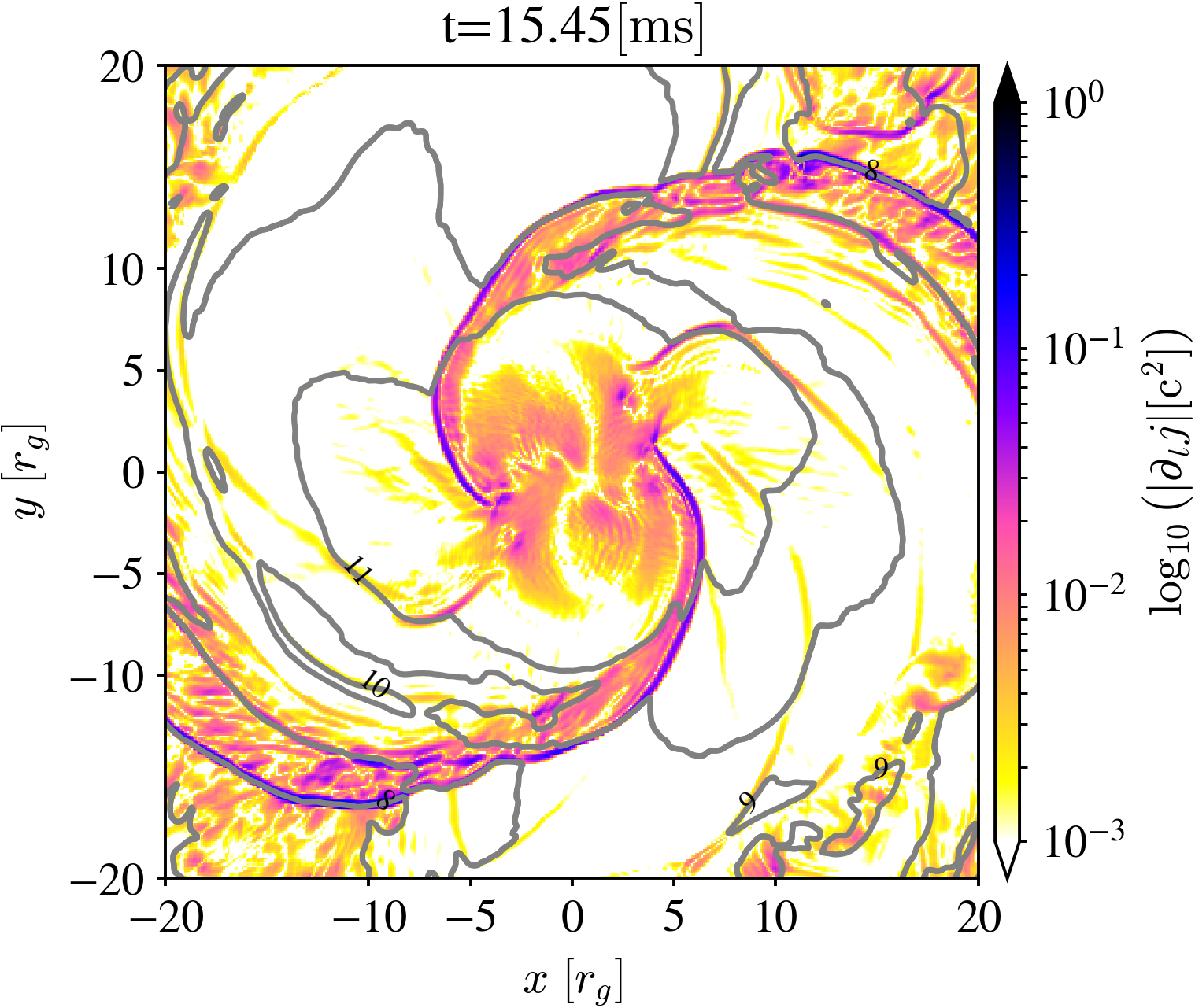} &%
    \includegraphics[width=0.44\textwidth]{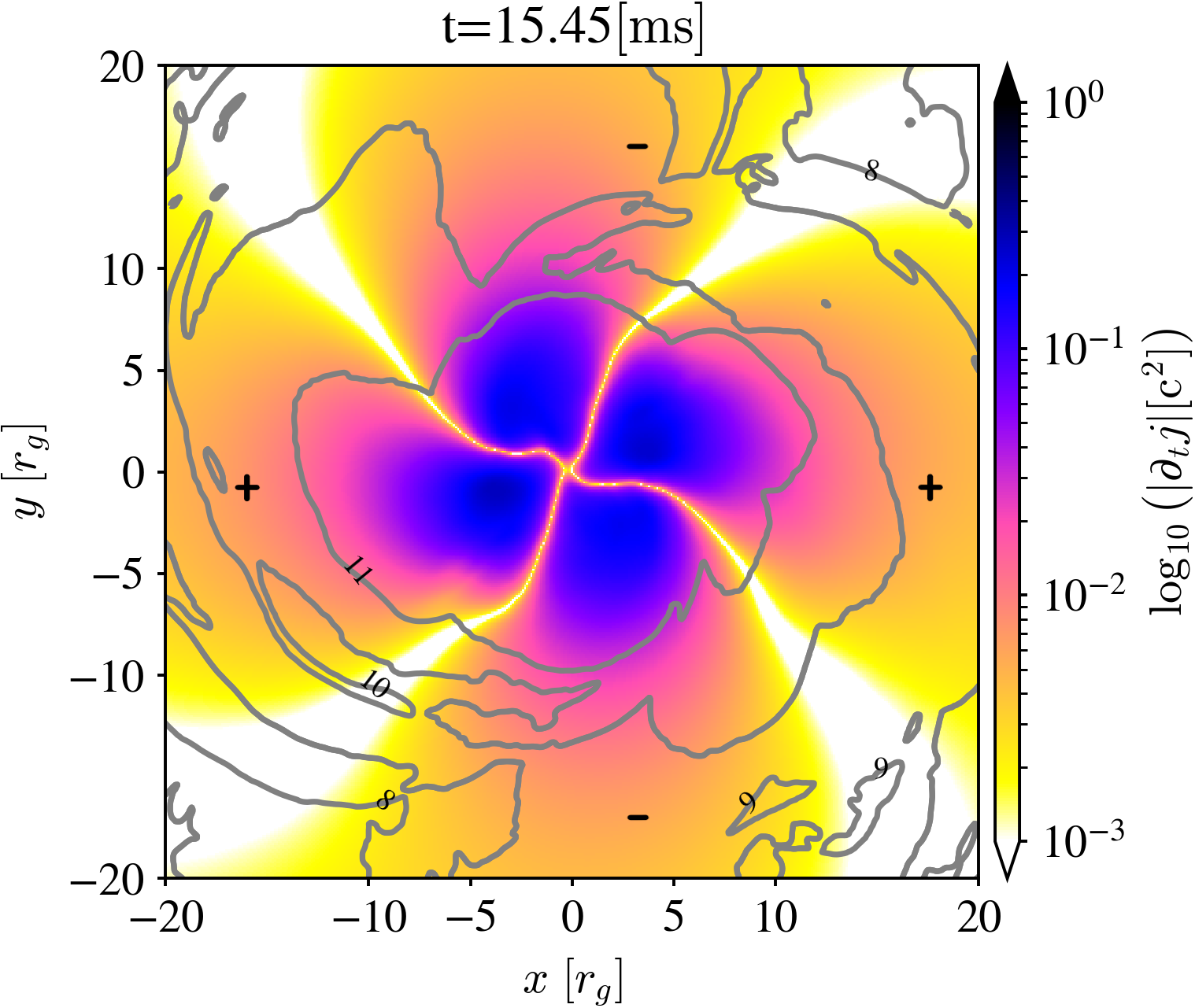} \\ 
    
    \includegraphics[width=0.43\textwidth]{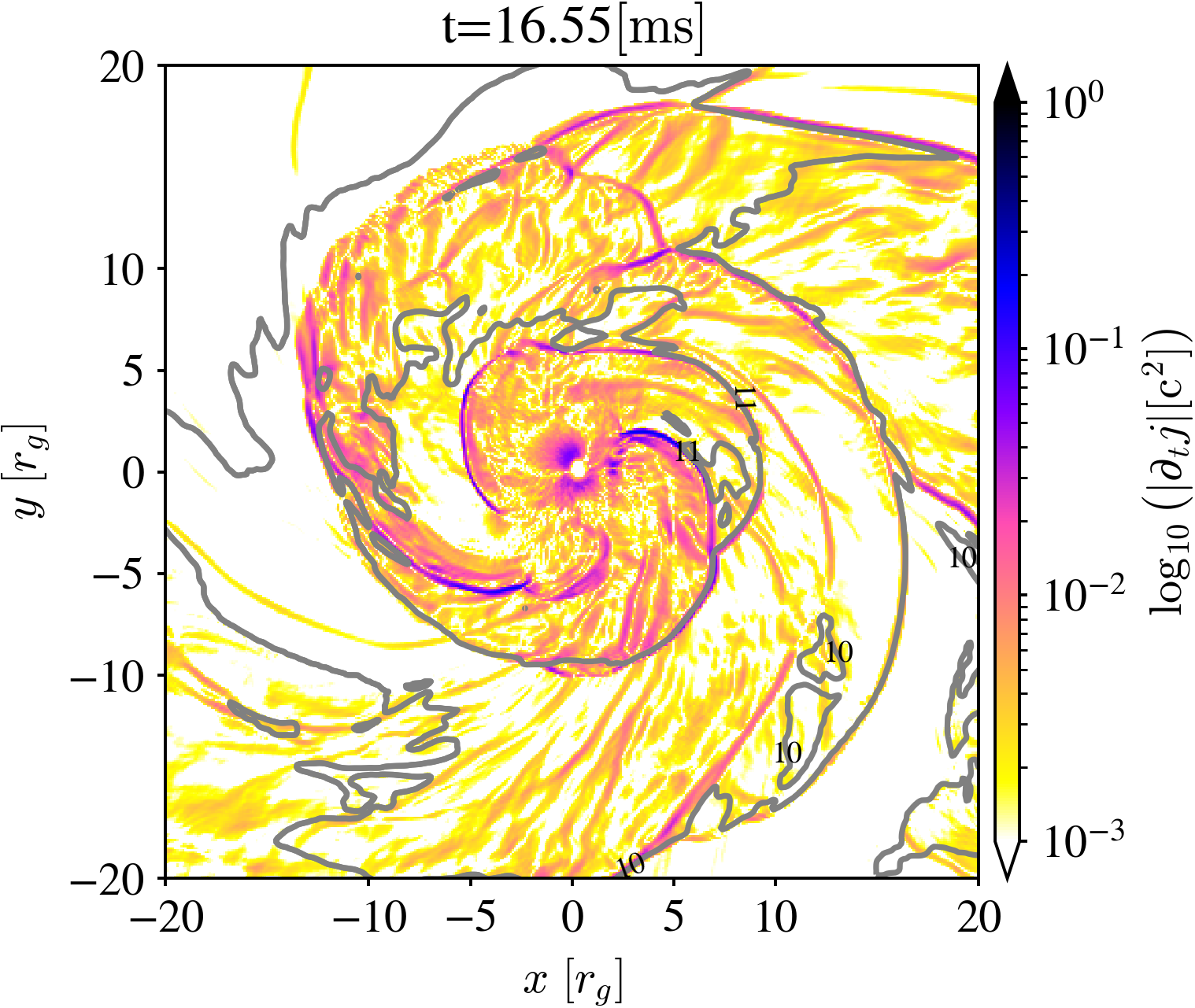} &%
    \includegraphics[width=0.43\textwidth]{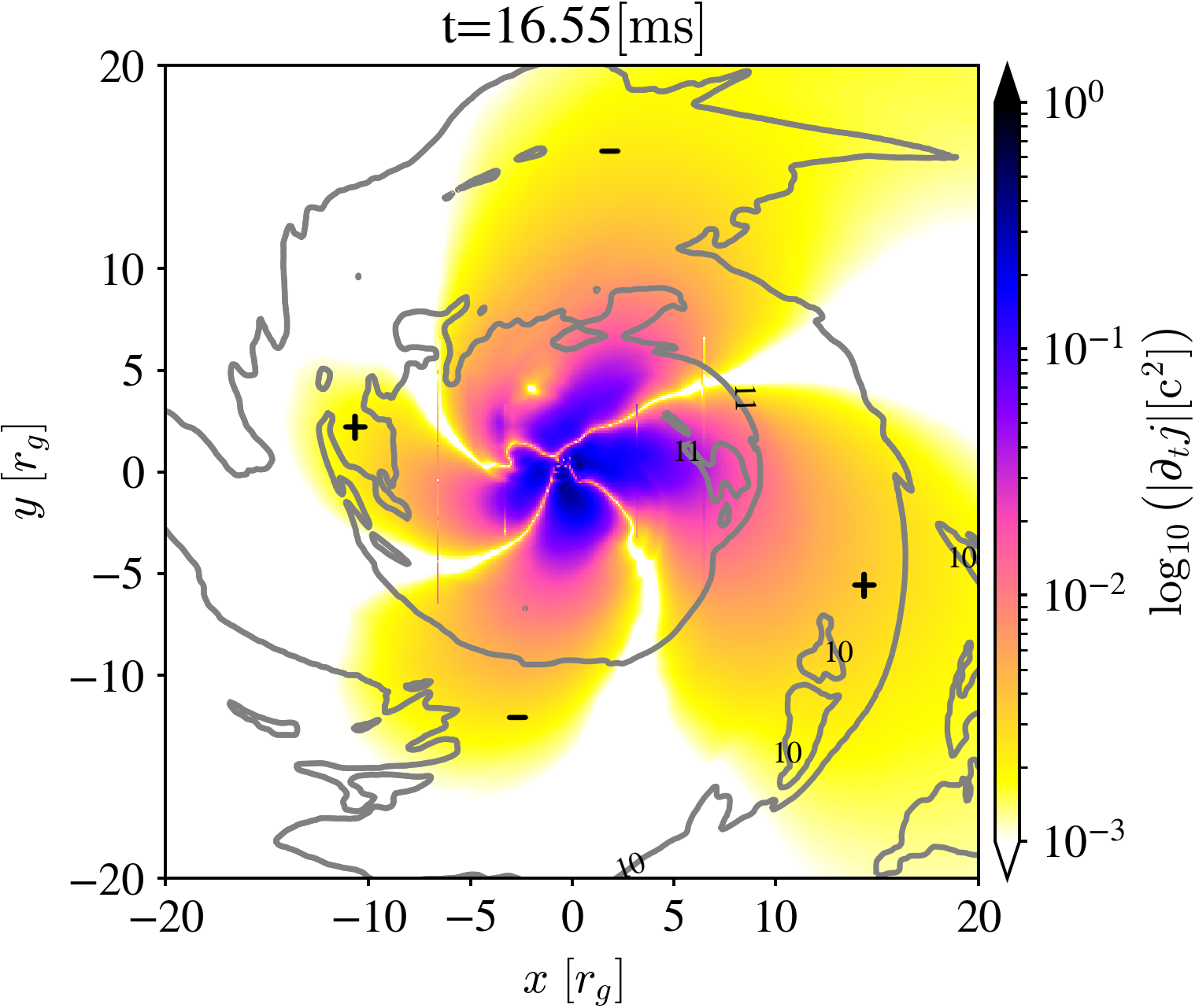} \\
  
  \end{tabular}
  \caption{Evolution of the absolute magnitudes of the gravitational and hydrodynamic torques per unit mass before the binary neutron star collapse to a black hole (color contours) and $\rho$ (gray contours; the separation between gray contour lines is a factor of 10 in density). The left panels show the hydrodynamic torque ($\log |\partial_t j_{\rm hydro}|$), and the right panels show the gravitational torque ($\log |\partial_t j_{\rm grav}|$). Both are in units of $c^2$. From top to bottom, the panels are at times \qty{14.82}{\ms}, \qty{15.45}{\ms}, and \qty{16.55}{\ms}.}
  \label{fig:dj_before}
\end{figure*}

\subsubsection{Angular momentum}

The three rows of images in Figure~\ref{fig:dj_before} illustrate the distribution within the orbital plane of the two sources of torque---azimuthal gravitational force arising from non-axisymmetry in the spacetime and azimuthal pressure force---during the period just before collapse to a black hole (at this stage in the debris evolution, the magnetic field remains too weak to exert significant stress). Given the comparison between the mass-weighted integrals of their absolute values (as shown in Fig.~\ref{fig:J_evolution}), it's not surprising that the gravitational torque is, in most areas, the dominant one. However, this is not the case everywhere.

One important locational distinction is that the merged neutron star occupies roughly the region inside the central density contour. Essentially all the volume at which the gravitational torque reaches its maximum lies within this boundary. Consequently, although gravitational torque clearly dominates inside the merged neutron stars, it is not so clearly dominant in the debris that will ultimately become the orbiting disk. Also, mergers of binary neutron stars and black hole–neutron star binaries claim how dominant the gravitational torque is \citep{Foucart+13, Foucart+19, Shibata_Kenta19, Most+21_FastEjecta}.

Another is that the gravitational torque varies with location much more smoothly than the hydrodynamic torque. The magnitude of the gravitational torque is close to axisymmetric, with the pattern broken only by four narrow channels of weak torque. 
By contrast, there is a very sharply-defined $m=2$ spiral wave pattern in the pressure gradient; at wave crests, the hydrodynamic torque is comparable to the maximum due to gravity. Moreover, whereas the peak hydrodynamic torque can be found primarily at the spiral wave crests, which extend to radii $\gtrsim 20 r_g$,
the greatest gravitational torque is restricted to a compact region inside  $\lesssim 10r_g$. In addition, as the bottom right panel demonstrates, the gravitational torque fades during the last ${\sim}\qty{1}{\ms}$ before collapse; although the hydrodynamic torque weakens somewhat shortly before collapse, the contrast with its strength ${\sim}1$--\qty{2}{\ms} earlier is smaller.

However, the most important contrast in the two torques' spatial distribution has to do with their signs. The hydrodynamic torque is, in all but a very few places, prograde. In sharp contrast, the sign of the gravitational torque switches from quadrant to quadrant (where the magnitude passes through zero), exhibiting strong inversion symmetry concerning the origin until very shortly before the system collapses to a black hole. The existence of these sign changes raises the possibility that the net rate of change in the angular momentum due to gravitational torques might be considerably less than the integral of the torque's absolute value.

To evaluate the effect of varying signs of torque on the matter that will eventually enter the orbiting debris disk, we compute the mean gravitational torque acting on the entire survivor subset of the tracer particles at several sample times.

We find that the {\it net} gravitational torque on the entire population of surviving tracer particles is smaller than the integral of its absolute value by an order of magnitude.

\begin{figure*}
\includegraphics[width=0.99\linewidth]{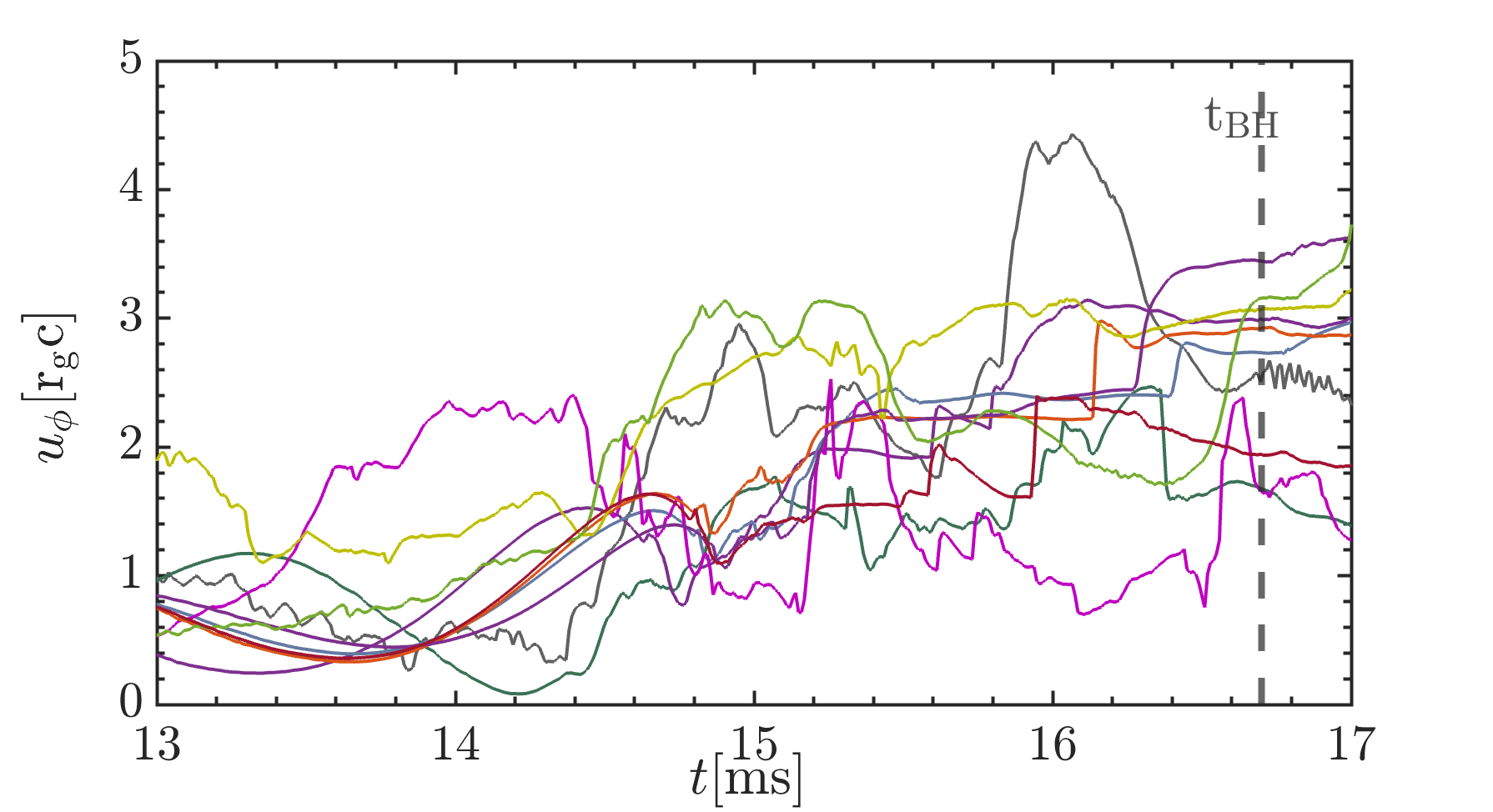}
\caption{Evolution of $u_{\phi}$ for 10 randomly selected survivor particles between 13 to 17 ms, including both bound and unbound particles. The vertical dashed line marks the moment the black hole forms.}

\label{fig:uphi_sur_13_17ms}
\end{figure*}

As a standard of comparison, we compute the population means of the hydrodynamic torque for the same set of survivor particles ( Fig.~\ref{fig:J_evolution}).
In the period before black hole collapse, the ratio between the net gravitational and hydrodynamic torques is considerably smaller than the ratio of their absolute magnitudes, ranging from ${\approx}0.5$ to ${\approx}4$ rather than having a consistent value ${\approx}10$ (both net torques are consistently positive). This ``Eulerian" result can also be understood in a ``Lagrangian'' sense by considering the time dependence of the torque on individual fluid elements. Viewed this way, the gravitational torque cancellation is due to the rotation of the merged neutron star. Whereas it rotates at close to the break-up rate, i.e., with a period ${\approx}\qty{0.5}{\ms}$, the angular speed of the orbiting debris is an order of magnitude slower. Consequently, over the ${\sim}\qty{1}{\ms}$ duration of strong gravitational torques, a fluid element sees the sign of the torque exerted on it flip four times per orbit, for a total of ${\sim}5$--10 times. 
Exactly this sort of behavior can be seen in Figure~\ref{fig:uphi_sur_13_17ms}, in which the angular momentum of the tracer particles changes in a roughly sinusoidal fashion, but with superposed shorter timescale fluctuations and an overall increasing trend.

The combination of all these effects---the locally strong, but alternating sign, gravitational torque, and the locally much weaker but consistently prograde sign of the hydrodynamic torque---accounts for the small net increase in angular momentum of the debris mass during this period. Although gravitational torque remains the single largest contributor, its strong internal cancellation makes the pressure torque nearly as large a contributor to the net change in angular momentum.

\begin{figure*}
  \begin{tabular}{ccc}

     \includegraphics[width=0.43\textwidth]{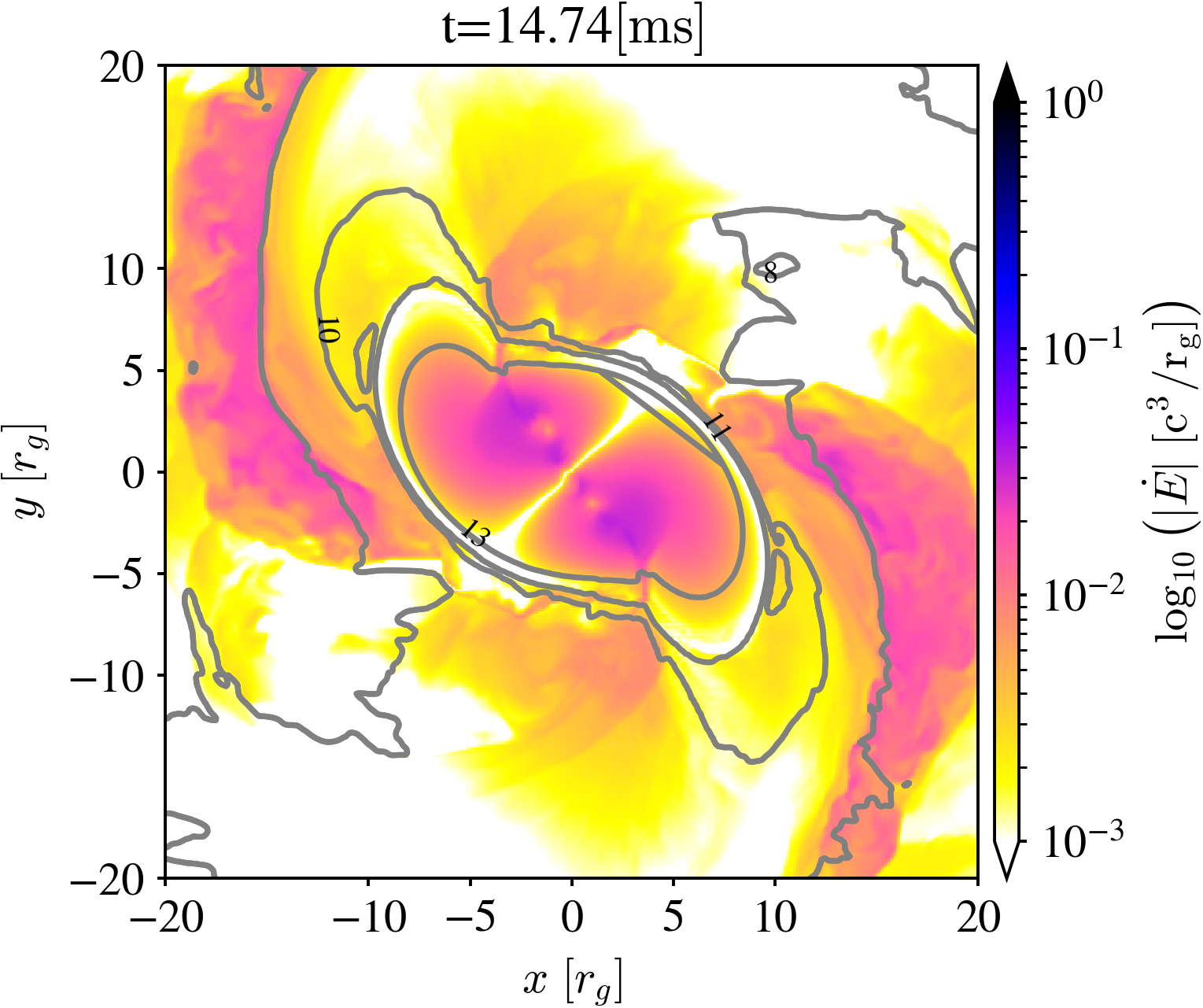}&%
     \includegraphics[width=0.44\textwidth]{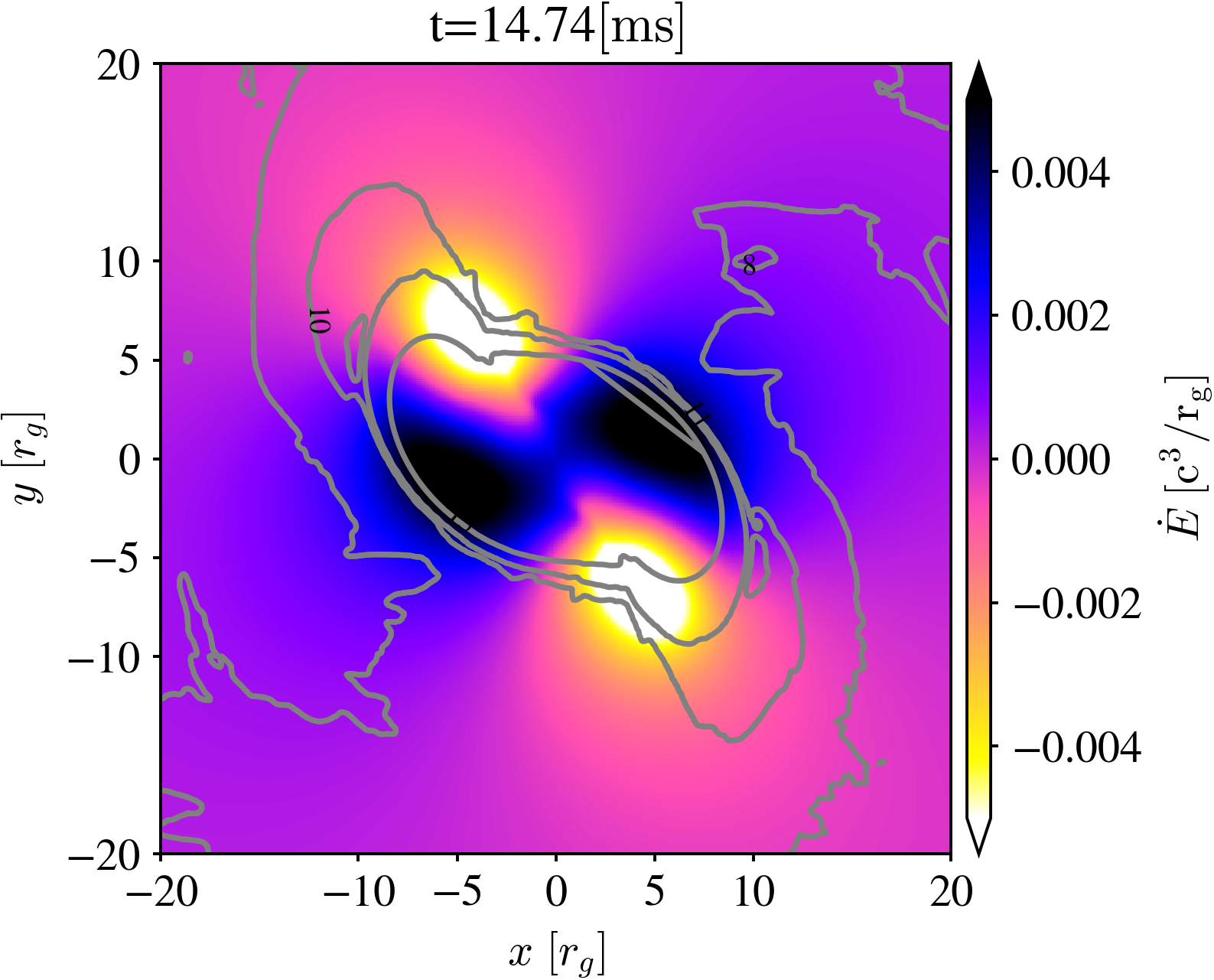}\\
     
     \includegraphics[width=0.43\textwidth]{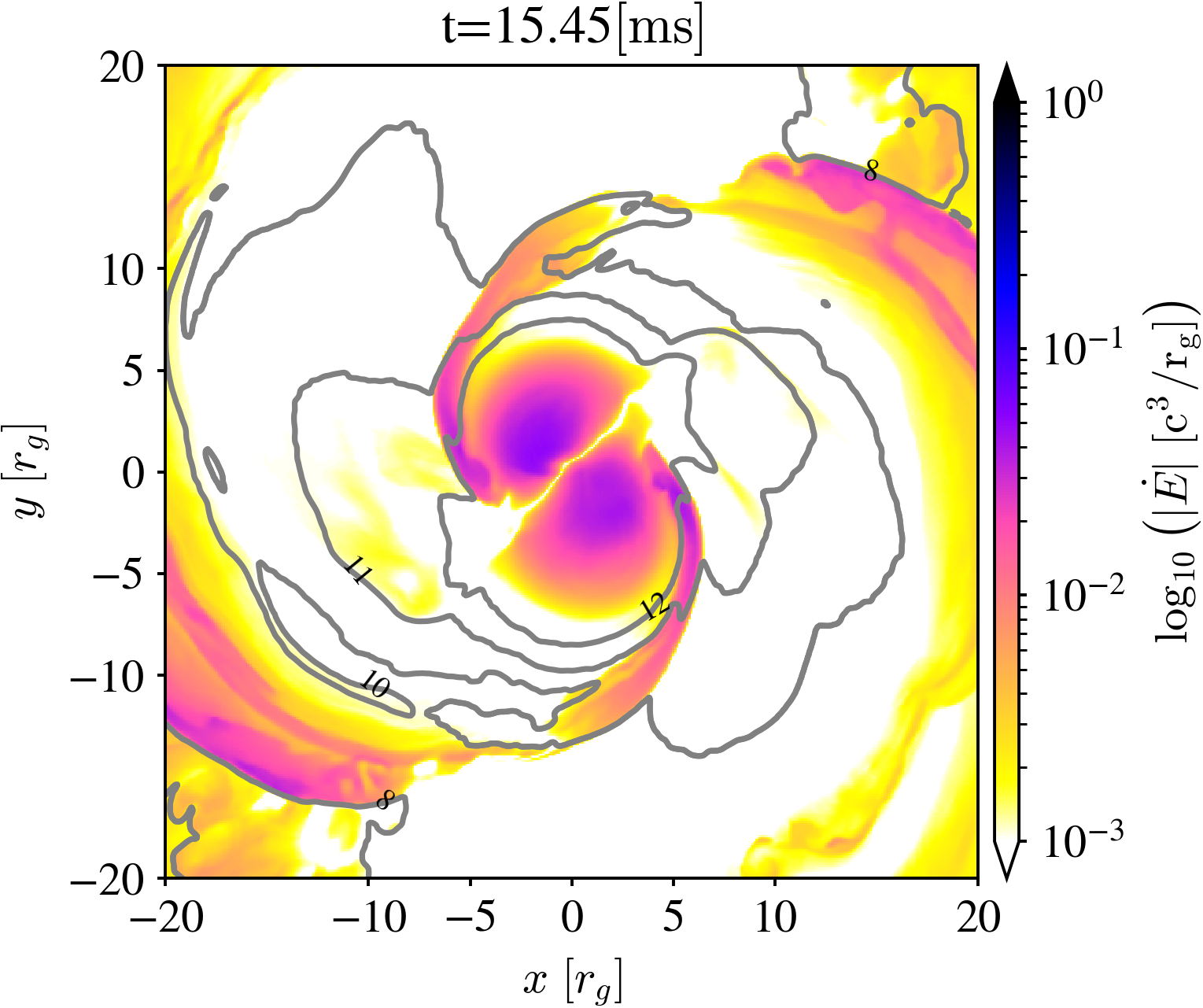}&%
     \includegraphics[width=0.44\textwidth]{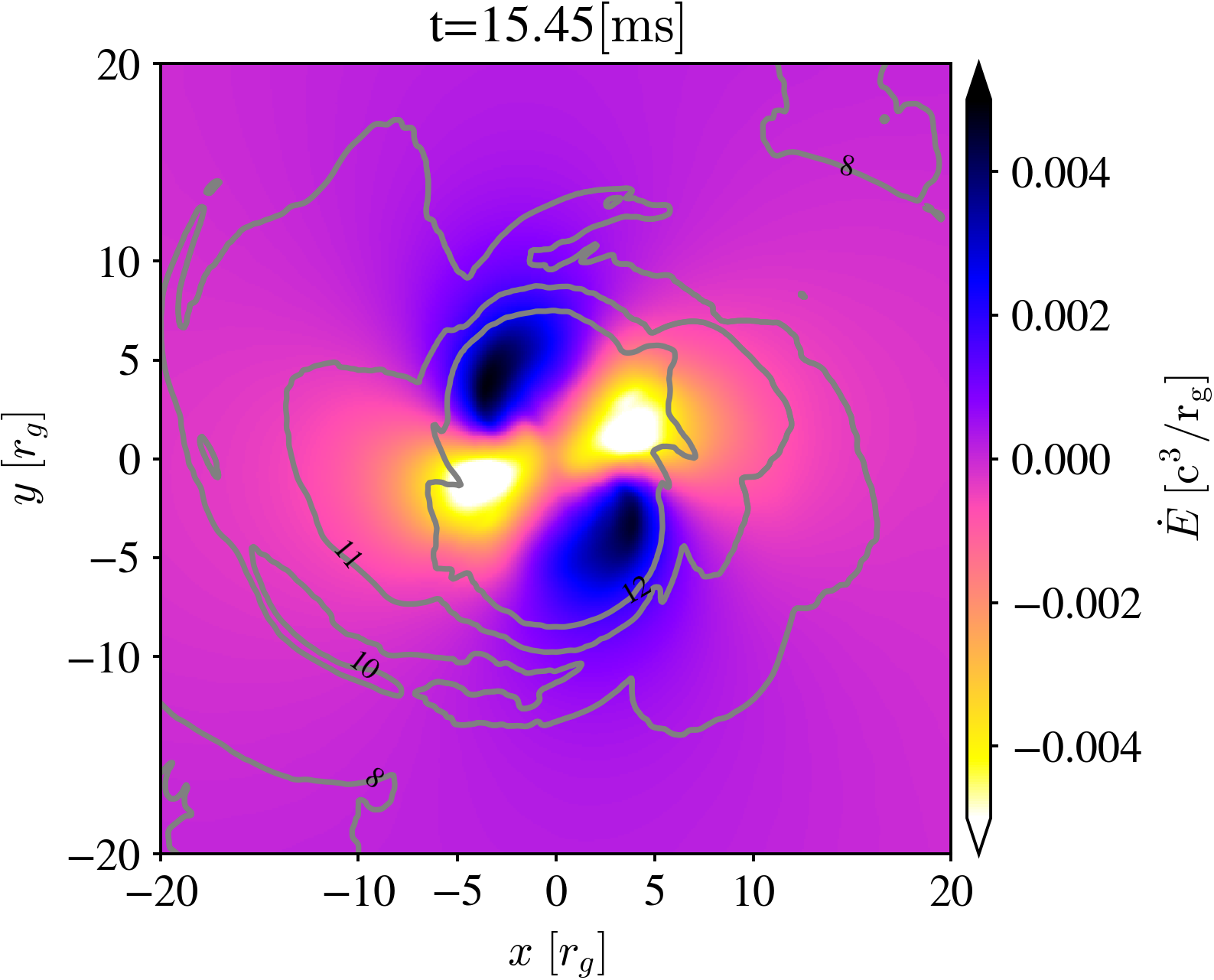} \\
    
    \includegraphics[width=0.43\textwidth]{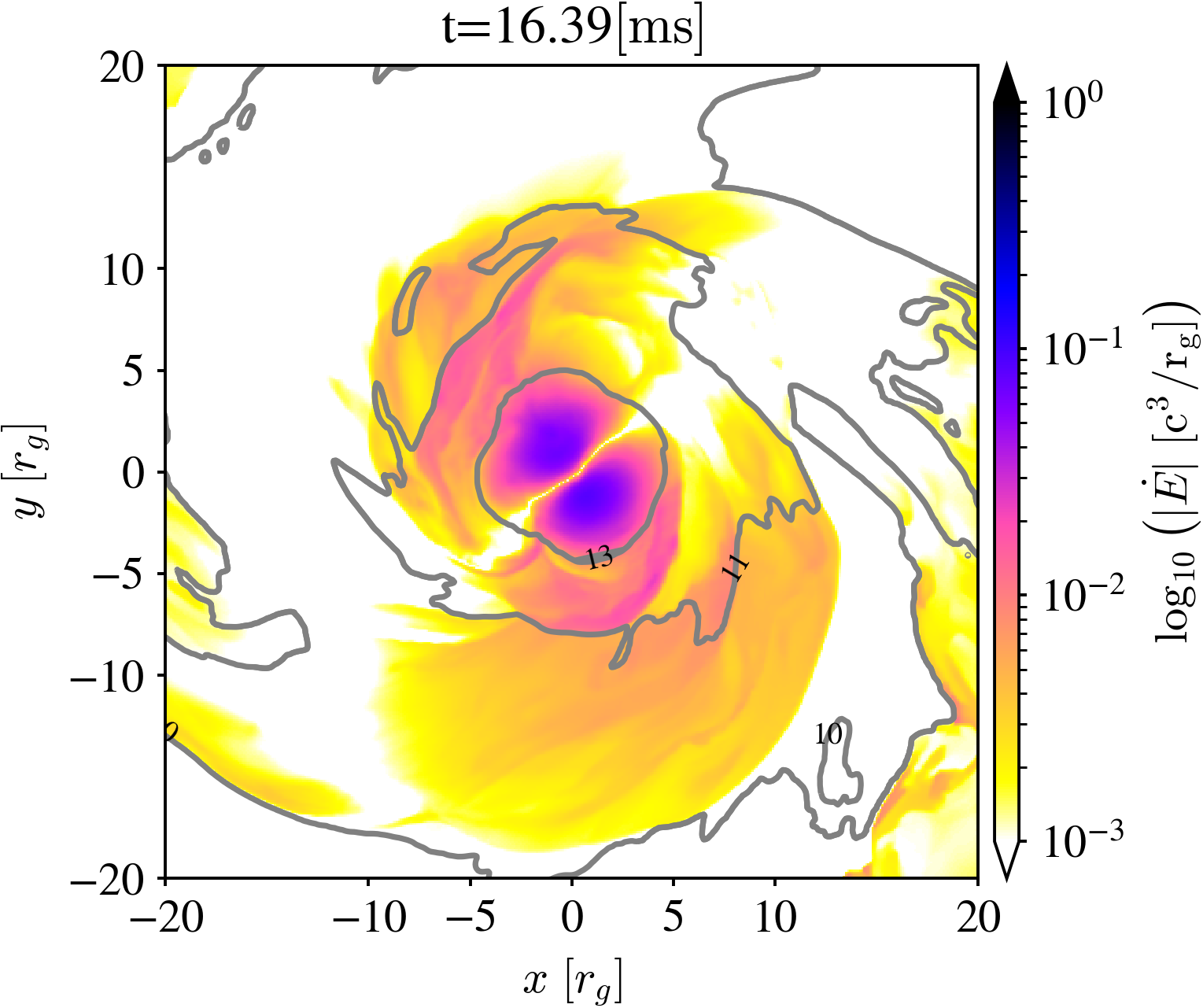}&%
    \includegraphics[width=0.44\textwidth]{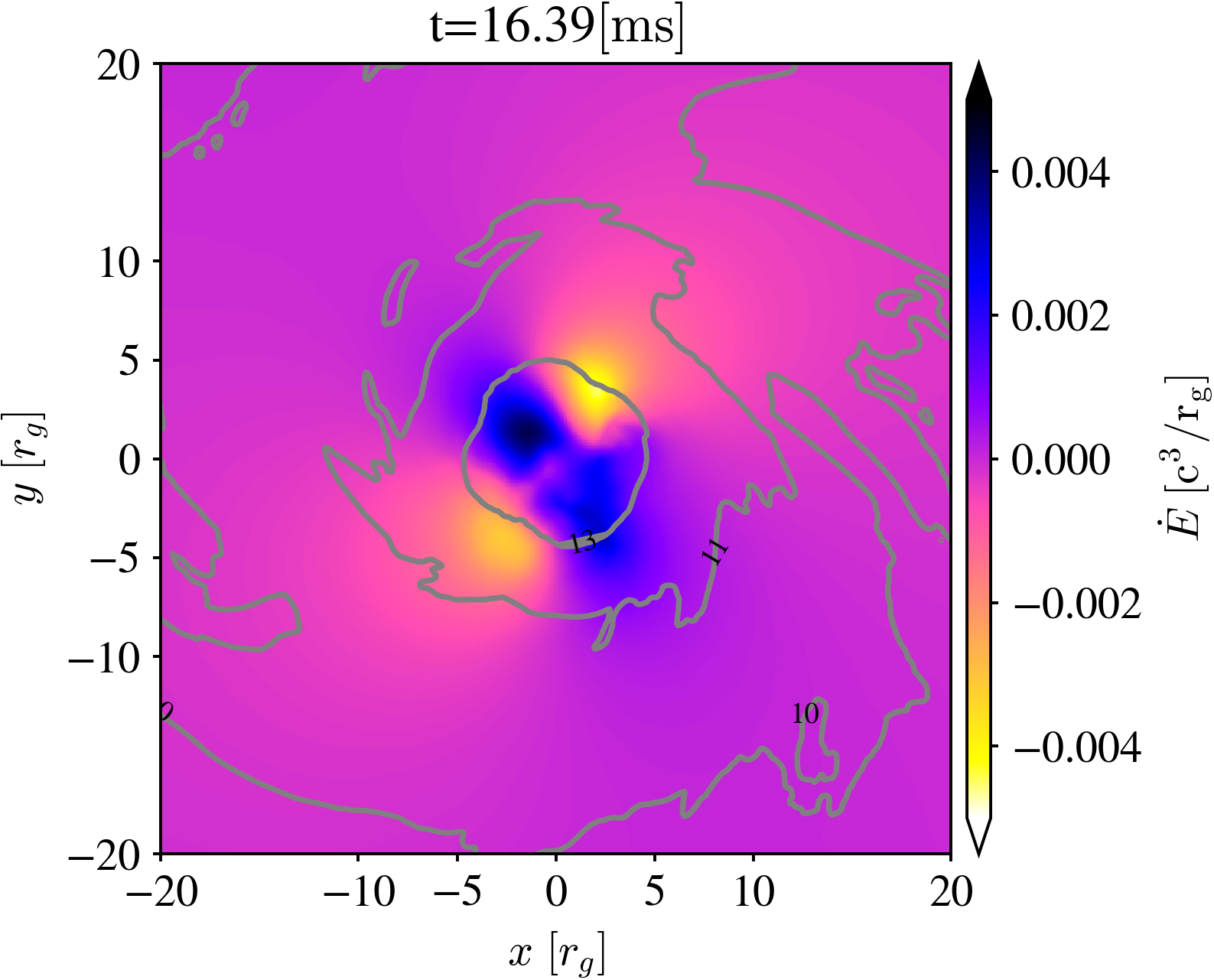} \\
   
  \end{tabular}
  \caption{Evolution of the rate per unit mass at which gravitational (right panels) and hydrodynamic (left panels) forces do work on the debris before the binary neutron star collapses to a black hole. Note that the left panels show $|\dot E|$, whereas the right panels show the signed quantity $\dot E$. The white diagonal line crossing the merged neutron star in the left panels separates positive $\dot E$ (upper-left) from negative (lower-right). Unlike Fig.~\ref{fig:dj_before} and the left panels, the color scale in the right panels is {\it linear}. All are in units of $c^3/r_g$. The time show \qty{14.74}{\ms}, \qty{15.45}{\ms}, and \qty{16.39}{\ms}.}
  \label{fig:de_before}
\end{figure*}

\subsubsection{Energy}

The source term for the energy-like primitive variable in \igm (see eqn.~\ref{eq:tausource}) is rather complex. In order to represent its principal features, we study a proxy, $\partial g_{tt}/\partial t$, for the rate of change of energy per unit rest-mass. This is effectively present within the source term through the relation $g_{tt} = -\alpha^2 + \gamma_{ij}\beta^i \beta^j$.
The rate of energy change due to hydrodynamic forces is $-(u^x\partial_x P + u^y\partial_y P)/\rho$.

Unlike the situation for torque, in terms of absolute value, the hydrodynamic portion is the dominant contributor to energy change throughout the merger: it is ${\sim}3$--$10$ times larger than the time-varying gravity portion before black hole collapse (see both Figure.~\ref{fig:uphi_ut_all}b and Fig.~\ref{fig:de_before}). Because the gravitational work, like the gravitational torque, changes sign between quadrants, the contrast between the net energy change due to pressure forces and that due to gravity is even larger than the contrast in absolute value, typically ${\sim}100$.

Also unlike the torque, the signs of both the hydrodynamic and the gravitational net energy change are functions of time; the overall net follows the sign of ${\dot E}_{\rm hydro}$ because its magnitude is so much larger. It is mostly, but not exclusively, negative from ${\approx}\qty{14}{\ms}$ to ${\approx}\qty{15.5}{\ms}$ (see Figure.~\ref{fig:Edot_evolution}), but---for the surviving particles--- consistently strongly positive for the last ${\approx}\qty{1}{\ms}$ before black hole collapse. For the particles swiftly captured by the black hole, the magnitude of ${\dot E}$ is similar but much more mixed in sign.

Interestingly, as can be seen in both Figure~\ref{fig:Edot_evolution} and Figure~\ref{fig:de_before}, the absolute magnitude of neither ${\dot E}_{\rm hydro}$ nor ${\dot E}_{\rm grav}$ changes appreciably from the moment of stellar contact to the time of black hole collapse. Because ${\dot E}_{\rm grav}$ has a highly symmetric spatial distribution, and the spatial distribution of ${\dot E}_{\rm hydro}$, while less symmetric, still exhibits rough parity in total volume between positive and negative portions, any net sign in ${\dot E}$ is due to a positional correlation between the survivor particles and the spatial pattern of ${\dot E}$. Put another way, the particles that survive are the ones that, integrated over time, have spent the most time in regions where ${\dot E}$ is both positive and relatively large in magnitude.

\subsection{Post-collapse}

\begin{figure*}
  \begin{tabular}{cccc}
    \includegraphics[width=0.3\textwidth]
    {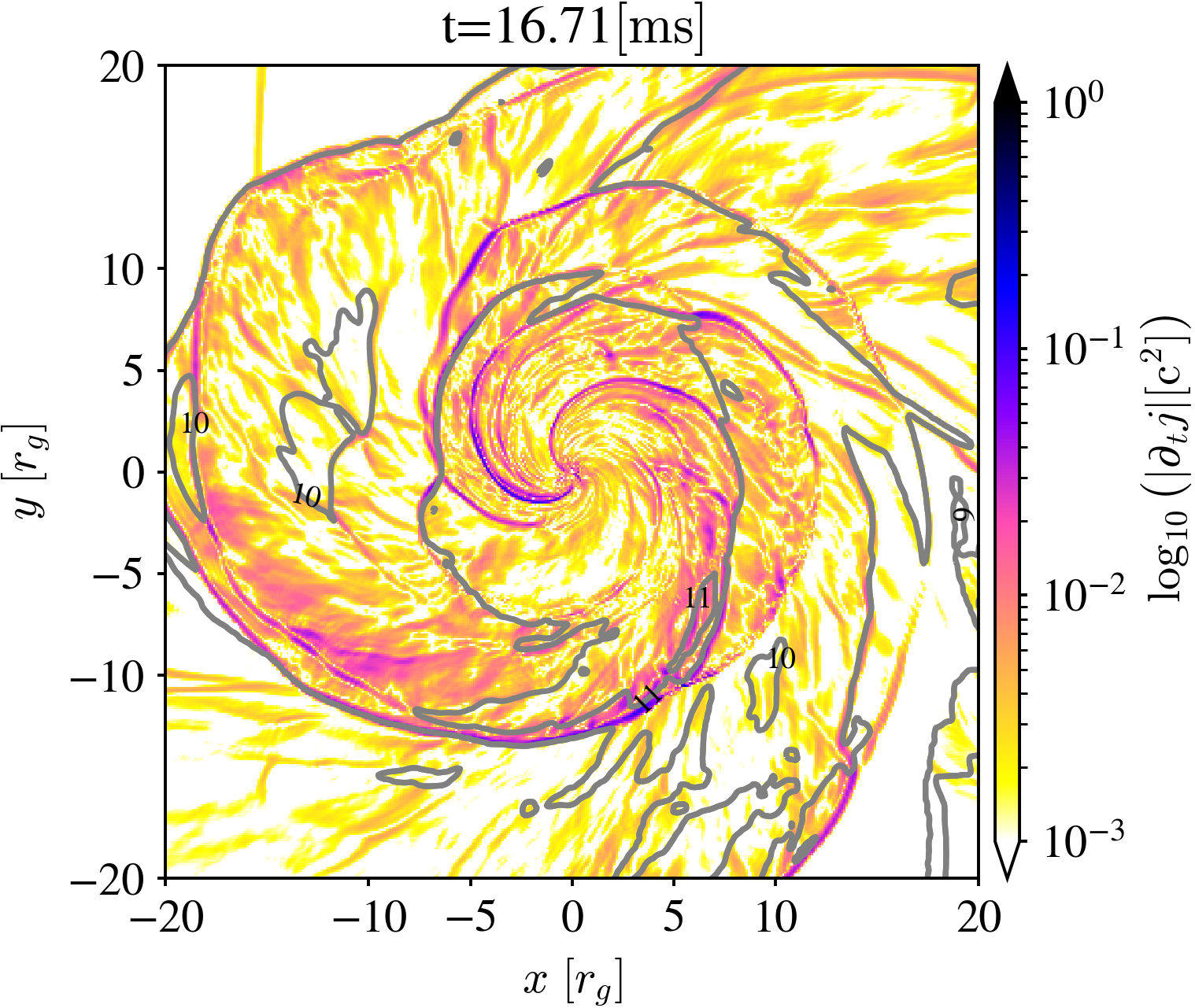} &%
    \includegraphics[width=0.3\textwidth]{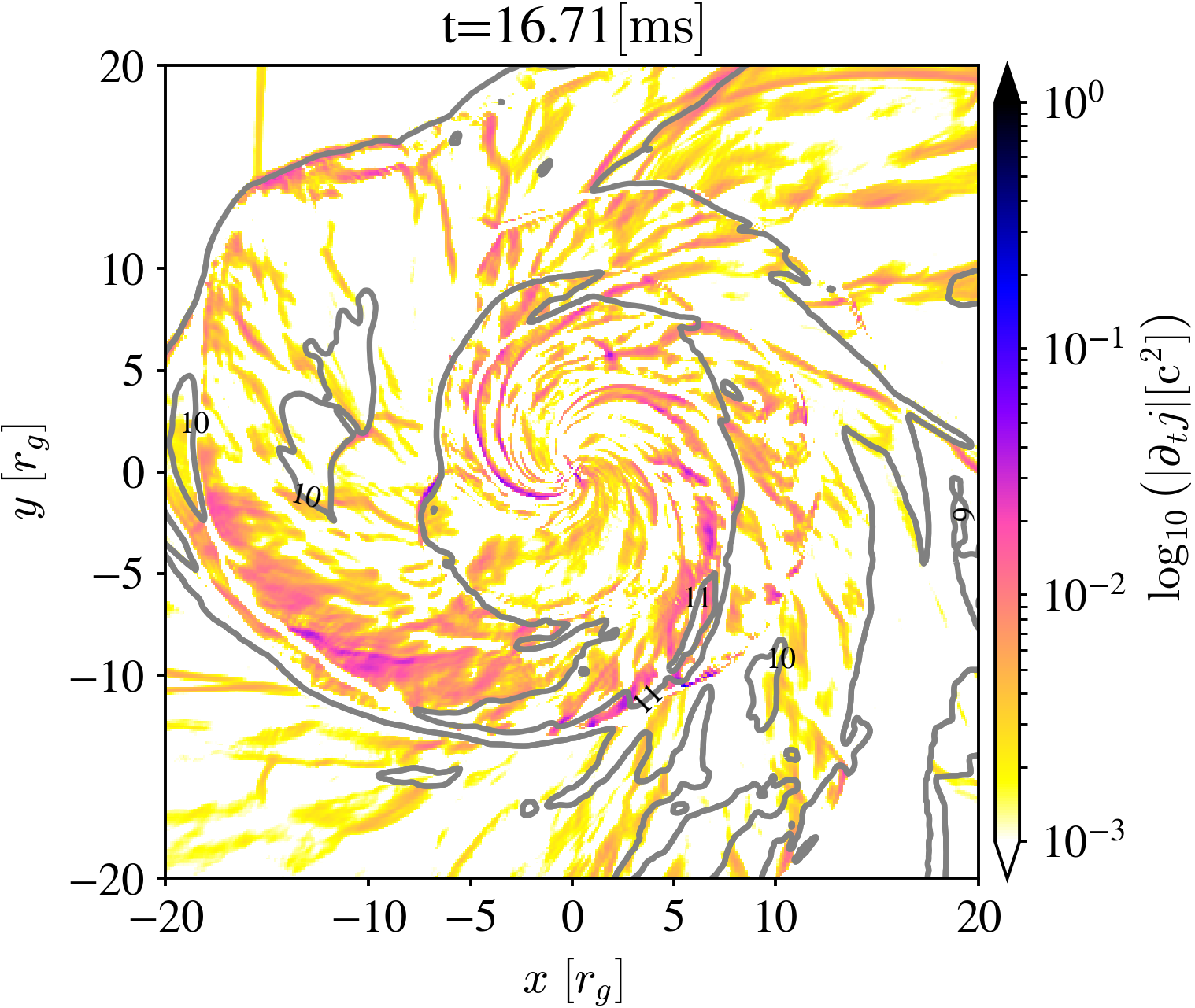} &%
    \includegraphics[width=0.3\textwidth]{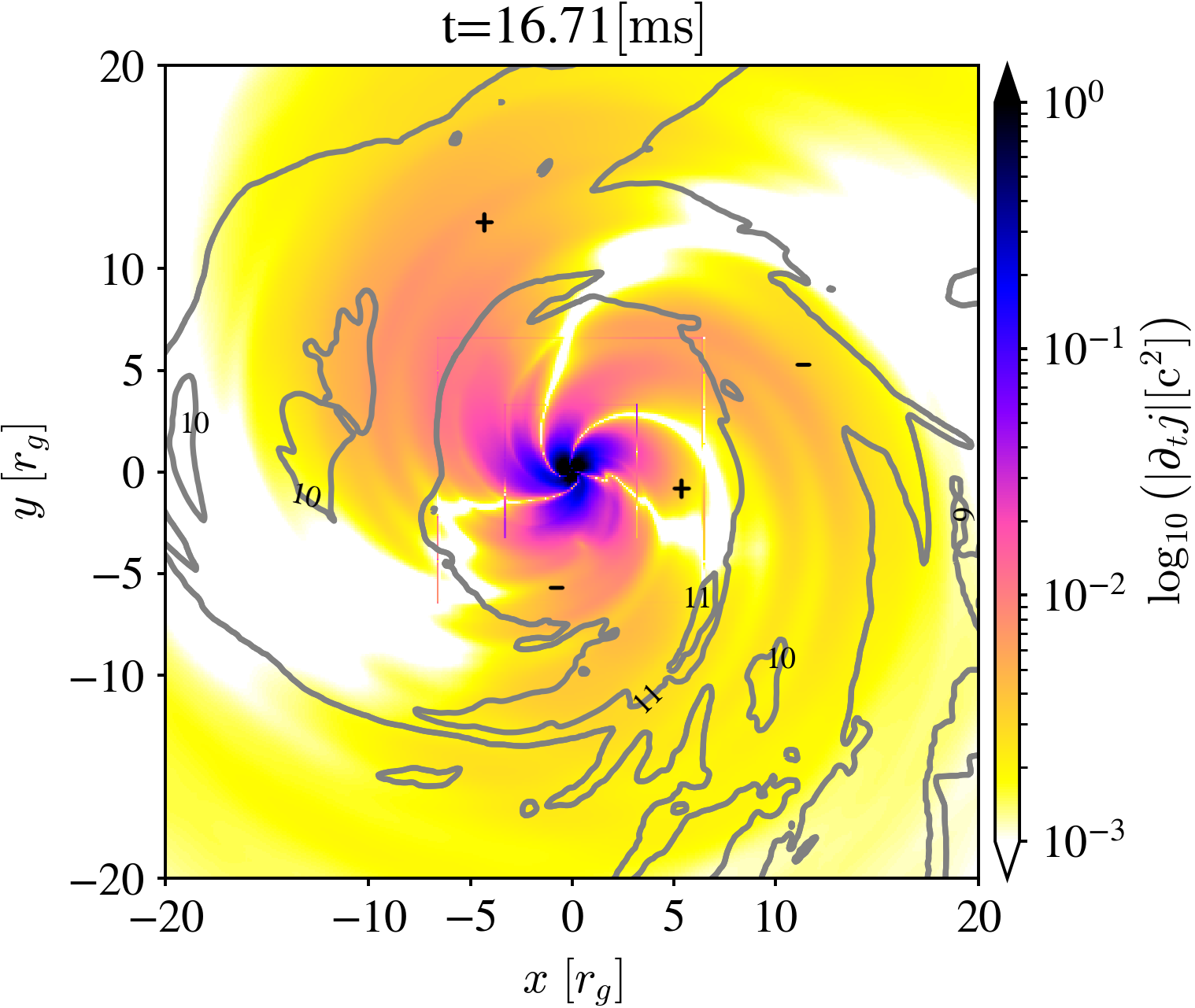} \\ 
    
   \includegraphics[width=0.3\textwidth]
   {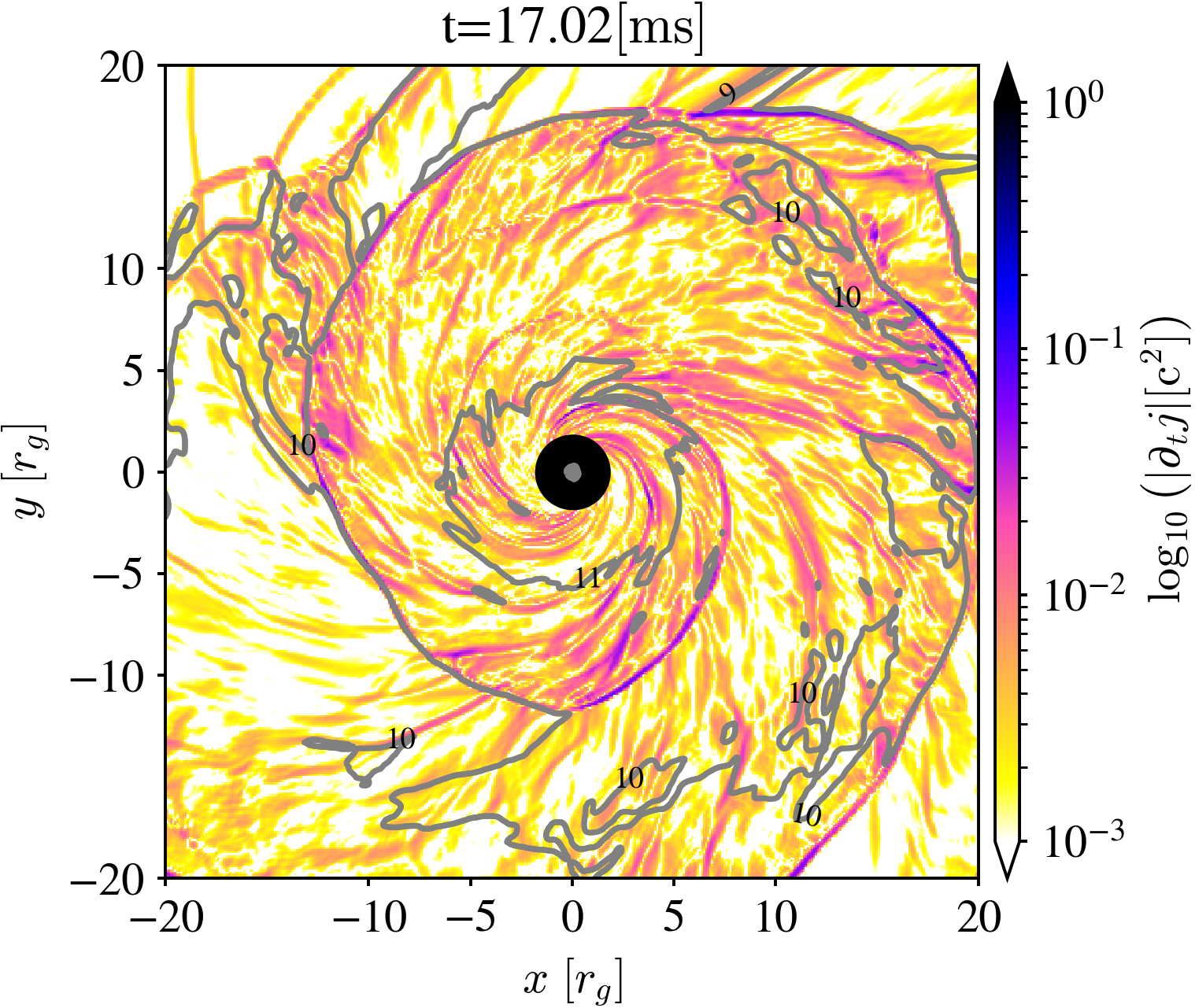} &%
   \includegraphics[width=0.3\textwidth]
   {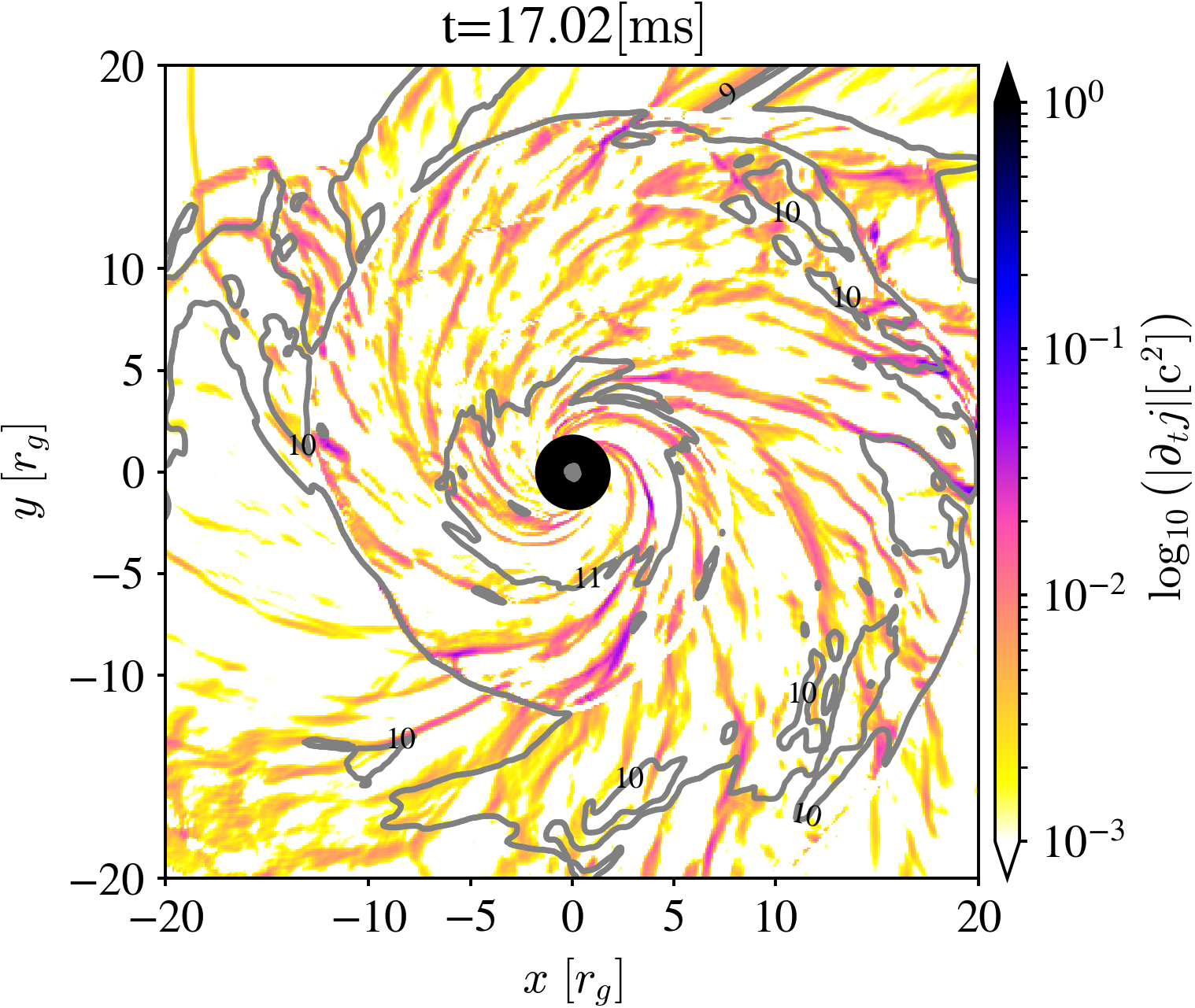} &%
    \includegraphics[width=0.3\textwidth]{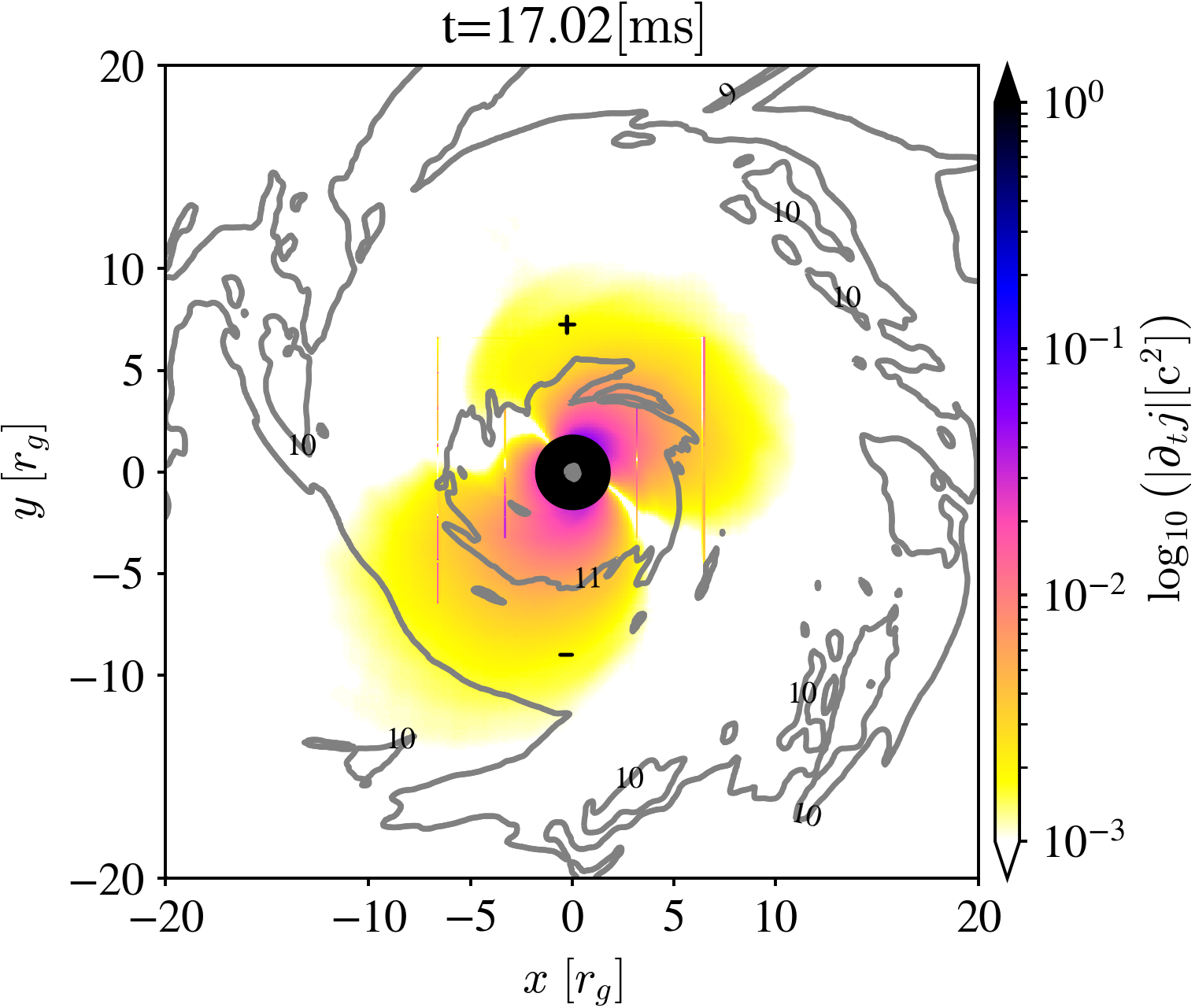} \\

    \includegraphics[width=0.3\textwidth]
    {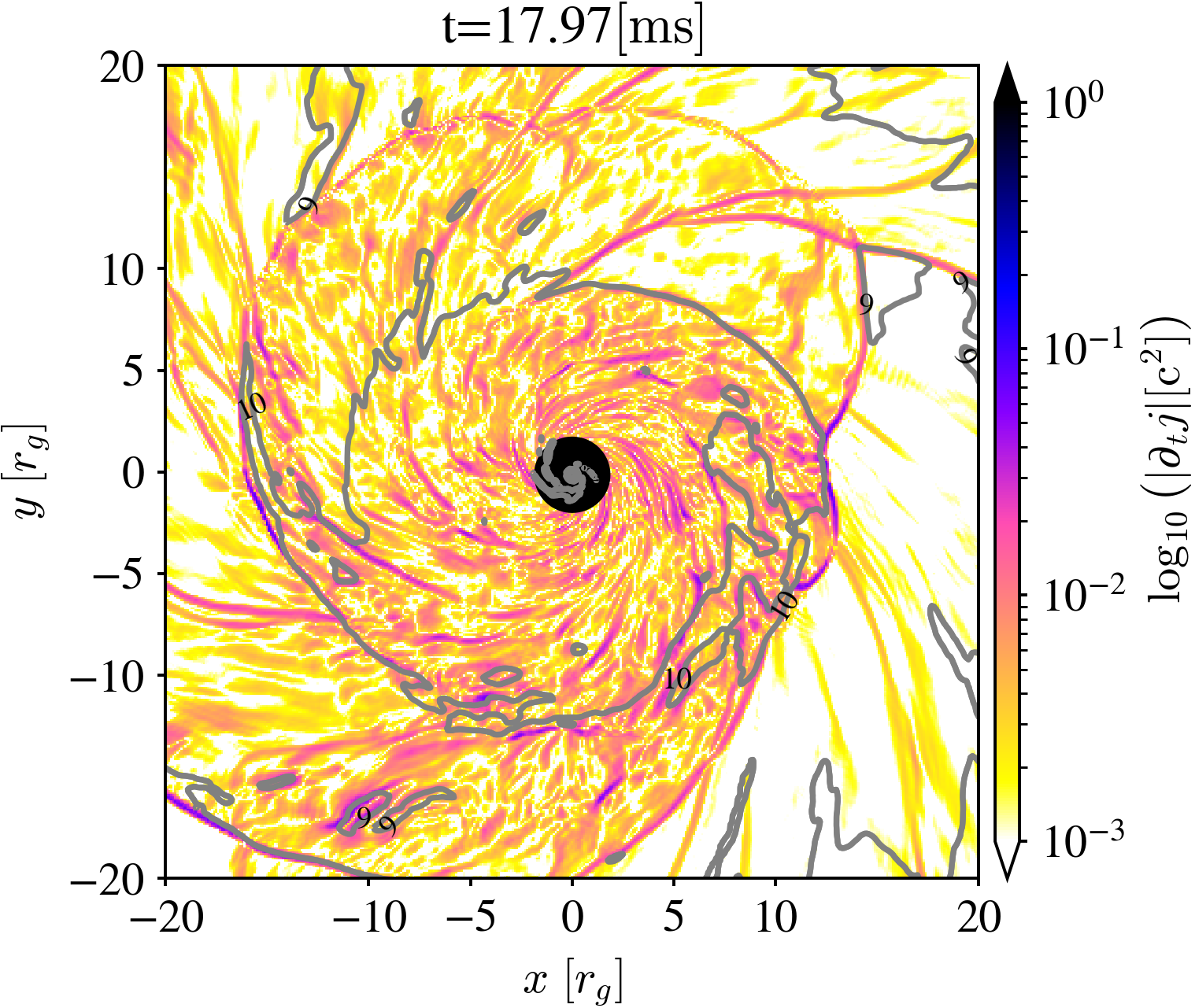} &%
    \includegraphics[width=0.3\textwidth]{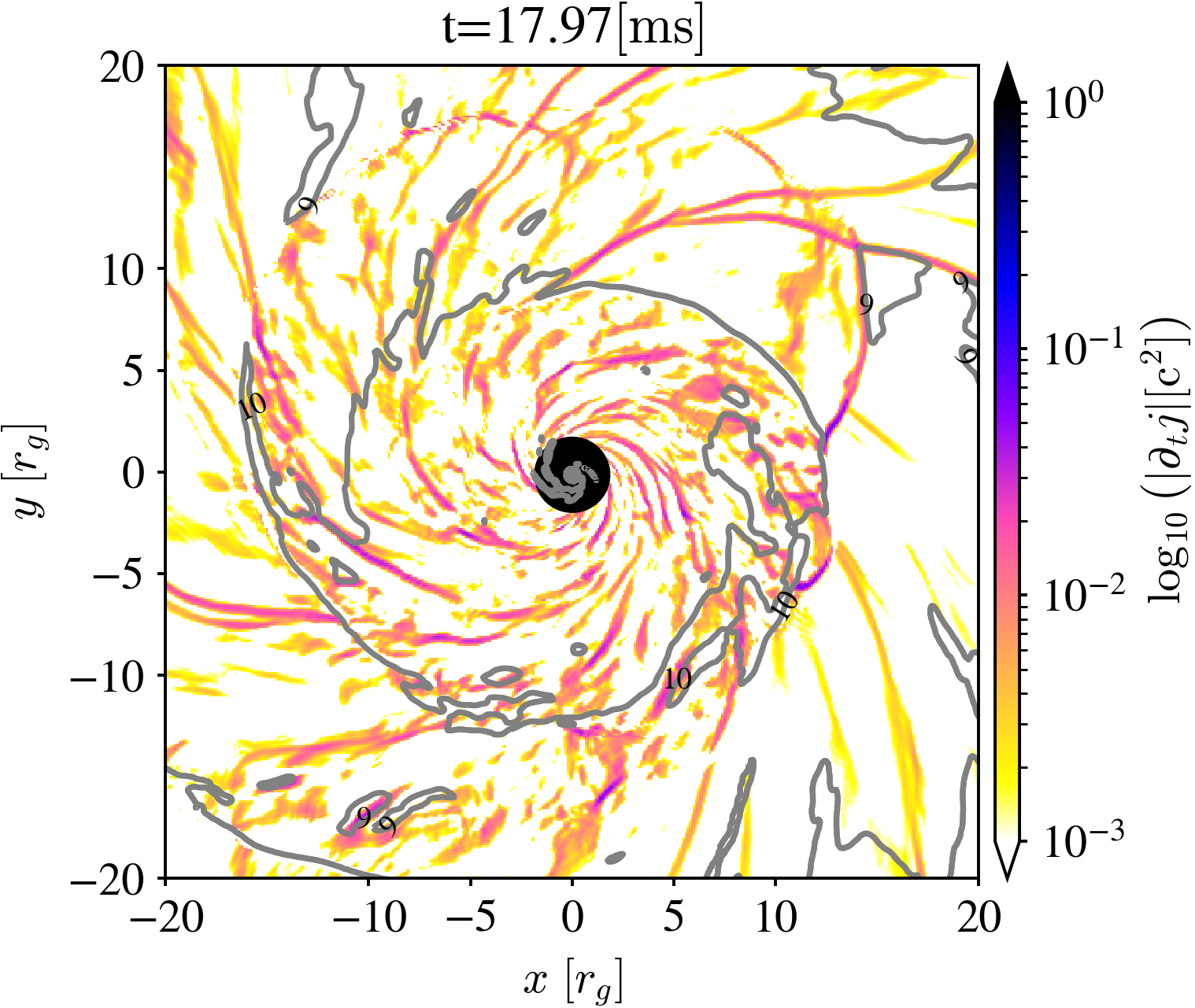} &%
    \includegraphics[width=0.3\textwidth]{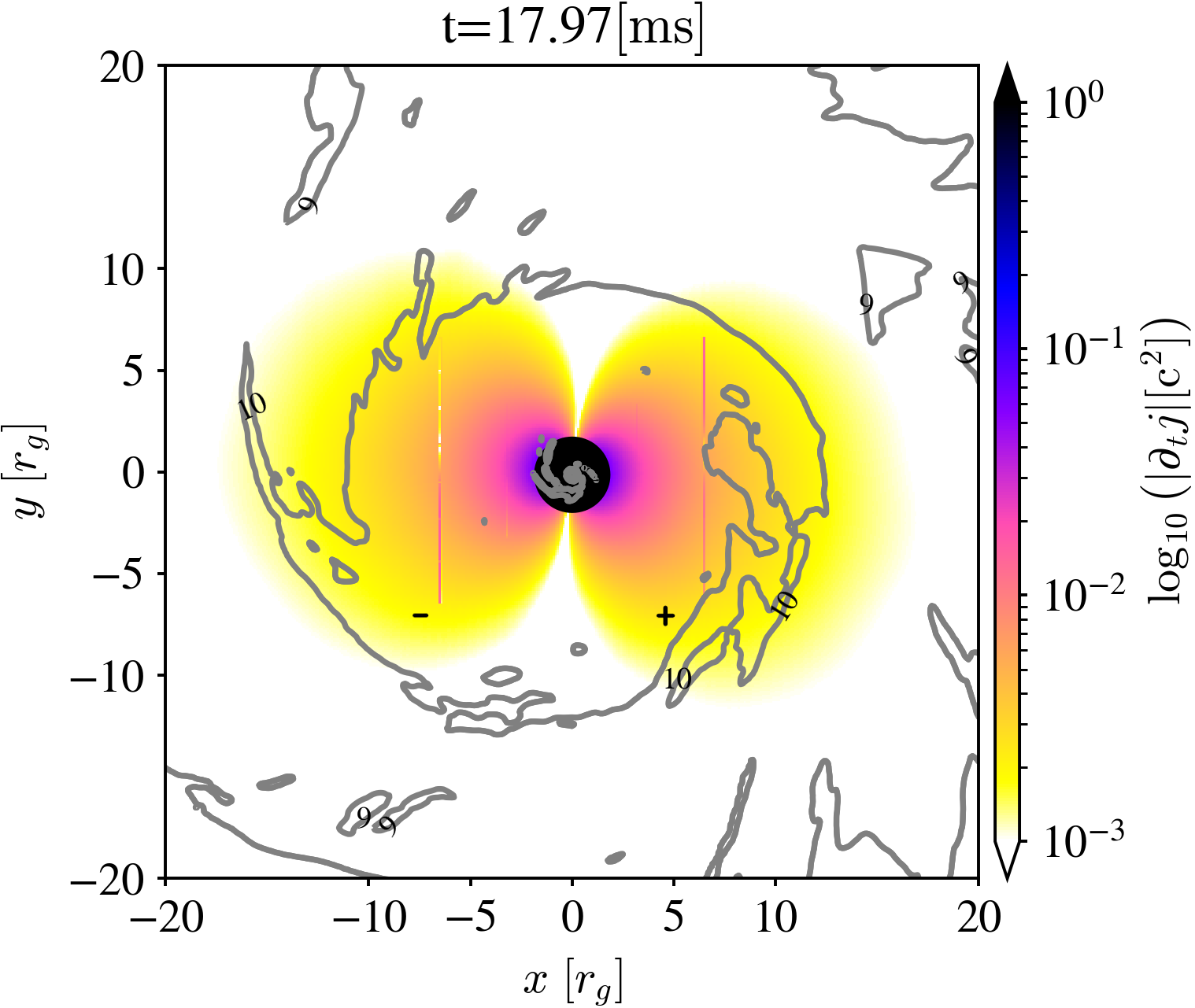} \\
    
  \end{tabular}
  \caption{Evolution of the magnitudes of the gravitational and hydrodynamic torques per unit mass after the binary neutron star collapse to a black hole (color contours) and $\rho$ (gray contours; the separation between gray contour lines is a factor of 10 in density). The left panels show the absolute value of the hydrodynamic torque ($\log |\partial_t j_{\rm hydro}|$), the middle panels show the absolute value of the hydrodynamic torque where the torque is negative (white denotes either negative torque with very small magnitude or positive torque),and the right panels show the absolute value of the gravitational torque ($\log |\partial_t j_{\rm grav}|$). All are in units of $c^2$. From top to bottom, the panels are at times \qty{16.71}{\ms} (the moment of horizon formation), \qty{17.02}{\ms}, \qty{17.50}{\ms}, and \qty{17.97}{\ms}.}
  \label{fig:dj_after}
\end{figure*}

\subsubsection{Angular momentum}

As shown in Figure~\ref{fig:dj_after}, both the gravitational torque and the pressure torque change qualitatively after the black hole is created. The black hole should relax to a pure Kerr spacetime on a timescale of a few tens of $GM/c^3$, i.e., a few tens of a millisecond in this case; it is then, by definition, exactly axisymmetric with respect to its spin axis, and therefore exactly preserves orbiting material's angular momentum component parallel to the spin \citep{EcheverriaFernando88, Berti15, Berti+18, Sarin_Lasky21}.
However, as can be seen in this figure, a small amount of nominal gravitational torque (${\sim}10^{-2}c^2$) remains post-collapse. We surmise that this
is a gauge effect.\footnote{It is unlikely to be due to mass remaining on the grid because even at 17.4~ms, the mass on the grid is only $\approx 1.5\%$ of the total mass, and to create this level of torque with a quadrupolar spatial distribution, the quadrupole moment of the mass would need to be order-unity. However, the density contours in Fig.~\ref{fig:dj_after} show no sign of such a large quadrupole moment.}

The coordinates associated with the numerical relativity solution are time-dependent in the sense that a coordinate basis (i.e., one in which the spatial metric is diagonal) rotates and also expands radially; put another way, the shift vectors correspond to an expanding rotating flow whose rotation rate is larger than that of the ZAMO frame for the Kerr spacetime \citep{SmarrL1973, Frolov_V_P2014, Braeck23}. Numerical effects may create an apparent asymmetry in the shift vectors compensated by symmetry in the coordinate definitions so that physical axisymmetry is preserved.

More importantly, the hydrodynamic torque is much larger, and both its spatial distribution and its overall magnitude change with time past the collapse. Although spiral waves are present both before and after collapse, they become much less distinct and symmetric as the black hole forms and in the milliseconds following (the beginning of this process can be seen in the \qty{16.55}{\ms} image in Figure.~\ref{fig:dj_before}). After black hole collapse, the region in which the torque is ${>}10^{-3}c^2$ becomes much larger, filling nearly the whole orbital plane within ${\approx}10r_g$. Simultaneously, the integrated absolute magnitude of ${\partial_t j}_{\rm hydro}$ grows (see inset in Fig.~\ref{fig:J_evolution}). Approximately \qty{3}{\ms} post-collapse, the hydrodynamic torque reaches a maximum at roughly ${\sim}10$ times the level seen at the time of the collapse. Beginning a few ms later, the magnitude of the hydrodynamic torque diminishes as some of the debris (${\sim}\qty{30}{\percent}$) falls into the black hole, while the remainder expands outward.

\begin{figure*}
\includegraphics[width=0.49\linewidth]{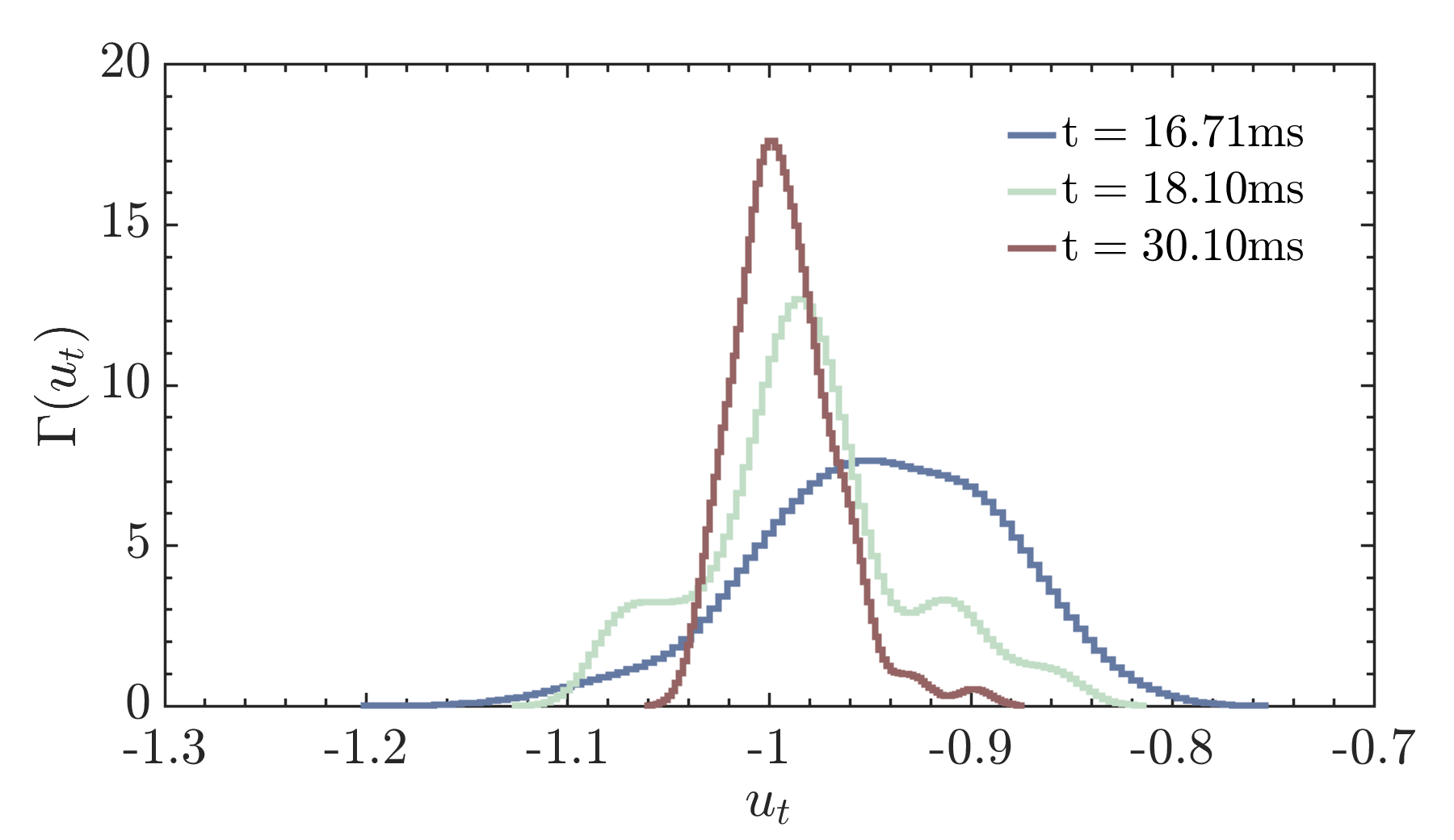}
\includegraphics[width=0.49\linewidth]{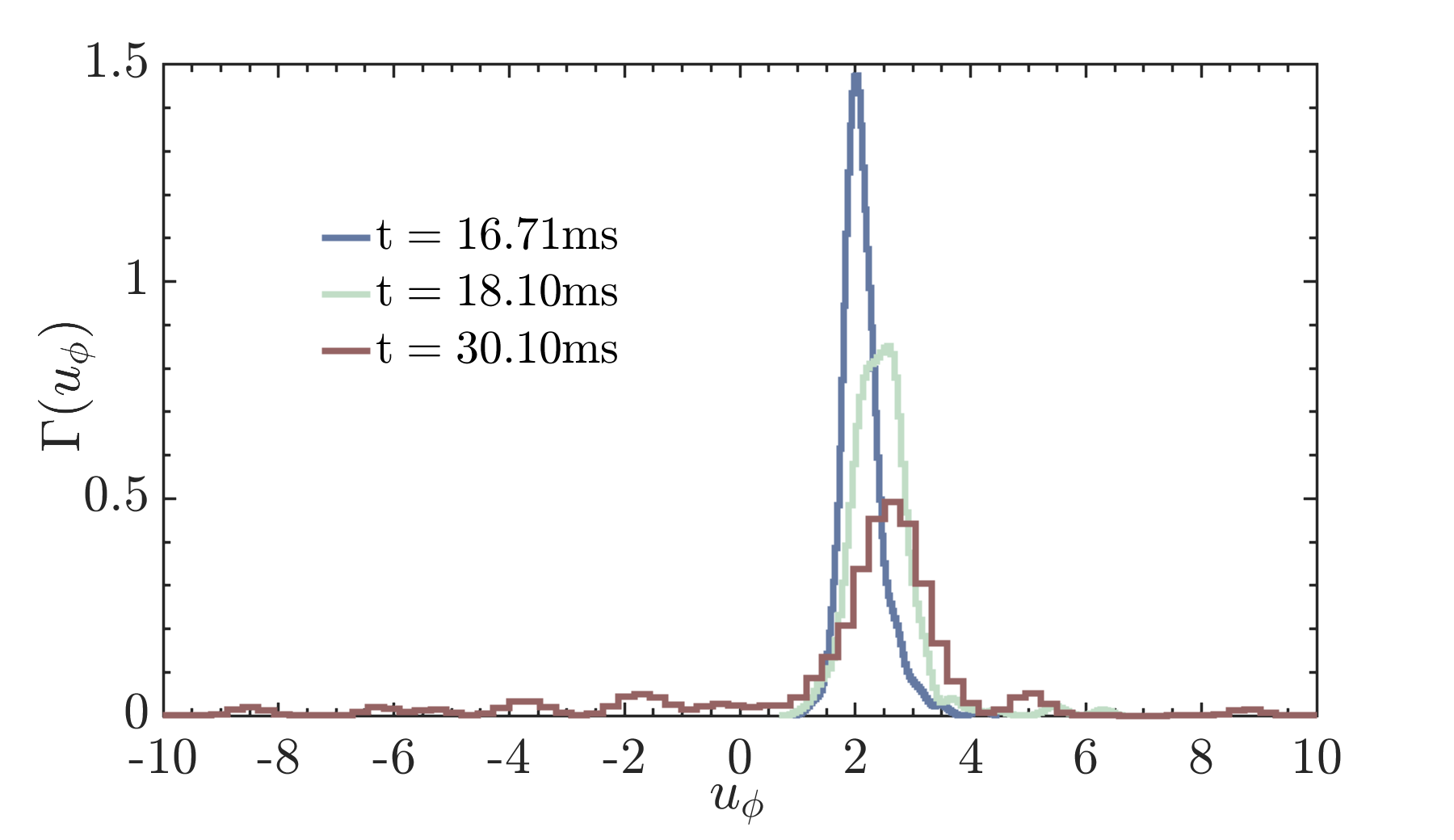}
\caption{Distribution of particle number per unit $u_t$ (left panel) and $u_{\phi}$ (right panel) for the particles within the problem volume and outside the black hole's ISCO at three times: \qty{16.71}{\ms} (the moment of black hole formation; red curve), \qty{18.10}{\ms} (green curve), and \qty{30.1}{\ms} (blue curve). The three distributions in each panel are normalized to have the same integral.}

\label{fig:ut_uphi_distribution}
\end{figure*}

Moreover, as shown in the middle panels of Figure~\ref{fig:dj_after}, the pressure torque per unit mass frequently goes negative, although it is significantly negative in a smaller volume than where significantly positive specific torque can be found. That it is frequently negative is, of course, an expression of angular momentum conservation because the end of gravitational torque means that there is no longer any angular momentum source for the orbiting matter. Comparison with density images (with finer contours than shown in Fig.~\ref{fig:dj_after}) demonstrates that regions of large positive specific torque lie on the leading edges of spiral wave density maxima---where the specific torque is negative.

Because the hydrodynamic torques can redistribute angular momentum within the gas, but cannot change its total amount, the total signed hydrodynamic torque can differ from zero only by the amount of angular momentum leaving the grid (e.g., by accretion onto the black hole). For this reason, if this were a competely closed system, fluid forces could change the shape of the angular momentum distribution, but not its mean value.  However, there actually is evolution in both the mean, which moves toward greater values, and the shape of the distribution, which widens over time (see Figures.~\ref{fig:ut_uphi_compare_distribution} and \ref{fig:ut_uphi_distribution}). Particles with small absolute values of angular momentum are quickly lost into the black hole, with the result that the mode of the distribution moves toward larger values, reaching $\approx 2.5 r_g c$ at $t=30$~ms.   At the same time, randomness in the sign of the azimuthal pressure gradients encountered by fluid elements broadens the distribution; it also creates the extended tail toward retrograde angular momentum.

Provided only that the pressure in the gas is small compared to the local gravitational potential, the distribution of mass with angular momentum maps directly onto the surface density profile of the disk:
\begin{equation}
    \Sigma(r) = \frac{1}{2\pi r}\frac{dm}{dr} = \frac{1}{2\pi r}\frac{dm}{dj}\frac{dj}{dr} \approx \left(\frac{GM_{\rm BH}}{16 \pi^{2} r^{3}}\right)^{1/2} \frac{dm}{dj}\, ,
\end{equation}
where the last semi-equality is exact in the Newtonian limit. Because $dm/dj$ remains fairly narrowly-peaked even at \qty{30.1}{\ms} (half-width at half-maximum ${\approx}1.25 \, r_g c$), the initial surface density profile of the debris is concentrated near the radial coordinate at which circular orbits have specific angular momentum ${\approx}2.5 r_g c$.
In other words, the narrowness of the angular momentum distribution given to the surviving debris places this debris into a narrow ring. The relatively small value of the debris' angular momentum means that this ring generally lies not far outside the ISCO; in fact, for this particular simulation, in which a black hole with spin parameter $\chi \approx 0.8$ was formed, the angular momentum of an ISCO orbit is ${\approx}2.38 r_g c$, so that almost half the debris rapidly spirals into the black hole. 

\subsubsection{Energy}

Just as the axisymmetry of Kerr spacetime forbids gravitational torque, Kerr spacetime's time-independence prohibits gravitational work. The consequences for the distribution function of $u_t$ are therefore analogous to those found for the distribution function of $u_\phi$: energy gained by one fluid element comes at the expense of another. Also like the angular momentum, because this is not a closed system, the distribution function of the remaining material can evolve. As shown in different ways in both Figure~\ref{fig:uphi_ut_all}b and Figure~\ref{fig:ut_uphi_distribution} (left panel), the mean orbital energy of the matter on the grid but outside the ISCO increases slightly after black hole collapse. Little of this change is accomplished in the first few milliseconds, but by ${\sim}\qty{13}{\ms}$ after the black hole forms, the mode of the binding energy per unit mass has decreased from ${\approx}0.05 c^2$ to ${\approx}0$ (note that our metric signature makes $u_t \leq -1$ for unbound particles and $u_t > -1$ for bound particles).

At the same time, this distribution, unlike the angular momentum distribution, narrows: whereas it extended
from $u_t \approx -1.15$ to almost $u_t \approx -0.75$ when the black hole formed, it spans only the range from $u_t \approx -1.05$ to $u_t \approx -0.9$ ${\sim}\qty{13}{\ms}$ later. The reason for this contrast is that particles at both energy extremes have left the grid: the strongly unbound material moves outward and has left the problem volume by the last time we record, while the more deeply bound matter falls into the black hole.

\section{Discussion}

\subsection{Debris mass}

The goal of our program is to identify the mechanisms regulating the initial state of the orbiting debris disk, the seat of so much of the phenomenology of binary neutron star mergers. The most basic quantity describing this disk is its total mass. As we have shown, not very surprisingly, the principal criterion determining whether neutron star material is left in the disk or captured is its orbital energy (see Figs.~\ref{fig:ut_uphi_compare_distribution} and \ref{fig:ut_uphi_distribution}).

The new element uncovered here is that when collapse to a black hole happens quickly, the orbital energy of matter changes dramatically in the ${\sim}\qty{2}{\ms}$ before the black hole forms; although there is a range of fluid element energy histories, on average, their energy first decreases, then increases. To escape the collapse, a fluid element's binding energy must be reduced to ${\lesssim}0.05$ in rest-mass units. This is largely accomplished by hydrodynamic interactions in the matter being squeezed out of the merging neutron stars. The surviving matter is the portion of the matter outside the merged star with the most net energy gain.

\subsection{Debris energy}

At the time of black hole collapse, the core of the energy distribution function is 
quite symmetrical around its mode at binding energy ${\approx}0.02$. After collapse, although nearly all debris with binding energy $> 0.05$ is captured into the new black hole, the wing of the distribution on the unbound side remains, stretching out to ${\sim}0.1$ rest-mass of kinetic energy at infinity.

Because achieving a time-independent spacetime means that gravity cannot alter particle energy, any further evolution of the distribution function is due to two mechanisms: pressure gradients doing work and removal of mass as the black hole captures additional mass.

\subsection{Angular momentum}

Perhaps the greatest surprise in our analysis is that even during the few milliseconds before collapse to a black hole when the spacetime is most dynamic and asymmetric, the net torque exerted on the fluid elements that survive collapse is delivered in roughly comparable amounts by gravity and hydrodynamic forces.

\begin{figure*}
\includegraphics[width=0.49\linewidth]{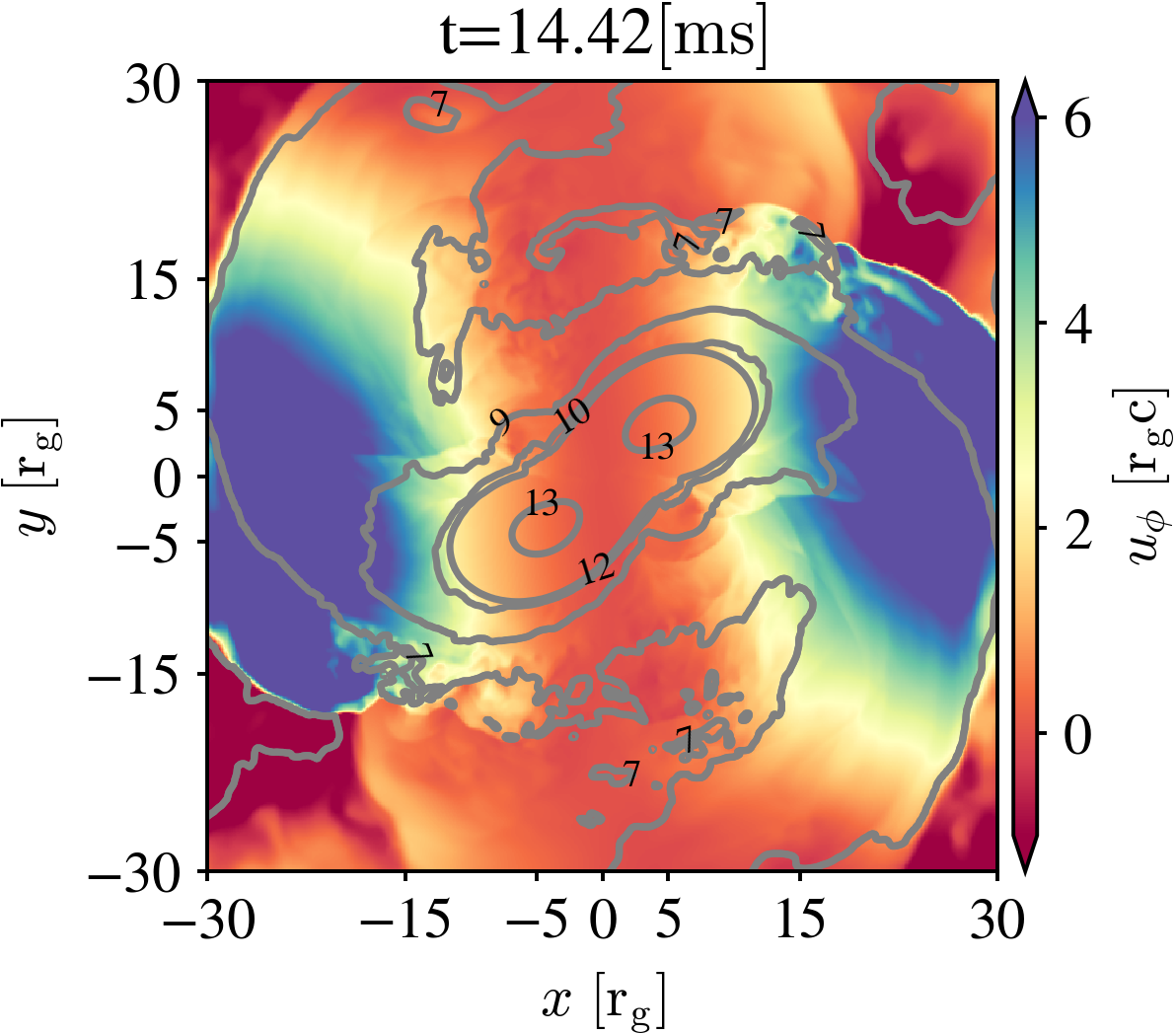}
\includegraphics[width=0.49\linewidth]{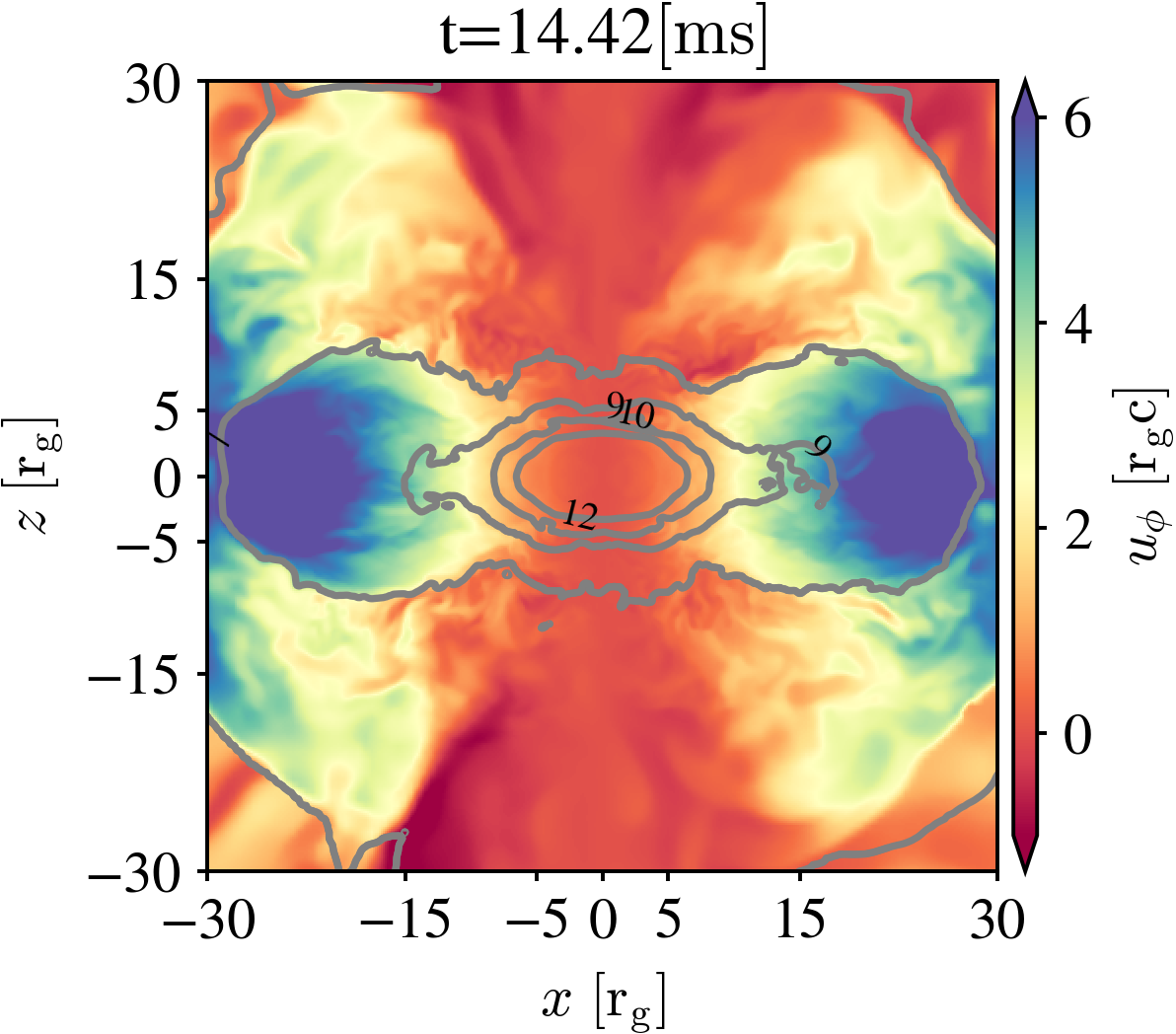}
\caption{Map of $u_{\phi}$  of the fluid (color contours) in the $x$-$y$ plane (left panel) and in the $x$-$z$ plane (right panel) at time \qty{14.42}{\ms}, when the contact surface between the two stars contains the $x$-axis. Rest-mass density $\rho$ shown in gray contours with attached numbers (e.g., 9, 12) giving $\log_{10}\rho$ in \unit{\g\cdot\cm\tothe{-3}}.}

\label{fig:uphi_14.42ms}
\end{figure*}

The key factor promoting the relative strength of pressure forces is that the torques due to pressure are almost entirely prograde, whereas the total retrograde torque due to gravity is almost as large as the total prograde torque. The preponderance of prograde hydrodynamic torque follows directly from the rotation of the newly---merged neutron star. As shown in the top-left and middle-left panels of Figure.~\ref{fig:dj_before} (at times ${\approx}1$--\qty{2}{\ms} before black hole collapse), tidal forces stretch the merged star's diameter along a particular axis in its frame. The same density data is shown in Figure~\ref{fig:uphi_14.42ms}, but contrasted with the specific angular momentum rather than torque. Because the merged pair of neutron stars rotate effectively as a solid body, the specific angular momentum increases in proportion to the square of the distance from the rotation axis. When $u_\phi \gtrsim 3$, its rotational kinetic energy is large enough to unbind it from the star (see Fig.~\ref{fig:uphi_14.42ms}; because the merged star's rotation rate is the orbital frequency at merger, it is automatically very close to the break-up rate). The spiral streams of high angular momentum material visible in the left panel of Figure~\ref{fig:uphi_14.42ms} show this material, which forms the tidal component of the dynamical ejecta.

On the other hand, as illustrated in the right panel of Figure~\ref{fig:uphi_14.42ms}, again because $u_\phi$ increases quadratically with distance from the rotation axis, the matter squeezed out of the merged stars near the contact surface has much lower angular momentum. This lower angular momentum material is shocked as the distended merged star rotates, receiving prograde angular momentum from the outer layers of the rotating star.

Angular momentum transfer between the star and its immediately surrounding gas ceases when the weakly-bound debris acquires a specific angular momentum matching that of the star's surface, which we identify with the sharpest density gradient, i.e., where the density falls below $\sim 10^{10}$~g~cm$^{-3}$; at that point, there is no difference between their angular velocities. We speculate that this effect regulates the mean specific angular momentum of the bound debris, a speculation supported by the close agreement between $u_\phi$ at the outer edge of the rotating star ${\approx}2.5 r_g c$ (left panel, Fig.~\ref{fig:uphi_14.42ms}) and the peak of the debris' angular momentum distribution (right panel, Fig.~\ref{fig:ut_uphi_distribution}).

The merged star's rotation also explains why, despite the self-cancellation of the gravitational torque, its net effect is prograde. Even though the gravitational torque alternates in sign, the acceleration a fluid element receives in a prograde torque region reduces the difference in angular velocity between it and the merged neutron star's rotation rate.
Because the shape of the gravitational torque distribution is tied to the figure of the asymmetric star, the fluid element then moves more slowly through the torque's sign changes.  Its exposure to prograde torque is thereby lengthened, while the deceleration suffered in a negative torque region does the opposite.

After the new black hole relaxes to a Kerr spacetime, genuine gravitational torques disappear.
From that point onward, the only evolution in the angular momentum distribution comes from redistribution by hydrodynamic forces. Because the only signal of a preferred rotation direction is spacetime frame-dragging, which both declines rapidly with distance from the black hole and exerts weak effects on matter whose angular momentum is parallel to the black hole's, the hydrodynamic torques are irregular, and their sign can be either positive or negative. The result is a broadening of the angular momentum distribution, including a tail stretching well out into the retrograde range.

Over the first several milliseconds after collapse, the magnitude of the hydrodynamic torque first increases and then decreases (see Figs.~\ref{fig:J_evolution} and \ref{fig:dj_after}). The increase is due to the expansion of higher angular momentum material toward the radial range at which it can, with sufficient energy loss, acquire a circular orbit. The subsequent decrease is due to the diffusive, and therefore smoothing, character of irregular fluid motions and, to a lesser degree, the loss of debris mass as the matter with the least angular momentum falls into the black hole.

\subsection{Long-term effects of the debris angular momentum distribution}

In formal solutions for accretion disks in inflow equilibrium, the angular momentum distribution is generally a power-law with a cut-off at the low end associated with a combination of general relativistic effects and behavior of internal stress near the ISCO. For example, if the stress/pressure ratio $\alpha$ is a constant and the fluid responds to compression like an ideal gas,
\begin{equation}\label{eq:alpha-model}
{dm/dj} \propto \alpha^{-4/5} j^{-1/5} (R_T/R_R)^{4/5}.
\end{equation}

Here $R_T$ and $R_R$ embody the low-$j$ cut-off and their ratio is generally close to unity \citep{Krolik1999}. The angular momentum distribution with which the disk around a freshly-made neutron star merger is furnished is very different: in rough terms, it is a narrow Gaussian peaking at a few $r_g c$ (in the example studied here, ${\approx}2.5 r_g c$) then a slowly-declining power-law.

We therefore conclude that debris disks around neutron star mergers are very far from a state of inflow equilibrium. Predictions of their subsequent evolution beginning from an equilibrium state \citep[see e.g,][]{Fernandez+15, Radice+16, Just+15, SiegelMetzger17PRL, Miller+19} are therefore very unlikely to reflect such disks' actual evolution. That this should be so is hardly a surprise. There is nothing in the structure of neutron stars or the dynamics of their merger that has any connection to the processes, such as magnetic stresses arising from orbital shear-correlated MHD turbulence, governing their local rates of mass flow.

In addition, because the gross surface density profile of a disk evolves on the same timescale as it accretes, it immediately follows that by the time the disk might relax toward an equilibrium state, it has also become depleted by accretion onto the black hole (or massive neutron star). Post-merger disks are therefore intrinsically transient. 

A further distinctly non-equilibrium feature of these disks is that these disks are born with a magnetic field that has yet to be amplified by the magnetorotational instability to its fully saturated intensity. Because the disk begins with its mass confined to a rather narrow annulus, one might then ask how the magnetic field develops at larger radii as the disk spreads out. As the gas moves outward, does it carry fully-developed turbulent magnetic field, or does the matter move outward and then wait for the magnetorotational instability to grow? We speculate that the first option is most likely on the ground that the surface density radial diffusion time is much longer than the time for the turbulence to reach saturation locally (${\sim}10$ orbits).\footnote{ Fluid elements in an accretion disk have a characteristic radial diffusion rate $\sim (h/r)^2 \Omega \langle B_r B_\phi\rangle/(4\pi P)$, where $h$ is the vertical scale height, $\Omega$ is the local orbital frequency, and $B_{r,\phi}$ represents the radial and azimuthal components of the magnetic field.  Because the field is turbulent, the net stress at a given location is the time-average of $B_r B_\phi/4\pi$.}

\subsection{Generality of these conclusions}

Because our investigation is based entirely on a single simulation, one might reasonably ask whether our conclusions are sensitive to either parameters (e.g., the neutron star masses) or choices about poorly-understood physics (e.g., the nuclear equation of state). We cannot confidently answer this question at a quantitative level regarding the relative contributions of the different forces. Nonetheless, there are several plausibility arguments supporting our conclusions' qualitative generality.

The importance of hydrodynamic torques relative to gravitational torques pre-collapse rests on a very general property: the quadrupolar character of the torque-inducing gravitational field. A rotationally symmetric torque distribution with alternating signs of the torque is an immediate consequence, and mutual cancellation directly follows.

That hydrodynamic forces dominate post-collapse evolution must certainly be general. Once the merged neutron star collapses to a black hole, its gravity must be both axisymmetric and time-steady, removing any possibility of gravitational change in the debris' angular momentum or energy. Even if the remnant neutron star is stable, it, too, quickly relaxes to an axisymmetric time-steady state. The only mechanism by which it could affect debris disk dynamics is through interaction with its magnetic field.

The narrowness of the angular momentum distribution follows from the existence of a characteristic value of the angular momentum, the value in the outer surface of the rotating merged neutron star. The particular value of this quantity is tied to the break-up specific angular momentum of the neutron star, $\propto (M_1 + M_2)^{1/2} R_{\rm ns}^{1/2} \eta$, where $\eta$ is the symmetric mass-ratio $M_1 M_2/(M_1 + M_2)^2$. Here the neutron star radius $R_{\rm ns}$ is taken to be independent of neutron star mass because it is a rather slowly-varying function of that quantity \citep[see e.g,][]{Lasota+96, Koranda+97, Douchin&Haensel01, ReadJ+09, Bauswein+12_EoS, Fryer+15, Ozel&Freire16, MargalitMetzger17, WeiJ+19, Lattimer19, Konstantinou&Morsink22}. Because neutron star masses have a comparatively narrow range, the mean specific angular momentum of the debris therefore varies over only a factor ${\sim}O(1)$. An extensive study of this quantity for many choices of total mass, mass ratio, and equation of state, including cases leading to a long-lived neutron star as well as rapid collapse to a black hole arrived at exactly this result: that $\langle u_\phi \rangle$ in the debris is always found between ${\approx}2.5$--$4 r_g c$ \citep{Camilletti+2024}.

\section{Summary}

We began this paper with a question: what determines the initial structure of the debris disk, the machine responsible
for nucleosynthesis and many of the EM signals associated with neutron star mergers?
There are numerous elements contributing to the answer to this question. The principal ones we have found include:

The mechanisms that produce the energy and angular momentum of the debris disk's matter act during two distinct periods. The first extends from the moment the neutron stars touch until the end of the dynamic spacetime, i.e., until a black hole forms and its spacetime relaxes to Kerr, or a long-lived neutron star similarly settles into a secularly evolving equilibrium. The mean values of orbital energy and angular momentum, $u_t$ and $u_\phi$, are determined during this period.

The second begins from the time the remnant, whether a black hole or a massive neutron star, relaxes to a nearly time-steady and axisymmetric state.  During this epoch, both the total angular momentum and the total energy of the debris outside the remnant are constant except as a result of debris leaving the region, whether by accretion onto the remnant or by becoming unbound and escaping the system. However, irregular pressure forces can exchange angular momentum and energy between fluid parcels, broadening their distributions, although the extent of the broadening can be curtailed by mass leaving the disk.

During the first period, although the absolute magnitude of torques due to non-axisymmetric gravity dominates hydrodynamic forces by roughly an order of magnitude, the geometric symmetry of sign changes in the gravitational torque substantially diminish its net effect, leaving its influence on bound debris angular momentum roughly comparable to that of pressure forces. The hydrodynamic torques transfer angular momentum from the merged neutron star to the surrounding gas because the gas escaping from the merged neutron star's interior has less angular momentum than the rotating outer layers of the merged neutron star, which drive shocks through the surrounding gas.

In this period, the net effect of hydrodynamic forces on the energy of the fluid is roughly two orders of magnitude greater than that of gravity. In the simulation analyzed here, for the first 2--\qty{3}{\ms} after contact the material tends to lose energy, while in the last ${\sim}\qty{1}{\ms}$ before collapse to a black hole, it gains energy. The matter that enters the debris disk is, not surprisingly, the matter having the greatest energy.

During the second period, gravity cannot change the fluid's total amount of either angular momentum or energy, but shocks and other events in the material can exchange both quantities, altering the distribution functions of both quantities.

The end-result is that the mass of the debris disk is initially strongly concentrated in radius because its angular momentum distribution is very narrow. The mean specific angular momentum of the orbiting matter is close to that of the outer regions of the rotating merged neutron star before it relaxes to axisymmetry. When the fluid's mean angular momentum has this value, if a black hole forms, the great majority of the disk is found close to its ISCO; if the remnant is a massive neutron star, the disk's material is just outside the star's surface. In both cases, its radial profile of surface density is far from equilibrium and must evolve substantially once the material develops internal stresses (presumably magnetic) capable of transporting angular momentum radially.

\section *{Acknowledgements}
\begin{acknowledgments} \label{Sec:ack}
This work was partially supported by NASA TCAN grants NNH17ZDA001N and 80NSSC24K0100. 
We all thank Yosef Zlochower for his insight about gauge effects in numerical relativity simulations. YZ thanks Paz Beniamini and Elias Most for helpful discussions. JK would like to thank Daniel Siegel for encouraging this inquiry at its start. L.R.W.\ and Z.B.E.\ gratefully acknowledge support from NSF awards PHY-1806596, PHY-2110352, OAC-2004311, as well as NASA award ISFM-80NSSC18K0538. 
This research made use of Idaho National Laboratory computing resources, which are supported by the Office of Nuclear Energy of the U.S. Department of Energy and the Nuclear Science User Facilities under Contract No. DE-AC07-05ID14517.
\end{acknowledgments}

\software{astropy \citep{Astropy+18}, 
          \igm \citep{Etienne+15}, Matplotlib \citep{Hunter07_matplotlib}, and Numpy \citep{2020NumPy-Array}.}

\bibliography{refBNS23}{}

\begin{thebibliography}{}
\expandafter\ifx\csname natexlab\endcsname\relax\def\natexlab#1{#1}\fi
\providecommand{\url}[1]{\href{#1}{#1}}
\providecommand{\dodoi}[1]{doi:~\href{http://doi.org/#1}{\nolinkurl{#1}}}
\providecommand{\doeprint}[1]{\href{http://ascl.net/#1}{\nolinkurl{http://ascl.net/#1}}}
\providecommand{\doarXiv}[1]{\href{https://arxiv.org/abs/#1}{\nolinkurl{https://arxiv.org/abs/#1}}}

\bibitem[{{Abbott} {et~al.}(2017{\natexlab{a}}){Abbott}, {Abbott}, {Abbott}, {Acernese}, {Ackley}, {Adams}, {Adams}, {Addesso}, {Adhikari}, {Adya}, \& et~al.}]{Abbott17:gw}
{Abbott}, B.~P., {Abbott}, R., {Abbott}, T.~D., {et~al.} 2017{\natexlab{a}}, \nat, 551, 85, \dodoi{10.1038/nature24471}

\bibitem[{{Abbott} {et~al.}(2017{\natexlab{b}}){Abbott}, {Abbott}, {Abbott}, {Acernese}, {Ackley}, {Adams}, {Adams}, {Addesso}, {Adhikari}, {Adya}, \& et~al.}]{Abbott17a}
---. 2017{\natexlab{b}}, \apjl, 848, L12, \dodoi{10.3847/2041-8213/aa91c9}

\bibitem[{{Abbott} {et~al.}(2017{\natexlab{c}}){Abbott}, {Abbott}, {Abbott}, {Acernese}, {Ackley}, {Adams}, {Adams}, {Addesso}, {Adhikari}, {Adya}, \& et~al.}]{Abbott17b}
---. 2017{\natexlab{c}}, \nat, 551, 85, \dodoi{10.1038/nature24471}

\bibitem[{{Akmal} {et~al.}(1998){Akmal}, {Pandharipande}, \& {Ravenhall}}]{Akmal+98}
{Akmal}, A., {Pandharipande}, V.~R., \& {Ravenhall}, D.~G. 1998, \prc, 58, 1804, \dodoi{10.1103/PhysRevC.58.1804}

\bibitem[{Alcubierre {et~al.}(2003)Alcubierre, Bruegmann, Diener, Koppitz, Pollney, Seidel, \& Takahashi}]{Alcubierre:2002kk}
Alcubierre, M., Bruegmann, B., Diener, P., {et~al.} 2003, Phys. Rev. D, 67, 084023, \dodoi{10.1103/PhysRevD.67.084023}

\bibitem[{{Alexander} {et~al.}(2017){Alexander}, {Berger}, {Fong}, {Williams}, {Guidorzi}, {Margutti}, {Metzger}, {Annis}, {Blanchard}, {Brout}, {Brown}, {Chen}, {Chornock}, {Cowperthwaite}, {Drout}, {Eftekhari}, {Frieman}, {Holz}, {Nicholl}, {Rest}, {Sako}, {Soares-Santos}, \& {Villar}}]{Alexander2017}
{Alexander}, K.~D., {Berger}, E., {Fong}, W., {et~al.} 2017, \apjl, 848, L21, \dodoi{10.3847/2041-8213/aa905d}

\bibitem[{{Alexander} {et~al.}(2018){Alexander}, {Margutti}, {Blanchard}, {Fong}, {Berger}, {Hajela}, {Eftekhari}, {Chornock}, {Cowperthwaite}, {Giannios}, {Guidorzi}, {Kathirgamaraju}, {MacFadyen}, {Metzger}, {Nicholl}, {Sironi}, {Villar}, {Williams}, {Xie}, \& {Zrake}}]{Alexander2018}
{Alexander}, K.~D., {Margutti}, R., {Blanchard}, P.~K., {et~al.} 2018, ArXiv e-prints.
\newblock \doarXiv{1805.02870}

\bibitem[{{Alford} {et~al.}(2019){Alford}, {Han}, \& {Schwenzer}}]{Alford_Han_Schwenzer19}
{Alford}, M.~G., {Han}, S., \& {Schwenzer}, K. 2019, Journal of Physics G Nuclear Physics, 46, 114001, \dodoi{10.1088/1361-6471/ab337a}

\bibitem[{{Astropy Collaboration} {et~al.}(2018){Astropy Collaboration}, {Price-Whelan}, {Sip{\H{o}}cz}, {G{\"u}nther}, {Lim}, {Crawford}, {Conseil}, {Shupe}, {Craig}, {Dencheva}, {Ginsburg}, {VanderPlas}, {Bradley}, {P{\'e}rez-Su{\'a}rez}, {de Val-Borro}, {Aldcroft}, {Cruz}, {Robitaille}, {Tollerud}, {Ardelean}, {Babej}, {Bach}, {Bachetti}, {Bakanov}, {Bamford}, {Barentsen}, {Barmby}, {Baumbach}, {Berry}, {Biscani}, {Boquien}, {Bostroem}, {Bouma}, {Brammer}, {Bray}, {Breytenbach}, {Buddelmeijer}, {Burke}, {Calderone}, {Cano Rodr{\'\i}guez}, {Cara}, {Cardoso}, {Cheedella}, {Copin}, {Corrales}, {Crichton}, {D'Avella}, {Deil}, {Depagne}, {Dietrich}, {Donath}, {Droettboom}, {Earl}, {Erben}, {Fabbro}, {Ferreira}, {Finethy}, {Fox}, {Garrison}, {Gibbons}, {Goldstein}, {Gommers}, {Greco}, {Greenfield}, {Groener}, {Grollier}, {Hagen}, {Hirst}, {Homeier}, {Horton}, {Hosseinzadeh}, {Hu}, {Hunkeler}, {Ivezi{\'c}}, {Jain}, {Jenness}, {Kanarek}, {Kendrew}, {Kern}, {Kerzendorf}, {Khvalko}, {King}, {Kirkby}, {Kulkarni},
  {Kumar}, {Lee}, {Lenz}, {Littlefair}, {Ma}, {Macleod}, {Mastropietro}, {McCully}, {Montagnac}, {Morris}, {Mueller}, {Mumford}, {Muna}, {Murphy}, {Nelson}, {Nguyen}, {Ninan}, {N{\"o}the}, {Ogaz}, {Oh}, {Parejko}, {Parley}, {Pascual}, {Patil}, {Patil}, {Plunkett}, {Prochaska}, {Rastogi}, {Reddy Janga}, {Sabater}, {Sakurikar}, {Seifert}, {Sherbert}, {Sherwood-Taylor}, {Shih}, {Sick}, {Silbiger}, {Singanamalla}, {Singer}, {Sladen}, {Sooley}, {Sornarajah}, {Streicher}, {Teuben}, {Thomas}, {Tremblay}, {Turner}, {Terr{\'o}n}, {van Kerkwijk}, {de la Vega}, {Watkins}, {Weaver}, {Whitmore}, {Woillez}, {Zabalza}, \& {Astropy Contributors}}]{Astropy+18}
{Astropy Collaboration}, {Price-Whelan}, A.~M., {Sip{\H{o}}cz}, B.~M., {et~al.} 2018, \aj, 156, 123, \dodoi{10.3847/1538-3881/aabc4f}

\bibitem[{{Baiotti}(2019)}]{Baiotti2019}
{Baiotti}, L. 2019, Progress in Particle and Nuclear Physics, 109, 103714, \dodoi{10.1016/j.ppnp.2019.103714}

\bibitem[{{Baiotti} \& {Rezzolla}(2017)}]{Baiotti&Rezzolla17}
{Baiotti}, L., \& {Rezzolla}, L. 2017, Reports on Progress in Physics, 80, 096901, \dodoi{10.1088/1361-6633/aa67bb}

\bibitem[{{Balasubramanian} {et~al.}(2021){Balasubramanian}, {Corsi}, {Mooley}, {Brightman}, {Hallinan}, {Hotokezaka}, {Kaplan}, {Lazzati}, \& {Murphy}}]{Balasubramanian+21_radio}
{Balasubramanian}, A., {Corsi}, A., {Mooley}, K.~P., {et~al.} 2021, \apjl, 914, L20, \dodoi{10.3847/2041-8213/abfd38}

\bibitem[{{Balberg} {et~al.}(1999){Balberg}, {Lichtenstadt}, \& {Cook}}]{Balberg+99}
{Balberg}, S., {Lichtenstadt}, I., \& {Cook}, G.~B. 1999, \apjs, 121, 515, \dodoi{10.1086/313196}

\bibitem[{Baumgarte \& Naculich(2007)}]{BaumgarteNaculich2007}
Baumgarte, T.~W., \& Naculich, S.~G. 2007, Phys. Rev. D, 75, 067502, \dodoi{10.1103/PhysRevD.75.067502}

\bibitem[{{Bauswein} {et~al.}(2013){Bauswein}, {Goriely}, \& {Janka}}]{Bauswein+13}
{Bauswein}, A., {Goriely}, S., \& {Janka}, H.~T. 2013, \apj, 773, 78, \dodoi{10.1088/0004-637X/773/1/78}

\bibitem[{{Bauswein} {et~al.}(2012){Bauswein}, {Janka}, {Hebeler}, \& {Schwenk}}]{Bauswein+12_EoS}
{Bauswein}, A., {Janka}, H.~T., {Hebeler}, K., \& {Schwenk}, A. 2012, \prd, 86, 063001, \dodoi{10.1103/PhysRevD.86.063001}

\bibitem[{{Baym} {et~al.}(2018){Baym}, {Hatsuda}, {Kojo}, {Powell}, {Song}, \& {Takatsuka}}]{Baym+18}
{Baym}, G., {Hatsuda}, T., {Kojo}, T., {et~al.} 2018, Reports on Progress in Physics, 81, 056902, \dodoi{10.1088/1361-6633/aaae14}

\bibitem[{{Beniamini} {et~al.}(2022){Beniamini}, {Gill}, \& {Granot}}]{Beniamini+22}
{Beniamini}, P., {Gill}, R., \& {Granot}, J. 2022, \mnras, 515, 555, \dodoi{10.1093/mnras/stac1821}

\bibitem[{{Berti} {et~al.}(2018){Berti}, {Yagi}, {Yang}, \& {Yunes}}]{Berti+18}
{Berti}, E., {Yagi}, K., {Yang}, H., \& {Yunes}, N. 2018, General Relativity and Gravitation, 50, 49, \dodoi{10.1007/s10714-018-2372-6}

\bibitem[{{Berti} {et~al.}(2015){Berti}, {Barausse}, {Cardoso}, {Gualtieri}, {Pani}, {Sperhake}, {Stein}, {Wex}, {Yagi}, {Baker}, {Burgess}, {Coelho}, {Doneva}, {De Felice}, {Ferreira}, {Freire}, {Healy}, {Herdeiro}, {Horbatsch}, {Kleihaus}, {Klein}, {Kokkotas}, {Kunz}, {Laguna}, {Lang}, {Li}, {Littenberg}, {Matas}, {Mirshekari}, {Okawa}, {Radu}, {O'Shaughnessy}, {Sathyaprakash}, {Van Den Broeck}, {Winther}, {Witek}, {Emad Aghili}, {Alsing}, {Bolen}, {Bombelli}, {Caudill}, {Chen}, {Degollado}, {Fujita}, {Gao}, {Gerosa}, {Kamali}, {Silva}, {Rosa}, {Sadeghian}, {Sampaio}, {Sotani}, \& {Zilhao}}]{Berti15}
{Berti}, E., {Barausse}, E., {Cardoso}, V., {et~al.} 2015, Classical and Quantum Gravity, 32, 243001, \dodoi{10.1088/0264-9381/32/24/243001}

\bibitem[{{Braeck}(2023)}]{Braeck23}
{Braeck}, S. 2023, Universe, 9, 120, \dodoi{10.3390/universe9030120}

\bibitem[{{Breschi} {et~al.}(2021){Breschi}, {Perego}, {Bernuzzi}, {Del Pozzo}, {Nedora}, {Radice}, \& {Vescovi}}]{Breschi+21}
{Breschi}, M., {Perego}, A., {Bernuzzi}, S., {et~al.} 2021, arXiv e-prints, arXiv:2101.01201.
\newblock \doarXiv{2101.01201}

\bibitem[{{Camilletti} {et~al.}(2024){Camilletti}, {Perego}, {Guercilena}, {Bernuzzi}, \& {Radice}}]{Camilletti+2024}
{Camilletti}, A., {Perego}, A., {Guercilena}, F.~M., {Bernuzzi}, S., \& {Radice}, D. 2024, arXiv e-prints, arXiv:2401.04102, \dodoi{10.48550/arXiv.2401.04102}

\bibitem[{{Ciolfi} \& {Kalinani}(2020)}]{Ciolfi&Kalinani20}
{Ciolfi}, R., \& {Kalinani}, J.~V. 2020, \apjl, 900, L35, \dodoi{10.3847/2041-8213/abb240}

\bibitem[{{Combi} \& {Siegel}(2023)}]{Combi_Siegel23}
{Combi}, L., \& {Siegel}, D.~M. 2023, \prl, 131, 231402, \dodoi{10.1103/PhysRevLett.131.231402}

\bibitem[{{Coulter} {et~al.}(2017){Coulter}, {Foley}, {Kilpatrick}, {Drout}, {Piro}, {Shappee}, {Siebert}, {Simon}, {Ulloa}, {Kasen}, {Madore}, {Murguia-Berthier}, {Pan}, {Prochaska}, {Ramirez-Ruiz}, {Rest}, \& {Rojas-Bravo}}]{Coulter17}
{Coulter}, D.~A., {Foley}, R.~J., {Kilpatrick}, C.~D., {et~al.} 2017, Science, 358, 1556, \dodoi{10.1126/science.aap9811}

\bibitem[{{Cowan} {et~al.}(2021){Cowan}, {Sneden}, {Lawler}, {Aprahamian}, {Wiescher}, {Langanke}, {Mart{\'\i}nez-Pinedo}, \& {Thielemann}}]{Cowan2021}
{Cowan}, J.~J., {Sneden}, C., {Lawler}, J.~E., {et~al.} 2021, Reviews of Modern Physics, 93, 015002, \dodoi{10.1103/RevModPhys.93.015002}

\bibitem[{{Damour} {et~al.}(2012){Damour}, {Nagar}, \& {Villain}}]{Damour_Nagar_Viilain+12}
{Damour}, T., {Nagar}, A., \& {Villain}, L. 2012, \prd, 85, 123007, \dodoi{10.1103/PhysRevD.85.123007}

\bibitem[{{Davies} {et~al.}(1994){Davies}, {Benz}, {Piran}, \& {Thielemann}}]{Davies+94}
{Davies}, M.~B., {Benz}, W., {Piran}, T., \& {Thielemann}, F.~K. 1994, \apj, 431, 742, \dodoi{10.1086/174525}

\bibitem[{{Dietrich} {et~al.}(2021){Dietrich}, {Hinderer}, \& {Samajdar}}]{Dietrich+21}
{Dietrich}, T., {Hinderer}, T., \& {Samajdar}, A. 2021, General Relativity and Gravitation, 53, 27, \dodoi{10.1007/s10714-020-02751-6}

\bibitem[{{Dietrich} {et~al.}(2017){Dietrich}, {Ujevic}, {Tichy}, {Bernuzzi}, \& {Br{\"u}gmann}}]{Dietrich+17}
{Dietrich}, T., {Ujevic}, M., {Tichy}, W., {Bernuzzi}, S., \& {Br{\"u}gmann}, B. 2017, \prd, 95, 024029, \dodoi{10.1103/PhysRevD.95.024029}

\bibitem[{{Douchin} \& {Haensel}(2001)}]{Douchin&Haensel01}
{Douchin}, F., \& {Haensel}, P. 2001, \aap, 380, 151, \dodoi{10.1051/0004-6361:20011402}

\bibitem[{{Echeverria}(1989)}]{EcheverriaFernando88}
{Echeverria}, F. 1989, \prd, 40, 3194, \dodoi{10.1103/PhysRevD.40.3194}

\bibitem[{{Eichler} {et~al.}(1989){Eichler}, {Livio}, {Piran}, \& {Schramm}}]{Eichler+89}
{Eichler}, D., {Livio}, M., {Piran}, T., \& {Schramm}, D.~N. 1989, \nat, 340, 126, \dodoi{10.1038/340126a0}

\bibitem[{Etienne {et~al.}(2015)Etienne, Paschalidis, Haas, M\"osta, \& Shapiro}]{Etienne+15}
Etienne, Z.~B., Paschalidis, V., Haas, R., M\"osta, P., \& Shapiro, S.~L. 2015, Class. Quant. Grav., 32, 175009, \dodoi{10.1088/0264-9381/32/17/175009}

\bibitem[{Feo {et~al.}(2017)Feo, De~Pietri, Maione, \& L\"offler}]{Feo+16}
Feo, A., De~Pietri, R., Maione, F., \& L\"offler, F. 2017, Class. Quant. Grav., 34, 034001, \dodoi{10.1088/1361-6382/aa51fa}

\bibitem[{{Fern{\'a}ndez} {et~al.}(2015){Fern{\'a}ndez}, {Kasen}, {Metzger}, \& {Quataert}}]{Fernandez+15}
{Fern{\'a}ndez}, R., {Kasen}, D., {Metzger}, B.~D., \& {Quataert}, E. 2015, \mnras, 446, 750, \dodoi{10.1093/mnras/stu2112}

\bibitem[{{Foucart} {et~al.}(2019){Foucart}, {Duez}, {Kidder}, {Nissanke}, {Pfeiffer}, \& {Scheel}}]{Foucart+19}
{Foucart}, F., {Duez}, M.~D., {Kidder}, L.~E., {et~al.} 2019, \prd, 99, 103025, \dodoi{10.1103/PhysRevD.99.103025}

\bibitem[{{Foucart} {et~al.}(2013){Foucart}, {Deaton}, {Duez}, {Kidder}, {MacDonald}, {Ott}, {Pfeiffer}, {Scheel}, {Szilagyi}, \& {Teukolsky}}]{Foucart+13}
{Foucart}, F., {Deaton}, M.~B., {Duez}, M.~D., {et~al.} 2013, \prd, 87, 084006, \dodoi{10.1103/PhysRevD.87.084006}

\bibitem[{{Foucart} {et~al.}(2016){Foucart}, {Haas}, {Duez}, {O'Connor}, {Ott}, {Roberts}, {Kidder}, {Lippuner}, {Pfeiffer}, \& {Scheel}}]{Foucart+16}
{Foucart}, F., {Haas}, R., {Duez}, M.~D., {et~al.} 2016, \prd, 93, 044019, \dodoi{10.1103/PhysRevD.93.044019}

\bibitem[{{Frolov} \& {Frolov}(2014)}]{Frolov_V_P2014}
{Frolov}, A.~V., \& {Frolov}, V.~P. 2014, \prd, 90, 124010, \dodoi{10.1103/PhysRevD.90.124010}

\bibitem[{{Fryer} {et~al.}(2015){Fryer}, {Belczynski}, {Ramirez-Ruiz}, {Rosswog}, {Shen}, \& {Steiner}}]{Fryer+15}
{Fryer}, C.~L., {Belczynski}, K., {Ramirez-Ruiz}, E., {et~al.} 2015, \apj, 812, 24, \dodoi{10.1088/0004-637X/812/1/24}

\bibitem[{{Fujibayashi} {et~al.}(2018){Fujibayashi}, {Kiuchi}, {Nishimura}, {Sekiguchi}, \& {Shibata}}]{Fujibayashi+18}
{Fujibayashi}, S., {Kiuchi}, K., {Nishimura}, N., {Sekiguchi}, Y., \& {Shibata}, M. 2018, \apj, 860, 64, \dodoi{10.3847/1538-4357/aabafd}

\bibitem[{{Goldstein} {et~al.}(2017){Goldstein}, {Veres}, {Burns}, {Briggs}, {Hamburg}, {Kocevski}, {Wilson-Hodge}, {Preece}, {Poolakkil}, {Roberts}, {Hui}, {Connaughton}, {Racusin}, {von Kienlin}, {Dal Canton}, {Christensen}, {Littenberg}, {Siellez}, {Blackburn}, {Broida}, {Bissaldi}, {Cleveland}, {Gibby}, {Giles}, {Kippen}, {McBreen}, {McEnery}, {Meegan}, {Paciesas}, \& {Stanbro}}]{Goldstein17}
{Goldstein}, A., {Veres}, P., {Burns}, E., {et~al.} 2017, \apjl, 848, L14, \dodoi{10.3847/2041-8213/aa8f41}

\bibitem[{{Gourgoulhon} {et~al.}(Accessed in 2022){Gourgoulhon}, {Grandcl{\'e}ment}, \& {Novak}}]{lorene_website}
{Gourgoulhon}, E., {Grandcl{\'e}ment}, P., \& {Novak}, J. Accessed in 2022, {L}{O}{R}{E}{N}{E}: {L}angage {O}bjet pour la {R}{E}lativit{\'e} {N}um{\'e}riqu{E}

\bibitem[{Gourgoulhon {et~al.}(2001)Gourgoulhon, Grandclement, Taniguchi, Marck, \& Bonazzola}]{Gourgoulhon+01}
Gourgoulhon, E., Grandclement, P., Taniguchi, K., Marck, J.-A., \& Bonazzola, S. 2001, Phys. Rev. D, 63, 064029, \dodoi{10.1103/PhysRevD.63.064029}

\bibitem[{{Guidorzi} {et~al.}(2017){Guidorzi}, {Margutti}, {Brout}, {Scolnic}, {Fong}, {Alexander}, {Cowperthwaite}, {Annis}, {Berger}, {Blanchard}, {Chornock}, {Coppejans}, {Eftekhari}, {Frieman}, {Huterer}, {Nicholl}, {Soares-Santos}, {Terreran}, {Villar}, \& {Williams}}]{Guidorzi17}
{Guidorzi}, C., {Margutti}, R., {Brout}, D., {et~al.} 2017, \apjl, 851, L36, \dodoi{10.3847/2041-8213/aaa009}

\bibitem[{Haas {et~al.}(2022)Haas, Cheng, Diener, Etienne, Ficarra, Ikeda, Kalyanaraman, Kuo, Leung, Tian, Tsao, Wen, Alcubierre, Alic, Allen, Ansorg, Babiuc-Hamilton, Baiotti, Benger, Bentivegna, Bernuzzi, Bode, Bozzola, Brandt, Brendal, Bruegmann, Campanelli, Cipolletta, Corvino, Cupp, Pietri, Dima, Dimmelmeier, Dooley, Dorband, Elley, Khamra, Faber, Font, Frieben, Giacomazzo, Goodale, Gundlach, Hawke, Hawley, Hinder, Huerta, Husa, Iyer, Ji, Johnson, Joshi, Kankani, Kastaun, Kellermann, Knapp, Koppitz, Laguna, Lanferman, Lasky, Löffler, Macpherson, Masso, Menger, Merzky, Miller, Miller, Moesta, Montero, Mundim, Nelson, Nerozzi, Noble, Ott, Paruchuri, Pollney, Price, Radice, Radke, Reisswig, Rezzolla, Richards, Rideout, Ripeanu, Sala, Schewtschenko, Schnetter, Schutz, Seidel, Seidel, Shalf, Sible, Sperhake, Stergioulas, Suen, Szilagyi, Takahashi, Thomas, Thornburg, Tobias, Tonita, Walker, Wan, Wardell, Werneck, Witek, Zilhão, Zink, \& Zlochower}]{roland_haas_2022_7245853}
Haas, R., Cheng, C.-H., Diener, P., {et~al.} 2022, The Einstein Toolkit, The "Sophie Kowalevski" release, ET\_2022\_11,  Zenodo, \dodoi{10.5281/zenodo.7245853}

\bibitem[{{Haensel}(2003)}]{Haensel2003}
{Haensel}, P. 2003, in EAS Publications Series, Vol.~7, EAS Publications Series, ed. C.~{Motch} \& J.-M. {Hameury}, 249, \dodoi{10.1051/eas:2003043}

\bibitem[{{Hajela} {et~al.}(2022){Hajela}, {Margutti}, {Bright}, {Alexander}, {Metzger}, {Nedora}, {Kathirgamaraju}, {Margalit}, {Radice}, {Guidorzi}, {Berger}, {MacFadyen}, {Giannios}, {Chornock}, {Heywood}, {Sironi}, {Gottlieb}, {Coppejans}, {Laskar}, {Cendes}, {Duran}, {Eftekhari}, {Fong}, {McDowell}, {Nicholl}, {Xie}, {Zrake}, {Bernuzzi}, {Broekgaarden}, {Kilpatrick}, {Terreran}, {Villar}, {Blanchard}, {Gomez}, {Hosseinzadeh}, {Matthews}, \& {Rastinejad}}]{Hajela+22}
{Hajela}, A., {Margutti}, R., {Bright}, J.~S., {et~al.} 2022, \apjl, 927, L17, \dodoi{10.3847/2041-8213/ac504a}

\bibitem[{{Hallinan} {et~al.}(2017){Hallinan}, {Corsi}, {Mooley}, {Hotokezaka}, {Nakar}, {Kasliwal}, {Kaplan}, {Frail}, {Myers}, {Murphy}, {De}, {Dobie}, {Allison}, {Bannister}, {Bhalerao}, {Chandra}, {Clarke}, {Giacintucci}, {Ho}, {Horesh}, {Kassim}, {Kulkarni}, {Lenc}, {Lockman}, {Lynch}, {Nichols}, {Nissanke}, {Palliyaguru}, {Peters}, {Piran}, {Rana}, {Sadler}, \& {Singer}}]{Hallinan17}
{Hallinan}, G., {Corsi}, A., {Mooley}, K.~P., {et~al.} 2017, Science, 358, 1579, \dodoi{10.1126/science.aap9855}

\bibitem[{Harris {et~al.}(2020)Harris, Millman, van~der Walt, Gommers, Virtanen, Cournapeau, Wieser, Taylor, Berg, Smith, Kern, Picus, Hoyer, van Kerkwijk, Brett, Haldane, Fernández~del Río, Wiebe, Peterson, Gérard-Marchant, Sheppard, Reddy, Weckesser, Abbasi, Gohlke, \& Oliphant}]{2020NumPy-Array}
Harris, C.~R., Millman, K.~J., van~der Walt, S.~J., {et~al.} 2020, Nature, 585, 357–362, \dodoi{10.1038/s41586-020-2649-2}

\bibitem[{{Hotokezaka} {et~al.}(2013){Hotokezaka}, {Kiuchi}, {Kyutoku}, {Okawa}, {Sekiguchi}, {Shibata}, \& {Taniguchi}}]{Hotokezaka+13a}
{Hotokezaka}, K., {Kiuchi}, K., {Kyutoku}, K., {et~al.} 2013, \prd, 87, 024001, \dodoi{10.1103/PhysRevD.87.024001}

\bibitem[{{Hotokezaka} {et~al.}(2011){Hotokezaka}, {Kyutoku}, {Okawa}, {Shibata}, \& {Kiuchi}}]{Hotokezaka+11PRD}
{Hotokezaka}, K., {Kyutoku}, K., {Okawa}, H., {Shibata}, M., \& {Kiuchi}, K. 2011, \prd, 83, 124008, \dodoi{10.1103/PhysRevD.83.124008}

\bibitem[{{Hotokezaka} {et~al.}(2018){Hotokezaka}, {Nakar}, {Gottlieb}, {Nissanke}, {Masuda}, {Hallinan}, {Mooley}, \& {Deller}}]{Hotokezaka18}
{Hotokezaka}, K., {Nakar}, E., {Gottlieb}, O., {et~al.} 2018, ArXiv e-prints.
\newblock \doarXiv{1806.10596}

\bibitem[{{Hotokezaka} {et~al.}(2017){Hotokezaka}, {Sari}, \& {Piran}}]{Hotokezaka_Sari_Piran17}
{Hotokezaka}, K., {Sari}, R., \& {Piran}, T. 2017, \mnras, 468, 91, \dodoi{10.1093/mnras/stx411}

\bibitem[{{Hunter}(2007)}]{Hunter07_matplotlib}
{Hunter}, J.~D. 2007, Computing in Science and Engineering, 9, 90, \dodoi{10.1109/MCSE.2007.55}

\bibitem[{{Just} {et~al.}(2015){Just}, {Bauswein}, {Ardevol Pulpillo}, {Goriely}, \& {Janka}}]{Just+15}
{Just}, O., {Bauswein}, A., {Ardevol Pulpillo}, R., {Goriely}, S., \& {Janka}, H.~T. 2015, \mnras, 448, 541, \dodoi{10.1093/mnras/stv009}

\bibitem[{{Kilpatrick} {et~al.}(2017){Kilpatrick}, {Foley}, {Kasen}, {Murguia-Berthier}, {Ramirez-Ruiz}, {Coulter}, {Drout}, {Piro}, {Shappee}, {Boutsia}, {Contreras}, {Di Mille}, {Madore}, {Morrell}, {Pan}, {Prochaska}, {Rest}, {Rojas-Bravo}, {Siebert}, {Simon}, \& {Ulloa}}]{Kilpatrick17:gw}
{Kilpatrick}, C.~D., {Foley}, R.~J., {Kasen}, D., {et~al.} 2017, Science.
\newblock \doarXiv{1710.05434}

\bibitem[{{Kiuchi} {et~al.}(2018){Kiuchi}, {Kyutoku}, {Sekiguchi}, \& {Shibata}}]{Kiuchi+18}
{Kiuchi}, K., {Kyutoku}, K., {Sekiguchi}, Y., \& {Shibata}, M. 2018, \prd, 97, 124039, \dodoi{10.1103/PhysRevD.97.124039}

\bibitem[{{Kiuchi} {et~al.}(2014){Kiuchi}, {Kyutoku}, {Sekiguchi}, {Shibata}, \& {Wada}}]{Kiuchi+14}
{Kiuchi}, K., {Kyutoku}, K., {Sekiguchi}, Y., {Shibata}, M., \& {Wada}, T. 2014, \prd, 90, 041502, \dodoi{10.1103/PhysRevD.90.041502}

\bibitem[{{Kiuchi} {et~al.}(2019){Kiuchi}, {Kyutoku}, {Shibata}, \& {Taniguchi}}]{Kiuchi+2019}
{Kiuchi}, K., {Kyutoku}, K., {Shibata}, M., \& {Taniguchi}, K. 2019, \apjl, 876, L31, \dodoi{10.3847/2041-8213/ab1e45}

\bibitem[{{Konstantinou} \& {Morsink}(2022)}]{Konstantinou&Morsink22}
{Konstantinou}, A., \& {Morsink}, S.~M. 2022, \apj, 934, 139, \dodoi{10.3847/1538-4357/ac7b86}

\bibitem[{{Koranda} {et~al.}(1997){Koranda}, {Stergioulas}, \& {Friedman}}]{Koranda+97}
{Koranda}, S., {Stergioulas}, N., \& {Friedman}, J.~L. 1997, \apj, 488, 799, \dodoi{10.1086/304714}

\bibitem[{{Korobkin} {et~al.}(2020){Korobkin}, {Wollaeger}, {Fryer}, {Hungerford}, {Rosswog}, {Fontes}, {Mumpower}, {Chase}, {Even}, {Miller}, {Misch}, \& {Lippuner}}]{Korobkin+20}
{Korobkin}, O., {Wollaeger}, R., {Fryer}, C., {et~al.} 2020, arXiv e-prints, arXiv:2004.00102.
\newblock \doarXiv{2004.00102}

\bibitem[{{Krolik}(1999)}]{Krolik1999}
{Krolik}, J.~H. 1999, {Active Galactic Nuclei. From the Central Black Hole to the Galactic Environment}

\bibitem[{{Kulkarni}(2005)}]{Kulkarni05}
{Kulkarni}, S.~R. 2005, arXiv e-prints, astro, \dodoi{10.48550/arXiv.astro-ph/0510256}

\bibitem[{{Lasota} {et~al.}(1996){Lasota}, {Haensel}, \& {Abramowicz}}]{Lasota+96}
{Lasota}, J.-P., {Haensel}, P., \& {Abramowicz}, M.~A. 1996, \apj, 456, 300, \dodoi{10.1086/176650}

\bibitem[{{Lattimer}(2019)}]{Lattimer19}
{Lattimer}, J.~M. 2019, Universe, 5, 159, \dodoi{10.3390/universe5070159}

\bibitem[{{Lattimer} \& {Prakash}(2007)}]{LattimerPrakash07_EoS}
{Lattimer}, J.~M., \& {Prakash}, M. 2007, \physrep, 442, 109, \dodoi{10.1016/j.physrep.2007.02.003}

\bibitem[{{Lattimer} \& {Schramm}(1974)}]{Lattimer_Schramm74}
{Lattimer}, J.~M., \& {Schramm}, D.~N. 1974, \apjl, 192, L145, \dodoi{10.1086/181612}

\bibitem[{{Lehner} {et~al.}(2016){Lehner}, {Liebling}, {Palenzuela}, {Caballero}, {O'Connor}, {Anderson}, \& {Neilsen}}]{Lehner+16}
{Lehner}, L., {Liebling}, S.~L., {Palenzuela}, C., {et~al.} 2016, Classical and Quantum Gravity, 33, 184002, \dodoi{10.1088/0264-9381/33/18/184002}

\bibitem[{{Li} \& {Paczy{\'n}ski}(1998)}]{Li98}
{Li}, L.-X., \& {Paczy{\'n}ski}, B. 1998, \apjl, 507, L59, \dodoi{10.1086/311680}

\bibitem[{Loffler {et~al.}(2012)}]{Loffler:2011ay}
Loffler, F., {et~al.} 2012, Class. Quant. Grav., 29, 115001, \dodoi{10.1088/0264-9381/29/11/115001}

\bibitem[{{Lovelace} {et~al.}(2008){Lovelace}, {Owen}, {Pfeiffer}, \& {Chu}}]{Lovelace+08}
{Lovelace}, G., {Owen}, R., {Pfeiffer}, H.~P., \& {Chu}, T. 2008, \prd, 78, 084017, \dodoi{10.1103/PhysRevD.78.084017}

\bibitem[{{Margalit} \& {Metzger}(2017)}]{MargalitMetzger17}
{Margalit}, B., \& {Metzger}, B.~D. 2017, \apjl, 850, L19, \dodoi{10.3847/2041-8213/aa991c}

\bibitem[{{Margutti} \& {Chornock}(2021)}]{Margutti_Chornock21}
{Margutti}, R., \& {Chornock}, R. 2021, \araa, 59, 155, \dodoi{10.1146/annurev-astro-112420-030742}

\bibitem[{{Margutti} {et~al.}(2017){Margutti}, {Berger}, {Fong}, {Guidorzi}, {Alexander}, {Metzger}, {Blanchard}, {Cowperthwaite}, {Chornock}, {Eftekhari}, {Nicholl}, {Villar}, {Williams}, {Annis}, {Brown}, {Chen}, {Doctor}, {Frieman}, {Holz}, {Sako}, \& {Soares-Santos}}]{Margutti+17}
{Margutti}, R., {Berger}, E., {Fong}, W., {et~al.} 2017, \apjl, 848, L20, \dodoi{10.3847/2041-8213/aa9057}

\bibitem[{{Margutti} {et~al.}(2018){Margutti}, {Alexander}, {Xie}, {Sironi}, {Metzger}, {Kathirgamaraju}, {Fong}, {Blanchard}, {Berger}, {MacFadyen}, {Giannios}, {Guidorzi}, {Hajela}, {Chornock}, {Cowperthwaite}, {Eftekhari}, {Nicholl}, {Villar}, {Williams}, \& {Zrake}}]{Margutti18}
{Margutti}, R., {Alexander}, K.~D., {Xie}, X., {et~al.} 2018, \apjl, 856, L18, \dodoi{10.3847/2041-8213/aab2ad}

\bibitem[{{Metzger} {et~al.}(2010){Metzger}, {Mart{\'\i}nez-Pinedo}, {Darbha}, {Quataert}, {Arcones}, {Kasen}, {Thomas}, {Nugent}, {Panov}, \& {Zinner}}]{Metzger+10_KN}
{Metzger}, B.~D., {Mart{\'\i}nez-Pinedo}, G., {Darbha}, S., {et~al.} 2010, \mnras, 406, 2650, \dodoi{10.1111/j.1365-2966.2010.16864.x}

\bibitem[{{Miller} {et~al.}(2019){Miller}, {Ryan}, {Dolence}, {Burrows}, {Fontes}, {Fryer}, {Korobkin}, {Lippuner}, {Mumpower}, \& {Wollaeger}}]{Miller+19}
{Miller}, J.~M., {Ryan}, B.~R., {Dolence}, J.~C., {et~al.} 2019, \prd, 100, 023008, \dodoi{10.1103/PhysRevD.100.023008}

\bibitem[{{Mooley} {et~al.}(2018{\natexlab{a}}){Mooley}, {Nakar}, {Hotokezaka}, {Hallinan}, {Corsi}, {Frail}, {Horesh}, {Murphy}, {Lenc}, {Kaplan}, {de}, {Dobie}, {Chandra}, {Deller}, {Gottlieb}, {Kasliwal}, {Kulkarni}, {Myers}, {Nissanke}, {Piran}, {Lynch}, {Bhalerao}, {Bourke}, {Bannister}, \& {Singer}}]{Mooley+18a_Nat}
{Mooley}, K.~P., {Nakar}, E., {Hotokezaka}, K., {et~al.} 2018{\natexlab{a}}, \nat, 554, 207, \dodoi{10.1038/nature25452}

\bibitem[{{Mooley} {et~al.}(2018{\natexlab{b}}){Mooley}, {Deller}, {Gottlieb}, {Nakar}, {Hallinan}, {Bourke}, {Frail}, {Horesh}, {Corsi}, \& {Hotokezaka}}]{Mooley2018b}
{Mooley}, K.~P., {Deller}, A.~T., {Gottlieb}, O., {et~al.} 2018{\natexlab{b}}, ArXiv e-prints.
\newblock \doarXiv{1806.09693}

\bibitem[{{Most} {et~al.}(2021){Most}, {Papenfort}, {Tootle}, \& {Rezzolla}}]{Most+21_FastEjecta}
{Most}, E.~R., {Papenfort}, L.~J., {Tootle}, S.~D., \& {Rezzolla}, L. 2021, \apj, 912, 80, \dodoi{10.3847/1538-4357/abf0a5}

\bibitem[{{Most} \& {Quataert}(2023)}]{Most-Quataert2023}
{Most}, E.~R., \& {Quataert}, E. 2023, \apjl, 947, L15, \dodoi{10.3847/2041-8213/acca84}

\bibitem[{{Murguia-Berthier} {et~al.}(2021){Murguia-Berthier}, {Noble}, {Roberts}, {Ramirez-Ruiz}, {Werneck}, {Kolacki}, {Etienne}, {Avara}, {Campanelli}, {Ciolfi}, {Cipolletta}, {Drachler}, {Ennoggi}, {Faber}, {Fiacco}, {Giacomazzo}, {Gupte}, {Ha}, {Kelly}, {Krolik}, {Lopez Armengol}, {Margalit}, {Moon}, {O'Shaughnessy}, {Rueda-Becerril}, {Schnittman}, {Zenati}, \& {Zlochower}}]{AriM+21}
{Murguia-Berthier}, A., {Noble}, S.~C., {Roberts}, L.~F., {et~al.} 2021, \apj, 919, 95, \dodoi{10.3847/1538-4357/ac1119}

\bibitem[{{Nakar}(2007)}]{Nakar07}
{Nakar}, E. 2007, \physrep, 442, 166, \dodoi{10.1016/j.physrep.2007.02.005}

\bibitem[{{Nedora} {et~al.}(2021){Nedora}, {Bernuzzi}, {Radice}, {Daszuta}, {Endrizzi}, {Perego}, {Prakash}, {Safarzadeh}, {Schianchi}, \& {Logoteta}}]{Nedora+21}
{Nedora}, V., {Bernuzzi}, S., {Radice}, D., {et~al.} 2021, \apj, 906, 98, \dodoi{10.3847/1538-4357/abc9be}

\bibitem[{Noble {et~al.}(2009)Noble, Krolik, \& Hawley}]{Noble:2008tm}
Noble, S.~C., Krolik, J.~H., \& Hawley, J.~F. 2009, Astrophys. J., 692, 411, \dodoi{10.1088/0004-637X/692/1/411}

\bibitem[{O'Connor \& Ott(2010)}]{OConnor:2009iuz}
O'Connor, E., \& Ott, C.~D. 2010, Class. Quant. Grav., 27, 114103, \dodoi{10.1088/0264-9381/27/11/114103}

\bibitem[{{Oechslin} \& {Janka}(2006)}]{Oechslin_Janka06}
{Oechslin}, R., \& {Janka}, H.~T. 2006, \mnras, 368, 1489, \dodoi{10.1111/j.1365-2966.2006.10238.x}

\bibitem[{{Oechslin} \& {Janka}(2007)}]{OechslinJanka07PRL}
---. 2007, \prl, 99, 121102, \dodoi{10.1103/PhysRevLett.99.121102}

\bibitem[{{{\"O}zel} \& {Freire}(2016)}]{Ozel&Freire16}
{{\"O}zel}, F., \& {Freire}, P. 2016, \araa, 54, 401, \dodoi{10.1146/annurev-astro-081915-023322}

\bibitem[{{Pian}(2023)}]{Pian2023}
{Pian}, E. 2023, Universe, 9, 105, \dodoi{10.3390/universe9020105}

\bibitem[{{Radice} {et~al.}(2020){Radice}, {Bernuzzi}, \& {Perego}}]{Radice+20_AnnRev}
{Radice}, D., {Bernuzzi}, S., \& {Perego}, A. 2020, Annual Review of Nuclear and Particle Science, 70, 95, \dodoi{10.1146/annurev-nucl-013120-114541}

\bibitem[{{Radice} {et~al.}(2016){Radice}, {Galeazzi}, {Lippuner}, {Roberts}, {Ott}, \& {Rezzolla}}]{Radice+16}
{Radice}, D., {Galeazzi}, F., {Lippuner}, J., {et~al.} 2016, \mnras, 460, 3255, \dodoi{10.1093/mnras/stw1227}

\bibitem[{{Radice} {et~al.}(2018{\natexlab{a}}){Radice}, {Perego}, {Zappa}, \& {Bernuzzi}}]{Radice+18_NSEoS}
{Radice}, D., {Perego}, A., {Zappa}, F., \& {Bernuzzi}, S. 2018{\natexlab{a}}, \apjl, 852, L29, \dodoi{10.3847/2041-8213/aaa402}

\bibitem[{{Radice} {et~al.}(2018{\natexlab{b}}){Radice}, {Perego}, {Zappa}, \& {Bernuzzi}}]{Radice_Bernuzzi+18}
---. 2018{\natexlab{b}}, \apjl, 852, L29, \dodoi{10.3847/2041-8213/aaa402}

\bibitem[{{Read} {et~al.}(2009){Read}, {Lackey}, {Owen}, \& {Friedman}}]{ReadJ+09}
{Read}, J.~S., {Lackey}, B.~D., {Owen}, B.~J., \& {Friedman}, J.~L. 2009, \prd, 79, 124032, \dodoi{10.1103/PhysRevD.79.124032}

\bibitem[{{Roberts} {et~al.}(2011){Roberts}, {Kasen}, {Lee}, \& {Ramirez-Ruiz}}]{Roberts+11_KN}
{Roberts}, L.~F., {Kasen}, D., {Lee}, W.~H., \& {Ramirez-Ruiz}, E. 2011, \apjl, 736, L21, \dodoi{10.1088/2041-8205/736/1/L21}

\bibitem[{{Rosswog} \& {Davies}(2002{\natexlab{a}})}]{Rosswog2002}
{Rosswog}, S., \& {Davies}, M.~B. 2002{\natexlab{a}}, \mnras, 334, 481, \dodoi{10.1046/j.1365-8711.2002.05409.x}

\bibitem[{{Rosswog} \& {Davies}(2002{\natexlab{b}})}]{Rosswog&Davies02}
---. 2002{\natexlab{b}}, \mnras, 334, 481, \dodoi{10.1046/j.1365-8711.2002.05409.x}

\bibitem[{{Rosswog} \& {Korobkin}(2024)}]{Rosswog&Korobkin24}
{Rosswog}, S., \& {Korobkin}, O. 2024, Annalen der Physik, 536, 2200306, \dodoi{10.1002/andp.202200306}

\bibitem[{{Rosswog} \& {Ramirez-Ruiz}(2002)}]{Rosswog_Ramirez-Ruiz02}
{Rosswog}, S., \& {Ramirez-Ruiz}, E. 2002, \mnras, 336, L7, \dodoi{10.1046/j.1365-8711.2002.05898.x}

\bibitem[{{Ruiz} {et~al.}(2016){Ruiz}, {Lang}, {Paschalidis}, \& {Shapiro}}]{Ruiz+16}
{Ruiz}, M., {Lang}, R.~N., {Paschalidis}, V., \& {Shapiro}, S.~L. 2016, \apjl, 824, L6, \dodoi{10.3847/2041-8205/824/1/L6}

\bibitem[{{Ruiz} {et~al.}(2021){Ruiz}, {Shapiro}, \& {Tsokaros}}]{Ruiz+21}
{Ruiz}, M., {Shapiro}, S.~L., \& {Tsokaros}, A. 2021, Frontiers in Astronomy and Space Sciences, 8, 39, \dodoi{10.3389/fspas.2021.656907}

\bibitem[{{Sarin} \& {Lasky}(2021)}]{Sarin_Lasky21}
{Sarin}, N., \& {Lasky}, P.~D. 2021, General Relativity and Gravitation, 53, 59, \dodoi{10.1007/s10714-021-02831-1}

\bibitem[{Schnetter {et~al.}(2004)Schnetter, Hawley, \& Hawke}]{Schnetter:2003rb}
Schnetter, E., Hawley, S.~H., \& Hawke, I. 2004, Class. Quant. Grav., 21, 1465, \dodoi{10.1088/0264-9381/21/6/014}

\bibitem[{{Sekiguchi} {et~al.}(2016){Sekiguchi}, {Kiuchi}, {Kyutoku}, {Shibata}, \& {Taniguchi}}]{Sekiguchi+16}
{Sekiguchi}, Y., {Kiuchi}, K., {Kyutoku}, K., {Shibata}, M., \& {Taniguchi}, K. 2016, \prd, 93, 124046, \dodoi{10.1103/PhysRevD.93.124046}

\bibitem[{{Shibata} {et~al.}(2023){Shibata}, {Fujibayashi}, {Hayashi}, {Kiuchi}, \& {Wanajo}}]{Shibata+23}
{Shibata}, M., {Fujibayashi}, S., {Hayashi}, K., {Kiuchi}, K., \& {Wanajo}, S. 2023, IAU Symposium, 362, 190, \dodoi{10.1017/S1743921322001351}

\bibitem[{{Shibata} \& {Hotokezaka}(2019)}]{Shibata_Kenta19}
{Shibata}, M., \& {Hotokezaka}, K. 2019, Annual Review of Nuclear and Particle Science, 69, 41, \dodoi{10.1146/annurev-nucl-101918-023625}

\bibitem[{{Siegel} \& {Metzger}(2017)}]{SiegelMetzger17PRL}
{Siegel}, D.~M., \& {Metzger}, B.~D. 2017, \prl, 119, 231102, \dodoi{10.1103/PhysRevLett.119.231102}

\bibitem[{{Smarr}(1973)}]{SmarrL1973}
{Smarr}, L. 1973, \prd, 7, 289, \dodoi{10.1103/PhysRevD.7.289}

\bibitem[{{Troja} {et~al.}(2017){Troja}, {Piro}, {van Eerten}, {Wollaeger}, {Im}, {Fox}, {Butler}, {Cenko}, {Sakamoto}, {Fryer}, {Ricci}, {Lien}, {Ryan}, {Korobkin}, {Lee}, {Burgess}, {Lee}, {Watson}, {Choi}, {Covino}, {D'Avanzo}, {Fontes}, {Gonz{\'a}lez}, {Khandrika}, {Kim}, {Kim}, {Lee}, {Lee}, {Kutyrev}, {Lim}, {S{\'a}nchez-Ram{\'\i}rez}, {Veilleux}, {Wieringa}, \& {Yoon}}]{Troja+17Nat}
{Troja}, E., {Piro}, L., {van Eerten}, H., {et~al.} 2017, \nat, 551, 71, \dodoi{10.1038/nature24290}

\bibitem[{{Villar} {et~al.}(2017){Villar}, {Guillochon}, {Berger}, {Metzger}, {Cowperthwaite}, {Nicholl}, {Alexander}, {Blanchard}, {Chornock}, {Eftekhari}, {Fong}, {Margutti}, \& {Williams}}]{Villar17}
{Villar}, V.~A., {Guillochon}, J., {Berger}, E., {et~al.} 2017, \apjl, 851, L21, \dodoi{10.3847/2041-8213/aa9c84}

\bibitem[{{Vincent} {et~al.}(2020){Vincent}, {Foucart}, {Duez}, {Haas}, {Kidder}, {Pfeiffer}, \& {Scheel}}]{Vincent+20}
{Vincent}, T., {Foucart}, F., {Duez}, M.~D., {et~al.} 2020, \prd, 101, 044053, \dodoi{10.1103/PhysRevD.101.044053}

\bibitem[{{Wei} {et~al.}(2019){Wei}, {Figura}, {Burgio}, {Chen}, \& {Schulze}}]{WeiJ+19}
{Wei}, J.~B., {Figura}, A., {Burgio}, G.~F., {Chen}, H., \& {Schulze}, H.~J. 2019, Journal of Physics G Nuclear Physics, 46, 034001, \dodoi{10.1088/1361-6471/aaf95c}

\bibitem[{{Werneck} {et~al.}(2023){Werneck}, {Etienne}, {Murguia-Berthier}, {Haas}, {Cipolletta}, {Noble}, {Ennoggi}, {Lopez Armengol}, {Giacomazzo}, {Assump{\c{c}}{\~a}o}, {Faber}, {Gupte}, {Kelly}, \& {Krolik}}]{Werneck+23}
{Werneck}, L.~R., {Etienne}, Z.~B., {Murguia-Berthier}, A., {et~al.} 2023, \prd, 107, 044037, \dodoi{10.1103/PhysRevD.107.044037}

\bibitem[{{Zenati} {et~al.}(2023){Zenati}, {Krolik}, {Werneck}, {Murguia-Berthier}, {Etienne}, {Noble}, \& {Piran}}]{zenati+23BNS}
{Zenati}, Y., {Krolik}, J.~H., {Werneck}, L.~R., {et~al.} 2023, \apj, 958, 161, \dodoi{10.3847/1538-4357/acf714}

\end{thebibliography}
\bibliographystyle{aasjournal}
\end{document}